\documentclass[12pt]{article}
\pdfoutput=1
\usepackage{amsmath, amsthm,comment}
\usepackage{amsfonts}
\usepackage{amssymb,graphics,psfrag}
\usepackage{array,epsfig,stmaryrd,graphicx}
\usepackage{cite}
\usepackage{graphicx}
\usepackage{makeidx}
\usepackage{multicol}
\usepackage{geometry}
\usepackage{amsfonts}
\usepackage{mathrsfs}
\usepackage{amssymb}
\usepackage{amsmath}
\usepackage{slashed}
\usepackage{heck}%
\providecommand{\U}[1]{\protect\rule{.1in}{.1in}}
\pdfoutput=1

\newcommand{\beq}{\begin{equation}}
\newcommand{\eeq}{\end{equation}}
\newcommand{\be}{\begin{equation}}
\newcommand{\ee}{\end{equation}}
\newcommand{\bea}{\begin{eqnarray}}
\newcommand{\eea}{\end{eqnarray}}
\newcommand{\ben}{\begin{eqnarray*}}
\newcommand{\een}{\end{eqnarray*}}
\newcommand{\ba}{\begin{aligned}}
\newcommand{\ea}{\end{aligned}}
\newcommand{\bt}{\begin{tabular}}
\newcommand{\et}{\end{tabular}}
\newcommand{\bc}{\begin{center}}
\newcommand{\ec}{\end{center}}

\newcommand{\cref}{{\bf [check ref]}}

\newcommand{\bs}{\begin{subarray}{c}}
\newcommand{\es}{\end{subarray}}

\numberwithin{equation}{section}
\begin{document}

\date{June, 2009}
\title{The Point of $E_{8}$ in F-theory GUTs}

\preprint{arXiv:0906.0581}

\institution{HarvardU}{\centerline{Jefferson Physical Laboratory, Harvard University, Cambridge,
MA 02138, USA}}

\authors{Jonathan J. Heckman\footnote{e-mail: {\tt
jheckman@fas.harvard.edu}},
Alireza Tavanfar\footnote{e-mail: {\tt
tavanfar@physics.harvard.edu}}, and Cumrun
Vafa\footnote{e-mail: {\tt
vafa@physics.harvard.edu}}}

\abstract{We show that in F-theory GUTs, a natural explanation of flavor hierarchies in the quark and
lepton sector requires a single point of $E_8$
enhancement in the internal geometry, from which all Yukawa couplings originate. The
monodromy group acting on the seven-brane configuration plays a key role in this analysis.
Moreover, the $E_8$ structure automatically leads to the existence of the additional fields and interactions needed
for minimal gauge mediated supersymmetry breaking, \textit{and almost nothing else}. Surprisingly, we find
that in all but one Dirac neutrino scenario the messenger fields in the gauge mediated supersymmetry breaking sector transform as vector-like pairs in the $10 \oplus \overline{10}$ of $SU(5)$.
We also classify dark matter candidates available from this enhancement point, and rule
out both annihilating and decaying dark matter scenarios as explanations for the recent
experiments PAMELA, ATIC and FERMI. In F-theory GUT models, a
$10-100$ MeV mass gravitino remains as the prime candidate for
dark matter, thus suggesting an astrophysical origin for recent experimental signals.}%

\maketitle
\tableofcontents

\section{Introduction}

The paradigm of unification provides a compelling and predictive framework for
high energy physics. In the context of string theory, this might at first
suggest that gauge and gravitational degrees of freedom should also unify.
Nevertheless, in constructions of gauge theories within string theory, there
is often a limit where gravity decouples. This is because gauge theory degrees
of freedom can localize on subspaces of the internal geometry. Moreover, the
fact that $M_{GUT}/M_{Planck}\ll1$ suggests that the existence of a limit
where gravity decouples may also be relevant for the gauge theory defined by
the Standard Model.

Flexibility in potential model building applications suggests branes with
maximal dimension such that a consistent decoupling limit exists. Such
considerations point to seven-branes, and thus type IIB string theory. On the
other hand, the requisite elements of Grand Unified Theories (GUTs), such as
the $5_{H}\times10_{M}\times10_{M}$ interaction require E-type gauge theory
structures, which is incompatible with perturbative IIB\ string theory.
Importantly, however, the strong coupling limit of IIB strings, namely
F-theory is flexible enough to accommodate such elements. Recent work on
F-theory GUTs in
\cite{BHVI,BHVII,HVGMSB,HVLHC,HVCKM,FGUTSCosmo,HKSV,BHSV,HVCP} (see also
\cite{DonagiWijnholt,WatariTATARHETF,HMSSNV,MarsanoGMSB,DonagiWijnholtBreak,MarsanoToolbox,Font:2008id,Blumenhagen:2008zz,Blumenhagen:2008aw,Bourjaily:2009vf,Hayashi:2009ge,Andreas:2009uf,Chen:2009me,DonagiWijnholtIII,Randall:2009dw,Bourjaily:2009ci,Tatar:2009jk,Jiang:2009za,Collinucci:2009uh,Blumenhagen:2009up}%
) has shown that many realistic features of particle phenomenology naturally
emerge, with potentially observable consequences for the LHC \cite{HVLHC,HKSV}
and upcoming neutrino experiments \cite{BHSV}.

\begin{figure}[ptb]
\begin{center}
\includegraphics[
natheight=5.807200in,
natwidth=9.751600in,
height=2.4007in,
width=4.0214in
]{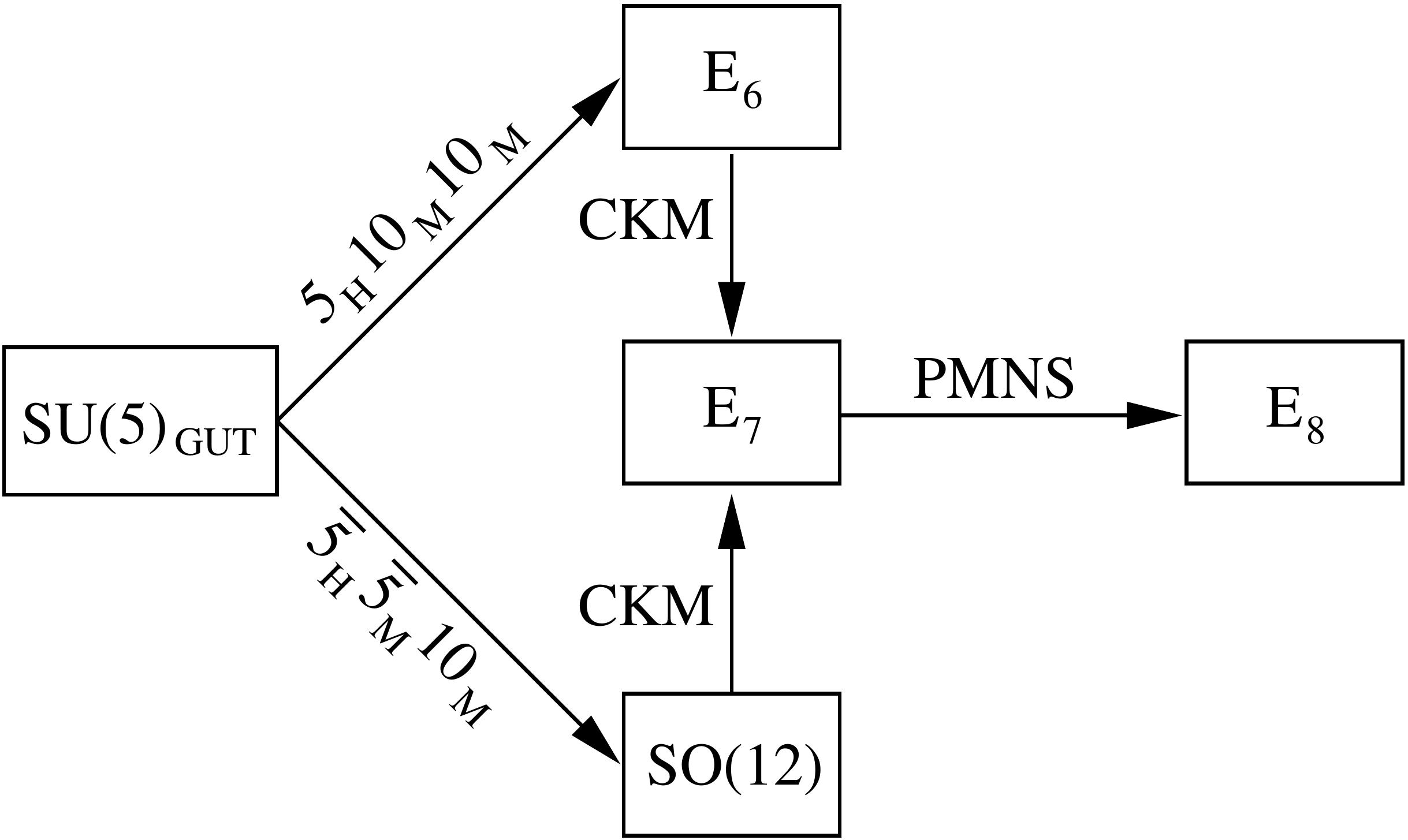}
\end{center}
\caption{Starting from a GUT seven-brane with $SU(5)$ symmetry, each
additional phenomenological condition leads to a further jump in the rank at a
point of the geometry. Including the $5_{H}10_{M}10_{M}$ interaction requires
an $E_{6}$ point, and the $\overline{5}_{H}\overline{5}_{M}10_{M}$ interaction
requires an $SO(12)$ point. A hierarchical CKM matrix then suggests that these
points should also unify to an $E_{7}$ point of enhancement. Incorporating
leptonic mixing structure pushes this all the way up to $E_{8}$.}%
\label{flowchart}%
\end{figure}

Low energy constraints impose important conditions that the internal geometry
must satisfy, and point towards the especially important role of exceptional
groups in F-theory GUTs. For example, the existence of an order one top quark
Yukawa requires a point where the singularity type of the geometry enhances to
$E_{6}$ \cite{BHVI} (see also \cite{TatarClassify}). The existence of higher
unification structures is also important in the context of flavor physics. For
example, the CKM\ matrix exhibits a hierarchical structure provided the up and
down type Yukawas localize at points which are sufficiently close
\cite{HVCKM}, which is suggestive of a single point of at least $E_{7}$
enhancement. As we show in this paper, incorporating a minimal neutrino sector
with a mildly hierarchical lepton mixing matrix (PMNS matrix) pushes this all
the way up to $E_{8}$. It is in principle possible to construct models where a
globally well-defined $E_{8}$ structure plays no special role, and in which
flavor hierarchies are solved through fine tuning. Even so, the most natural
option, and the option which flavor hierarchies dictate is the existence of a
\textit{single} $E_{8}$ \textit{enhanced symmetry point in the internal
geometry from which all the interactions descend}. In this paper we will
assume this is indeed the case. In fact in this paper we classify
all the minimal F-theory GUT scenarios which descend from a single $E_{8}$
point, which turn out to be interestingly predictive. See figure
\ref{flowchart} for a depiction of how various enhancement points each demand
a higher unification structure. 

Cosmological considerations provide another window into the physics of
F-theory GUTs. The cosmology of such models neatly avoids many of the problems
which sometimes afflict supersymmetric models \cite{FGUTSCosmo}. For example, assuming the supersymmetry breaking
scenario of \cite{HVGMSB}, the mass of the gravitino is $10-100$ MeV.\footnote{As in early work on local
F-theory GUTs such as \cite{HVGMSB}, we assume that the dominant contribution to supersymmetry breaking originates
from gauge mediated supersymmetry, and ingredients present in the local model.}
It is known that in other contexts this typically leads to
over-production of such particles. This issue is bypassed in F-theory GUTs
because the saxion comes to dominate the energy density of the Universe. The
subsequent decay of the saxion dilutes this relic abundance, leaving
gravitinos instead as a prominent component of dark matter. In a certain
regime of parameters, axionic dark matter is also possible. Both the gravitino
and axion interact very weakly with the Standard Model, and are therefore
unlikely to be detected. On the other hand, recent results from experiments
such as PAMELA \cite{Adriani:2008zq,Adriani:2008zr}, ATIC \cite{ATIC},
PPB-BETS \cite{PPB}, HESS \cite{HESSone} and FERMI \cite{FERMI} can
potentially be explained by weak to TeV scale dark matter which interacts more
strongly with the Standard Model. One of the aims of this paper is to address
whether there are any additional dark matter candidates in minimal F-theory
GUT models besides the gravitino and axion.

\begin{figure}
[ptb]
\begin{center}
\includegraphics[
natheight=9.784500in,
natwidth=13.311200in,
height=4.3708in,
width=5.9361in
]%
{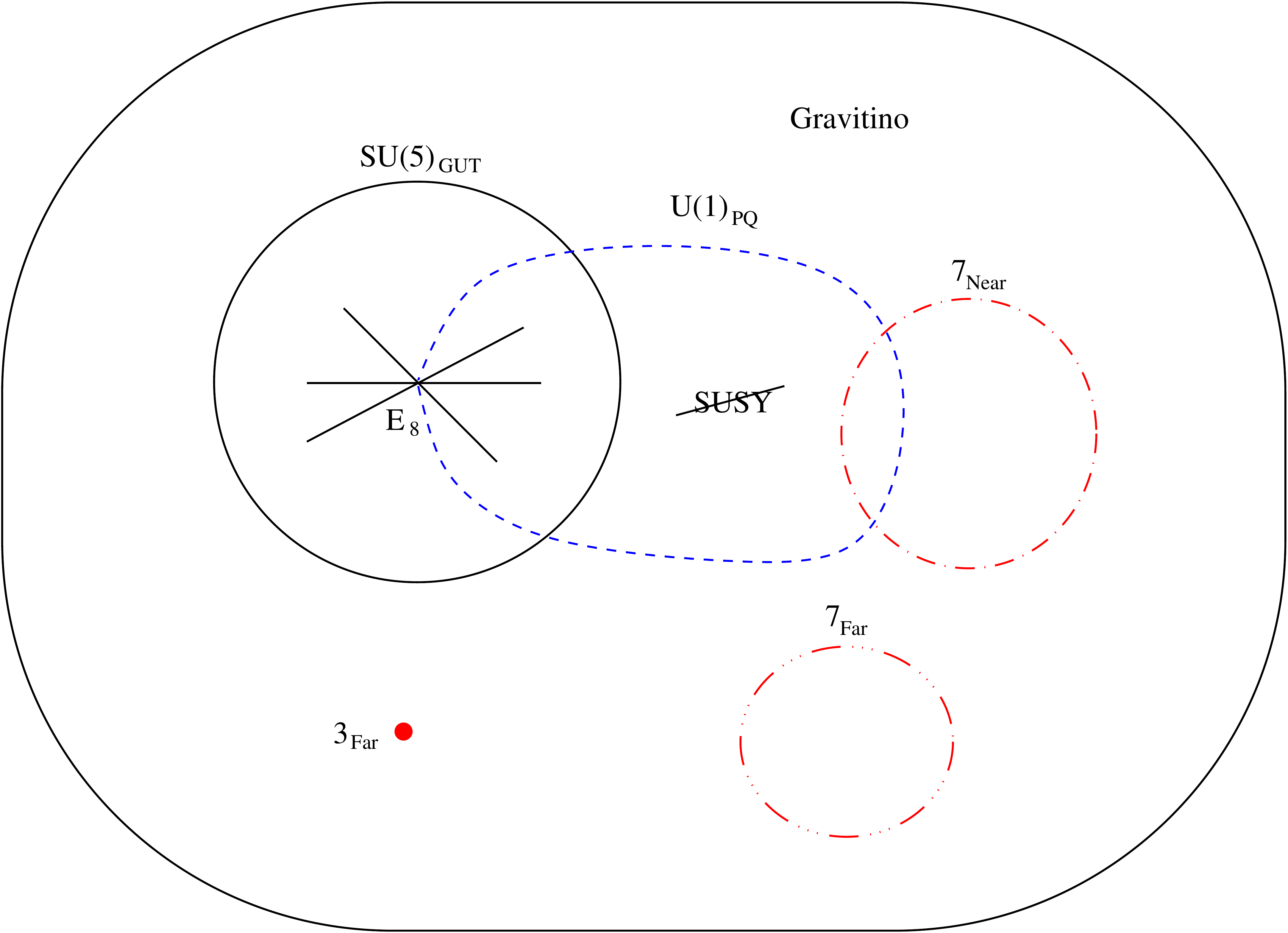}%
\caption{Depiction of the $SU(5)_{GUT}$\ seven-brane, the $U(1)_{PQ}$
seven-brane, and other possible branes and bulk modes which can in principle
appear as \textquotedblleft dark objects\textquotedblright\ in F-theory GUTs.
Here, the $7_{\text{Near}}$-branes share a mutual intersection with the
$SU(5)_{GUT}$\ seven-brane along the PQ seven-brane. As in \cite{HVGMSB},
supersymmetry breaking occurs due to dynamics localized on the PQ seven-brane.
In addition, we have also included the possibility of additional three-branes
and seven-branes, $3_{\text{Far}}$ and $7_{\text{Far}}$, as well as bulk
modes, such as the gravitino. These latter possibilities are less directly
connected to the supersymmetry breaking sector, and so typically do not
contain weak to TeV scale dark matter candidates.}%
\label{efarnear}%
\end{center}
\end{figure}

The matter content of F-theory GUTs roughly divides into the degrees of
freedom on the GUT\ seven-brane, nearby seven-branes which share a mutual
intersection with the GUT seven-brane, and branes which are far away in the
sense that they do not directly couple to the GUT\ sector. In addition, there
are also degrees of freedom which do not localize on a brane but instead
propagate in the bulk of the geometry. Since the dynamics of supersymmetry
breaking takes place due to the dynamics of a\ seven-brane with Peccei-Quinn
(PQ) gauge symmetry, weak to TeV scale dark matter candidates must also be
relatively nearby. See figure \ref{efarnear} for a depiction of the possible
ingredients which can in principle participate in this construction.%

Assuming a single $E_{8}$ enhancement point in our local patch, we classify all visible and dark
matter which can descend from the adjoint of $E_{8}$. Phenomenological
requirements then lead to a rich interplay between group theoretic and
geometric conditions considerations. In particular, having available suitable
curves which accommodate the Higgs and the matter fields severely restricts
the possibilities. Moreover, even though $E_{8}$ contains the maximal subgroup
$SU(5)_{GUT}\times SU(5)_{\bot}$, the presence of non-trivial monodromies in
the seven-brane configuration required by phenomenology cuts down
$SU(5)_{\bot}$ to either $U(1)_{PQ}\times U(1)_{\chi}$, as in Dirac neutrino
scenarios, or $U(1)_{PQ}$ in Majorana scenarios. As the notation suggests, in
both cases, one of the $U(1)$ factors can be identified with a Peccei-Quinn
symmetry. Here, $U(1)_{\chi}$ of the Dirac scenario corresponds to a
non-anomalous symmetry and is a linear combination of $U(1)_{Y}$ and
$U(1)_{B-L}$. We find that only a few possible monodromy groups lead to
consistent flavor physics. The full list of possible monodromy groups are
$\mathbb{Z}_{2},\mathbb{Z}_{2}\times\mathbb{Z}_{2},\mathbb{Z}_{3},S_{3}$ and
$Dih_{4}$ (the symmetry group of the square). A quite surprising outcome of
the classification of visible matter fields is that all the fields needed for
a successful implementation of gauge mediated supersymmetry breaking
automatically follows from the existence of this $E_{8}$ point. Moreover, we
find that in all but one Dirac neutrino scenario, the messenger fields of the
minimal gauge mediated supersymmetry breaking (mGMSB) sector are
\textit{forced} to transform as vector-like pairs in the $10\oplus
\overline{10}$ of $SU(5)$. In fact, in two of the Majorana neutrino scenarios
all the charged matter that descends from the $E_{8}$ point are necessary and
sufficient for mGMSB and the interactions of the MSSM! This rigid structure
also extends to the list of available dark matter candidates, providing only a
few options with very specific $U(1)_{PQ}$\ (as well as $U(1)_{\chi}$ for
Dirac scenarios) charge assignments.

Having classified extra matter fields available from the $E_{8}$ point, we
next turn to whether any of these can serve as dark matter candidates for the
purpose of explaining the PAMELA, ATIC and FERMI experiments. Many of these
experiments require some additional component of electrons and positrons
generated by either dark matter physics or astrophysics, and our focus will be
on whether the former possibility can be realized in F-theory GUTs. In the
context of dark matter scenarios there are two main types of models based on
either annihilating or decaying dark matter. In annihilating dark matter
scenarios, dark matter states annihilate and produce electrons and positrons.
In the case of decaying dark matter, GUT\ scale suppressed higher dimension
operators trigger a decay of the dark matter into electrons and positions. The
decaying scenario is more in line with the idea of F-theory GUTs, where
unification plays a key role in low energy phenomenology due to GUT scale
suppressed operators. Nevertheless, we find that none of the available fields
provide a viable dark matter candidate for either decaying, or annihilating
scenarios. There are multiple obstacles (though fewer obstacles in the
decaying scenario). The overarching problem is that in both scenarios the
available TeV scale dark matter candidates decay too rapidly.

There are also other issues, even if one imposes extra structures to avoid
these decays. For example, there is a cosmological issue: In scenarios where
the dark matter relic abundance is generated non-thermally from the decay of
the saxion, the saxion mass needs to be bigger than $1$ TeV and this is in
conflict with the fact that the saxion mass cannot be that large without
significant fine tuning to avoid the stau mass from becoming
tachyonic\footnote{This is especially problematic when the messengers must
transform in the $10\oplus\overline{10}$, as we have found to be the generic
case with a single $E_{8}$ point of enhancement.}. In addition, the saxion
decay also overproduces dark matter candidates in non-thermal scenarios. In
scenarios where the dark matter is generated thermally, the decay of the
saxion overdilutes the relic abundance. This issue can be overcome in decaying
scenarios by a mild fine tuning, but is especially problematic in annihilating
scenarios. Also in the annihilating scenario one needs a light field to
communicate between the visible and the hidden sector. One can rule out a
gauge boson playing this role, thanks to the classification of allowed gauge
factors. Light scalar mediators from inside the $E_{8}$ do exist that can in
principle do the job, but even in this case one needs to assume many
additional ingredients for this to work.

One could ask if there are other possible charged matter which can communicate
with our sector by gauge interactions. For example if we consider matter which
is charged under our $E_{8}$ as well as some other group $G$, this could in
principle provide another class of dark matter candidates. In this regard,
E-type singularities resist the intuition derived from quiver diagrams.
Indeed, matter charged under $E_{8}\times G$ is not allowed in string theory!
There are two ways to state why this is not possible. One way of arguing for
the absence of such particle states is that matter on colliding branes can be
explained by locally Higgsing a higher singularity of a simple group
\cite{KatzVafa}, so that $E_{8}$ places an upper cap on the allowed
singularity type. More directly, it is also known that the collision of E-type
branes with one another in higher dimensions lead to tensionless strings! The
four-dimensional reflection of such tensionless strings are conformal theories
with E-type symmetry. If we are to avoid a tower of nearly massless particles,
we can consider a sector with badly broken conformal symmetry. Even though
this is logically allowed we find it to be somewhat exotic from the
perspective of F-theory GUT constructions. Of course one can also speculate
about the potential implications of having a nearly conformal sector with an
E-type symmetry, and we offer some speculations along this line later in the paper.

It is in principle possible to also consider dark matter candidates which
originate from bulk gravitational modes, or modes associated with other branes
of the compactification. In this case, the main issue is that generating a
weak scale mass for the candidate involves a non-trivial extension of the
model, which appears quite ad hoc. All of this reinforces the idea that if one
takes seriously the notion of unifying geometric structures in F-theory GUTs,
the gravitino remains as the main candidate for dark matter. This also points
to an astrophysical origin for the signals observed in the PAMELA, ATIC and
FERMI experiments.

The rest of this paper is organized as follows. We first review the main
building blocks of F-theory GUTs in section \ref{sec:massscale}, and in
particular provide in crude terms a characterization of the expected mass
scales for matter within, close to, and disjoint from the GUT seven-brane. In
section \ref{sec:gravDM} we review the fact that in F-theory GUTs, gravitinos
provide a natural dark matter candidate. In section \ref{sec:FLAVOR} we
demonstrate that hierarchical structures in the CKM and PMNS matrix require a
single point of $E_{8}$ enhancement. This is followed in section
\ref{sec:DarkE} by a classification of F-theory GUTs which respect this
property, as well as other mild phenomenological conditions. In sections
\ref{sec:ESHIELD} and \ref{sec:FULLDARK}, we respectively consider matter
which intersects a part of $E_{8}$, and matter disjoint from $E_{8}$. Having
catalogued potential dark matter candidates, in section \ref{sec:Review} we
review the main features of current experiments and their potential dark
matter and astrophysical explanations. Section \ref{sec:FVS} contains our
analysis of annihilating and decaying scenarios, where we find obstructions in
all cases to realizing a viable interacting dark matter scenario. In section
\ref{sec:CONC} we present our conclusions and potential directions for future
investigation. Appendix A contains the classification of viable monodromy
groups associated with a single $E_{8}$ interaction point.

\section{Building Blocks and Mass Scales of F-theory GUTs\label{sec:massscale}%
}

Before proceeding to a specific dark matter candidate, in this section we
briefly review the main ingredients and mass scales which enter into F-theory
GUTs. The discussion roughly separates into those objects which are part of
the GUT seven-brane, those which communicate directly with the supersymmetry
breaking sector localized in the seven-brane with Peccei-Quinn (PQ) gauge
symmetry, and those degrees of freedom which are not directly connected with
the PQ seven-brane. We argue that weak to TeV scale matter must either possess
a non-trivial PQ charge, or interact closely with such a field. Fields cutoff
from the dynamics of the PQ seven-brane have far lower mass, on the order of
the gravitino mass $m_{3/2}\sim10-100$ MeV.

In F-theory GUTs, the relevant degrees of freedom of the GUT\ model localize
on matter curves or propagate in the complex surface wrapped by the
GUT\ seven-brane. All of the matter content of F-theory GUTs is controlled by
the local enhancement in the singularity type of the geometry. For example,
the bulk seven-brane corresponds to a locus where an $SU(5)$ singularity is
present. Matter trapped on Riemann surfaces or \textquotedblleft
curves\textquotedblright\ corresponds to places where the local singularity
type of the geometry enhances further. This matter can then be described in
terms of a local Higgsing operation. At points of the geometry, a further
enhancement is possible, and this is where matter curves meet, and Yukawa
couplings localize.

Many of the necessary ingredients localize within the GUT\ seven-brane. On the
other hand, the existence of matter curves indicates the presence of other
seven-branes which intersect the GUT\ brane. The gauge symmetries of these
additional branes provide additional nearly exact global symmetries for the
low energy theory. In principle, extra matter fields which are GUT\ singlets
can also localize on curves inside of such \textquotedblleft$7_{\bot}%
$-branes\textquotedblright. Such matter fields will typically interact with
the charged matter of the GUT model. For example, the field responsible for
supersymmetry breaking, $X$ is neutral under the GUT\ seven-brane, but is
nevertheless charged under a $U(1)$ Peccei-Quinn seven-brane. This field
develops a vev of the form:%
\begin{equation}
\left\langle X\right\rangle =x+\theta^{2}F_{X}\text{,}%
\end{equation}
where:
\begin{align}
x  &  \sim10^{12}\text{ GeV,}\label{scalvev}\\
F_{X}  &  \sim10^{17}\text{ GeV}^{2}\text{.} \label{Fvev}%
\end{align}
Within this framework, it is possible to accommodate a $\mu$ term correlated
with supersymmetry breaking through the higher-dimension operator $X^{\dag
}H_{u}H_{d}/\Lambda_{\text{UV}}$, as well as a minimal gauge mediated
supersymmetry breaking sector through the F-term $XYY^{\prime}$, where the
$Y$'s denote messenger fields \cite{HVGMSB}. See
\cite{BHVI,BHVII,HVGMSB,HVLHC,HVCKM,FGUTSCosmo,HKSV,BHSV,HVCP} for other
recent work on the phenomenology of F-theory GUTs. In principle, there can be additional contributions
to supersymmetry breaking effects in passing from local to global compactifications. In this paper we
assume that the dominant source of supersymmetry breaking effects in the visible sector descends from
a gauge mediation scenario, with degrees of freedom present in the local model.

So far, our discussion has been purely local in the sense that we have
decoupled the effects of gravity, and have focussed on degrees of freedom
which are geometrically nearby the GUT\ seven-brane. In a globally consistent
model, additional degrees of freedom will necessarily be present. Tadpole
cancellation conditions will likely require the presence of additional
seven-branes and three-branes. Moreover, bulk gravitational modes which
propagate in the threefold base could also be present.

In this paper we shall assume that the masses of all degrees of freedom are
either specified by high scale supersymmetric dynamics, or are instead
correlated with the effects of supersymmetry breaking. Since supersymmetry
breaking originates from the $X$ field, it follows that proximity to the PQ
seven-brane strongly influences the mass scale for the corresponding degree of
freedom. Very massive modes can always be integrated out, so we shall focus on
the mass scales of objects with mass correlated with either the Higgs or $X$
field vev. Generating masses for scalars is typically more straightforward
than for fermions, and so we shall focus on the ways in which fermions can
develop mass.

First consider objects which are charged under the GUT seven-brane. Such
matter fields can develop a mass by coupling to the Higgs fields. Once the
Higgs develops a suitable vacuum expectation value, this induces a weak scale
mass for the corresponding particles. Here, we are using a loose notion of
\textquotedblleft weak scale\textquotedblright\ so that for example the mass
of the electron is effectively controlled by the same dynamics. In gauge
mediation scenarios, the soft scalars and gauginos all develop mass through
loop suppressed interactions with the messenger fields. Communicating
supersymmetry breaking to the MSSM\ then leads to a sparticle spectrum with
masses in the range of $\sim100-1000$ GeV.

The situation is far different for matter fields which are not charged under
the GUT group. Such matter fields subdivide into those which are charged under
$U(1)_{PQ}$, and those which are uncharged. Assuming that the relevant degrees
of freedom are relatively \textquotedblleft nearby\textquotedblright\ the
GUT\ seven-brane, the suppression scale $\Lambda_{\text{UV}}$ for the relevant
higher dimension operators will be close to the GUT scale. In this case, the
relevant F-terms and D-terms which can generate masses for all components of
chiral superfields $\Phi_{1}$ and $\Phi_{2}$ with bosonic and fermionic
components $\phi_{1},\phi_{2}$ and $\psi_{1},\psi_{2}$ are:\footnote{While it
is tempting to include contributions of the form $\int d^{4}\theta
\log\left\vert X\right\vert ^{2}\Phi^{\dag}\Phi$, as would be present in gauge
mediated supersymmetry breaking, note that this generates mass for the scalars
of the chiral multiplet, but not the fermions.}%
\begin{align}
\int d^{4}\theta\frac{X^{\dag}\Phi_{1}\Phi_{2}}{\Lambda_{\text{UV}}} &
\rightarrow\mu\int d^{2}\theta\Phi_{1}\Phi_{2}\\
\int d^{4}\theta\frac{X^{\dag}X\Phi_{1}\Phi_{2}}{\Lambda_{\text{UV}}^{2}} &
\rightarrow\left\vert \mu\right\vert ^{2}\phi_{1}\phi_{2}+\frac{\mu\cdot
x}{\Lambda_{\text{UV}}}\psi_{1}\psi_{2}+\cdots\label{reduced}\\
\int d^{2}\theta X\Phi_{1}\Phi_{2} &  \rightarrow F_{X}\phi_{1}\phi
_{2}-|x|^{2}|\phi_{i}|^{2}+x\psi_{1}\psi_{2}\text{,}\label{FTERM}%
\end{align}
where in the first two lines we have used the rough relation \cite{HVGMSB}:%
\begin{equation}
\mu\sim\frac{\overline{F_{X}}}{\Lambda_{\text{UV}}}\text{.}%
\end{equation}
Note that the first and third lines both require non-trivial PQ charge for at
least one of the chiral superfields $\Phi_{i}$. In particular, we conclude
that PQ charged objects, or direct interactions with PQ objects lead to weak
scale masses, while uncharged chiral multiples have lower fermionic masses.

Integrating out the heavy PQ\ gauge boson generates operators of the form:%
\begin{equation}
4\pi\alpha_{PQ}e_{X}e_{\Phi}\int d^{4}\theta\frac{X^{\dag}X\Phi^{\dag}\Phi
}{M_{U(1)_{PQ}}^{2}}\rightarrow4\pi\alpha_{PQ}e_{X}e_{\Phi}\left\vert
\frac{F_{X}}{M_{U(1)_{PQ}}}\right\vert ^{2}\left\vert \phi\right\vert
^{2}\text{,}\label{PQdefdef}%
\end{equation}
where $\alpha_{PQ}$ denotes the fine structure constant of the PQ$\ $gauge
theory, and $e_{X}$ and $e_{\Phi}$ denote the charges of $X$ and $\Phi$ under
$U(1)_{PQ}$. This term generates a contribution to the mass squared of the
scalar component, but does not produce a fermionic mass. An interesting
feature of this \textquotedblleft PQ\ deformation\textquotedblright:%
\begin{equation}
\Delta_{PQ}\propto\left\vert \frac{F_{X}}{M_{U(1)_{PQ}}}\right\vert
\end{equation}
is that depending on the relative charges of $X$ and $\Phi$, this can either
produce a positive or tachyonic contribution to the overall mass squared.

The presence of the PQ deformation actually provides another potential means
by which fields could develop weak scale masses. For appropriate PQ charges,
the PQ deformation induces a tachyonic mass squared, so that the corresponding
field can develop a non-zero vev on the order of the size of the PQ
deformation. Couplings between this singlet and other fields can then generate
weak scale masses, either through cubic superpotential terms, or through
couplings to vector multiplets.

The fermions of vector multiplets can also develop mass through couplings to
$X$:
\begin{align}
\int d^{2}\theta\log X\cdot Tr\left(  W^{\alpha}W_{\alpha}\right)   &
\rightarrow\frac{F_{X}}{x}\lambda^{\alpha}\lambda_{\alpha}\label{firstgaugino}%
\\
\int d^{4}\theta\frac{X^{\dag}X}{\Lambda_{\text{UV}}^{2}}\cdot Tr\left(
W^{\alpha}W_{\alpha}\right)   &  \rightarrow\frac{\overline{\mu}\cdot
\overline{x}}{\Lambda_{\text{UV}}}\cdot\lambda^{\alpha}\lambda_{\alpha
}\text{.}\label{secondgaugino}%
\end{align}
Let us comment on the two mass scales present in lines (\ref{firstgaugino})
and (\ref{secondgaugino}). The first case is the characteristic value expected
in gauge mediated supersymmetry breaking, and is due to integrating out heavy
messenger fields. This type of coupling requires, however, that the
corresponding vector multiplet couple to fields which are charged under
$U(1)_{PQ}$. Indeed, in the absence of such fields, line (\ref{secondgaugino})
establishes a far lower mass for the corresponding gaugino, which is in line
(up to factors of the gauge coupling) with the estimate of line (\ref{reduced}).

Returning to the explicit values of $x$ and $F_{X}$ given in equations
(\ref{scalvev}) and (\ref{Fvev}), it follows that the F-term coupling of line
(\ref{FTERM}) produces masses for the scalars and bosons far above the weak
scale. On the other hand, we see that generic couplings to the $X$ field
always generate weak scale to TeV scale masses for the bosons. Note, however,
that while the operators $X^{\dag}\Phi_{1}\Phi_{2}/\Lambda_{\text{UV}}$ and
$\log X\cdot Tr\left(  W^{\alpha}W_{\alpha}\right)  $ leads to fermion masses
in the same range, in the case of the operators $X^{\dag}X\Phi_{1}\Phi
_{2}/\Lambda_{\text{UV}}^{2}$ and $X^{\dag}XTr\left(  W^{\alpha}W_{\alpha
}\right)  /\Lambda_{\text{UV}}^{2}$, the corresponding operators have far
lower mass. Indeed, in this case, the fermion masses are:%
\begin{equation}
\left\vert m_{\text{fermion}}^{(\Lambda_{\text{UV}})}\right\vert
\sim\left\vert \frac{\mu\cdot x}{\Lambda_{\text{UV}}}\right\vert
\sim\left\vert \mu\right\vert \cdot10^{-3}\sim100-1000\text{ MeV.}%
\label{mfermGUT}%
\end{equation}
In all cases, the essential point is that generating a weak or TeV scale mass
for the fermions requires the degree of freedom to closely interact with PQ
charged fields.

Next consider degrees of freedom which propagate in the bulk of the threefold
base, or which are completely sequestered from the PQ\ seven-brane. In such
cases, the relevant suppression scale for all higher dimension operators is
more on the order of the reduced Planck scale $M_{PL}\sim2.4\times10^{18}$ GeV
rather than the GUT\ scale. The absence of PQ\ charged objects significantly
limits the available couplings of $X$ between vector and chiral multiplets.
Here, the expected mass scales are:
\begin{align}
\int d^{4}\theta\frac{X^{\dag}X\Phi_{1}\Phi_{2}}{M_{PL}^{2}} &  \rightarrow
\frac{\left\vert \mu\right\vert ^{2}\Lambda_{\text{UV}}^{2}}{M_{PL}^{2}}%
\phi_{1}\phi_{2}+\frac{\mu\cdot x}{\Lambda_{\text{UV}}}\frac{\Lambda
_{\text{UV}}}{M_{PL}}\psi_{1}\psi_{2}\\
\int d^{4}\theta\frac{X^{\dag}X}{\Lambda_{\text{UV}}^{2}}\cdot Tr\left(
W^{\alpha}W_{\alpha}\right)   &  \rightarrow\frac{\Lambda_{\text{UV}}}{M_{PL}%
}\frac{\overline{\mu}\cdot\overline{x}}{\Lambda_{\text{UV}}}\cdot
\lambda^{\alpha}\lambda_{\alpha}\text{.}%
\end{align}
In all cases then, the relevant mass scale is related to that of equation
(\ref{mfermGUT}) as:%
\begin{equation}
\left\vert m_{\text{fermion}}^{(M_{pl})}\right\vert \sim\frac{\Lambda
_{\text{UV}}}{M_{PL}}\left\vert m_{\text{fermion}}^{(\Lambda_{\text{UV}}%
)}\right\vert \sim1-100\text{ keV.}%
\end{equation}
Bulk gravitational degrees of freedom have similar mass to that of the
gravitino. In the context of F-theory GUTs, the mass of the gravitino is:%
\begin{equation}
m_{3/2}\sim\frac{F_{X}}{M_{PL}}\sim10-100\text{ MeV.}%
\end{equation}

To summarize, weak to TeV scale degrees of freedom must interact with the PQ
seven-brane. This implies that the most natural TeV scale dark matter
candidates are effectively \textquotedblleft nearby\textquotedblright\ the
other ingredients of the local F-theory GUT.

\section{Gravitino Dark Matter and the Cosmology of F-theory
GUTs\label{sec:gravDM}}

In this section we review some of the main features of the cosmology of
F-theory GUTs, and in particular, the fact that the gravitino already provides
a very natural dark matter candidate \cite{FGUTSCosmo}. In certain scenarios,
the axion can also contribute towards the dark matter relic abundance. As
shown in \cite{FGUTSCosmo}, this is due to a rich interplay between the
cosmology of various components of the axion supermultiplet in F-theory
GUTs.\footnote{Although ultimately different, see also
\cite{Ibe:2006rc,KitanoIbeSweetSpot} for features of gravitino dark matter in
the context of the \textquotedblleft sweet spot\textquotedblright\ model of
supersymmetry breaking.} We now review the primary features of the analysis in
\cite{FGUTSCosmo} which naturally evades some of the typical problems present
in the cosmology of gauge mediated supersymmetry breaking scenarios. As in
\cite{FGUTSCosmo}, we shall focus on the cosmology of F-theory GUTs, treated
as a model of particle physics at temperatures $T<T_{RH}^{0}$, where
$T_{RH}^{0}$ denotes the \textquotedblleft initial reheating
temperature\textquotedblright\ of the Universe, which corresponds to the
temperature at which the Universe transitions to an era of radiation domination.

Since F-theory GUTs correspond to a deformation away from the minimal gauge
mediation scenario, the gravitino corresponds to the lightest supersymmetric
partner with mass:%
\begin{equation}
m_{3/2}\sim\frac{F_{X}}{M_{PL}}\sim10-100\text{ MeV}%
\end{equation}
where $\sqrt{F_{X}}$ denotes the scale of supersymmetry breaking. In the
context of supersymmetric models with a stable gravitino in this mass range,
there is a strong tendency to over-produce gravitinos because the gravitino
decouples from the thermal bath quite early on in the thermal history of the
Universe. Indeed, in the context of F-theory GUTs, the freeze out temperature
for gravitinos is:%
\begin{equation}
T_{3/2}^{f}\sim10^{10}\text{ GeV.}%
\end{equation}
One common way to solve the \textquotedblleft gravitino
problem\textquotedblright\ is to lower the reheating temperature $T_{RH}^{0}$
to the point where the relic abundance of gravitinos is truncated to a
sufficiently low level. For example, in the parameter range preferred by a
$10-100$ MeV mass gravitino, this translates into the upper bound $T_{RH}%
^{0}<10^{6}$ GeV \cite{Moroi:1993mb}. This is problematic for generating a
suitable baryon asymmetry because aside from the Affleck-Dine mechanism,
mechanisms such as leptogenesis and GUT\ scale baryogenesis typically require
thermal processes at much higher temperatures between $10^{12}-10^{16}$ GeV to
be available. As we now explain, the cosmology of the saxion in F-theory GUTs
plays a crucial role in bypassing this constraint.

The scalar part of the axion supermultiplet corresponds to two real degrees of
freedom, given by the axion, and the saxion. The mass of the saxion is
sensitive to a stringy effect which also shifts the soft scalar
masses of the MSSM. With respect to this parameter $\Delta_{PQ}$, the
corresponding mass of the saxion is:%
\begin{equation}
m_{sax}\propto\Delta_{PQ}\text{,}%
\end{equation}
where the actual constant of proportionality depends on details of the saxion potential, and how $SU(5)\times
U(1)_{PQ}$ embeds in $E_{8}$ \cite{HVGMSB,BHSV}.

The cosmological history of the saxion leads to a significant shift in the
expected relic abundance of the gravitino. Below the initial reheating
temperature $T_{RH}^{0}$, the saxion field will typically be displaced from
its origin by some initial amplitude. At a temperature $T_{osc}^{sax}$, this
field begins to oscillate. An interesting numerical coincidence is that in
F-theory GUTs:%
\begin{equation}
T_{osc}^{sax}\sim T_{3/2}^{f}\sim10^{10}\text{ GeV.} \label{confluence}%
\end{equation}
At somewhat lower temperatures, the energy stored in the oscillation of the
saxion comes to dominate the energy density of the Universe.

The era of saxion domination terminates when the saxion decays. This typically
occurs above the starting temperature for BBN, $T_{BBN}\sim1-10$ MeV.
Efficient reheating of the Universe from saxion decay further requires that
the saxion decay primarily to Higgs fields, which imposes a lower bound on its
mass. Remarkably, this translates into a lower bound on the size of the PQ
deformation parameter on the order of $50$ GeV.

The decay of the saxion significantly dilutes the relic abundance of all
thermally produced relics. The relation between the relic abundances before
and after the decay of the saxion are:%
\begin{equation}
\Omega_{\text{after}}=D_{\text{sax}}\Omega_{\text{before}}\text{,}
\label{dilution}%
\end{equation}
where in the context of F-theory GUTs the saxion dilution factor is roughly:%
\begin{equation}
D_{\text{sax}}\sim\frac{M_{PL}^{2}}{s_{0}^{2}}\frac{T_{RH}^{sax}}{\min
(T_{osc}^{sax},T_{RH}^{0})}\text{,}%
\end{equation}
with $T_{osc}^{sax}$ the temperature of saxion oscillation, and $T_{RH}^{sax}$
is the temperature at which the saxion reheats the Universe. The initial
amplitude $s_{0}$ is naturally in the range:%
\begin{equation}
s_{0}\sim\Lambda_{\text{UV}}\sim10^{15.5}\text{ GeV,}%
\end{equation}
while the saxion reheating temperature is set by the total saxion decay rate:%
\begin{equation}
T_{RH}^{sax}\sim0.5\sqrt{M_{PL}\Gamma_{sax}}\sim0.1-10\text{ GeV.}
\label{TSAX}%
\end{equation}
The typical size of the dilution factor is roughly \cite{FGUTSCosmo}:%
\begin{equation}
D_{\text{sax}}\sim10^{-4}%
\end{equation}
in the range of maximal interest when $T_{RH}^{0}>T_{osc}^{sax}$.

The decay of the saxion dilutes the thermally produced gravitinos, while
introducing additional gravitinos as decay products. The confluence of
temperatures in equation (\ref{confluence}) actually causes the resulting
relic abundance of thermally produced gravitinos to remain independent of
$T_{RH}^{0}$, provided there is still an era where the oscillations of the
saxion dominates the energy density of the Universe. As shown in
\cite{FGUTSCosmo}, the resulting relic abundance for thermally produced
gravitinos is roughly given as:%
\begin{equation}
\Omega_{3/2}^{TP}h^{2}\sim0.01-0.1\text{,}%
\end{equation}
which is in the required range for gravitinos to constitute a substantial
component of dark matter. In addition, gravitinos produced through the decay
of the saxion can, in certain parameter regimes can account for at most $10\%$
of the gravitino relic abundance. In this regime, the gravitino relic
abundance is independent of the reheating temperature $T_{RH}^{0}$. This
reintroduces possible high temperature mechanisms for generating a suitable
baryon asymmetry, such as leptogenesis, or GUT scale baryogenesis.

At lower temperature scales where the reheating temperature $T_{RH}^{0}$ is
below the required temperature for saxion domination, it is also in principle
possible for oscillations of the axion to constitute a component of the dark
matter relic abundance. In F-theory GUTs, the axion decay constant is roughly
$f_{a}\sim10^{12}$\ GeV, which is the requisite range for the axion to play a
prominent role. Here it is important to note that the oscillations of the
saxion can sometimes disrupt coherent oscillations of the axion.

As we have already mentioned, the confluence of various temperature scales
suggests gravitinos as a natural dark matter component which can in suitable
circumstances be supplemented by an axionic dark matter component. On the
other hand, both the gravitino and axion only interact with the matter content
of the Standard Model through higher derivative terms. This in particular
means that such candidates are unlikely to be observed either by direct, or
indirect dark matter detection experiments. Given the recent influx of
tantalizing hints at dark matter detection, in this paper we focus on whether
there are any TeV scale dark matter candidates available in minimal F-theory
GUT\ models which could potentially interact more directly with the Standard Model.

\section{Flavor Implies $E_{8}$\label{sec:FLAVOR}}

Up to this point, we have simply reviewed many of the ingredients which have
figured in previous F-theory GUT\ constructions. This leaves open the
question, however, as to how many of these ingredients are required, and how
many are additional inputs necessary to solve a particular phenomenological
problem. Here we show that assuming only that there is a flavor hierarchy in
the CKM\ and PMNS matrix \textit{requires} an $E_{8}$ point of enhancement! In
fact, many of the extra fields left over can then play a specific role in the
phenomonology of F-theory GUT scenarios, and in section \ref{sec:DarkE} we
classify these options.

The only assumptions we shall make in this section are that the following
interaction terms:%
\begin{equation}
\int d^{2}\theta\text{ }5_{H}\times10_{M}\times10_{M}+\int d^{2}\theta\text{
}\overline{5}_{H}\times\overline{5}_{M}\times10_{M}+\text{Neutrinos}
\label{intint}%
\end{equation}
be present, where we shall assume that the neutrino sector can correspond to
either a Dirac \cite{BHSV}, or Majorana scenario of the form:%
\begin{align}
\text{Dirac}  &  \text{: \ }\int d^{4}\theta\text{ }\frac{H_{d}^{\dag}LN_{R}%
}{\Lambda_{\text{UV}}}\label{Dirac}\\
\text{Majorana}  &  \text{: \ }\int d^{2}\theta\text{ }H_{u}LN_{R}%
+M_{\text{maj}}N_{R}N_{R}\text{.} \label{Majorana}%
\end{align}
In addition, we assume that all of the matter fields $5_{H}$, $\overline
{5}_{H}$, $10_{M}$ and $N_{R}$ localize on curves of the geometry. We will not
even assume the mechanism of doublet triplet splitting via fluxes proposed in
\cite{BHVII}, but will instead simply require a much milder genericity
statement from index theory that if a four-dimensional zero mode localizes on
a curve, then the conjugate representation cannot have any zero modes
localized on the same curve.\footnote{After this paper appeared, there have been various
claims made in the literature, starting with \cite{Marsano:2009gv} that activating GUT breaking by a hyperflux is incompatible with doublet
triplet splitting, in the sense that it induces exotics in the low energy spectrum. This analysis
applies to a very limited class of local models which can be treated using the spectral cover description. It is still an open problem
to realize a consistent local model with no charged exotics, but we stress that the obstruction \cite{Marsano:2009gv} only rules
out a small fraction of possible ways of building a local model. See section \ref{sec:DarkE} for further discussion on this point.}

Flavor hierarchy imposes a number of significant restrictions on the available
enhancements in the singularity type. First, the presence of the interaction
term $5_{H}\times10_{M}\times10_{M}$ requires a point where $SU(5)$ enhances
to at least $E_{6}$. Moreover, the presence of the interaction term
$\overline{5}_{H}\times\overline{5}_{M}\times10_{M}$ requires an enhancement
to at least $SO(12)$. As found in \cite{HVCKM}, the CKM\ matrix will exhibit a
hierarchical structure provided these points of enhancement are close
together, so that there is at least an $E_{7}$ point of enhancement.

In this section we show that generating a mildly hierarchical PMNS\ matrix
forces this enhancement up to $E_{8}$. Our strategy for obtaining this result
will be to ask whether an $E_{7}$ point of enhancement is sufficient for
realizing all of the required interaction terms. The obstructions we encounter
will imply that only $E_{8}$ is available. Since $E_{7}$ is a subgroup of
$E_{8}$, we can phrase our analysis in terms of interaction terms inside of
$E_{8}$. What we will effectively show is that enough of $E_{8}$ is used that
it cannot all be fit inside of $E_{7}$.

The maximal subgroup of $E_{8}$ which contains the GUT\ group $SU(5)$ is
$SU(5)_{GUT}\times SU(5)_{\bot}$. \ The adjoint representation of $E_{8}$
decomposes into irreducible representations of $SU(5)_{GUT}\times SU(5)_{\bot
}$ as:%
\begin{align}
E_{8}  &  \supset SU(5)_{GUT}\times SU(5)_{\bot}\\
248  &  \rightarrow(1,24)+(24,1)+(5,\overline{10})+(\overline{5}%
,10)+(10,5)+(\overline{10},\overline{5})\text{.}%
\end{align}
Although this would appear to provide a large number of additional ingredients
for model building, consistency with qualitative phenomenological features
imposes a number of identifications in the low energy effective field theory.
For example, the interaction term $5_{H}\times10_{M}^{(1)}\times10_{M}^{(2)}$
would at first appear to involve three matter curves. This is problematic,
because if the $10_{M}$'s localize on distinct curves, then the resulting mass
matrix will without fine tuning lead to at least two massive generations
\cite{BHVII}. At the level of the effective field theory, achieving one heavy
generation then requires the existence of an interchange symmetry:%
\begin{equation}
10_{M}^{(1)}\leftrightarrow10_{M}^{(2)}\text{.}%
\end{equation}
The presence of such symmetries may appear as an extra level of fine tuning.
In fact, as noted in \cite{Hayashi:2009ge}, such identifications will
generically occur. This is essentially because the positions of the
seven-branes are dictated by the locations of singularities in the fibration
structure of the geometry, which are in turn controlled by polynomial
equations in several variables. For example, polynomials with rational numbers
for coefficients will often contain non-trivial symmetry groups which
interchange the various roots. A similar phenomenon is present in the
factorization of polynomials of more than one variable. This means in
particular that rather than being a special feature of the geometry, it is to
be expected on quite general grounds that such identifications will occur.

In the context of $SU(5)_{\bot}$, this discrete symmetry group is maximally
given by its Weyl group $S_{5}$, the symmetric group on five letters. The
group $S_{5}$ acts by identifying directions in the $5$ of $SU(5)_{\bot}$.
From the perspective of the effective field theory, this occurs because inside
of $SU(5)_{\bot}$, there are additional discrete symmetries which act as
identifications on the representations of $SU(5)_{\bot}$. Thus, there will
generically be identifications of some of the $5$'s and $10$'s. In the
physical \textquotedblleft quotient theory\textquotedblright\ where all
identifications have taken place, the actual number of distinct matter curves
will be greatly reduced compared to the \textquotedblleft covering
theory\textquotedblright. For example, there will be five distinct $10_{M}$
curves in the covering theory, since the $10_{M}$ transforms as a $5$ of
$SU(5)_{\bot}$. Acting with the discrete symmetry group will then identify
some of these fields.

As we show in section \ref{sec:DarkE}, this analysis leaves us with just a few
additional ingredients which can also interact at the same point of
enhancement. Quite remarkably, these also play a significant role: They are
the fields of the supersymmetry breaking sector utilized in \cite{HVGMSB}!
Thus, simply achieving the correct flavor structure will automatically include
the necessary ingredients for supersymmetry breaking.

The rest of this section is organized as follows. Before proceeding to the
main result of this section, we first setup notation, and introduce the
features of monodromy groups which will be important for our analysis. Using
some minimal facts about such monodromy groups coupled with the existence of
all interactions in line (\ref{intint}) will then allow us to deduce that
flavor implies $E_{8}$.

\subsection{Monodromy Groups: Generalities\label{revmon}}

Since the action of monodromy groups play such a crucial role in the analysis
to follow, here we explain in more precise terms how this group acts on the
available matter curves. Geometrically, matter fields localize along loci
where elements in the Cartan of $SU(5)_{\bot}$ combine with $SU(5)_{GUT}$ so
that the singularity type enhances. The precise location of each matter curve
can be analyzed by introducing a weight space decomposition of the various
representations. \ The Cartan of $SU(5)_{\bot}$ can be parameterized by the
coordinates $t_{1},...,t_{5}$ subject to the constraint:%
\begin{equation}
t_{1}+...+t_{5}=0\text{.}%
\end{equation}
The weights of the $5_{\bot}$, $10_{\bot}$ and $24_{\bot}$ of $SU(5)_{\bot}$
are then given as:%
\begin{align}
5_{\bot}  &  :t_{i}\\
10_{\bot}  &  :t_{i}+t_{j}\\
24_{\bot}  &  :\pm\left(  t_{i}-t_{j}\right)  +4\times(0\text{ weights})
\end{align}
for $1\leq i,j\leq5$ such that $i\neq j$. Using the decomposition of the
adjoint of $E_{8}$ into irreducible representations of $SU(5)_{GUT} \times
SU(5)_{\bot}$, the matter fields with appropriate $SU(5)_{GUT}$ representation
content localize along the following curves:%
\begin{align}
5_{GUT}  &  :-t_{i}-t_{j}=0\\
10_{GUT}  &  :t_{i}=0\\
1_{GUT}  &  :t_{i}-t_{j}=0\text{,}%
\end{align}
so that the vanishing loci then correspond to local enhancements in the
singularity type of the compactification. A matter field with a given weight
will necessarily also be charged under a $U(1)$ subgroup of $SU(5)_{\bot}$,
dictated by its weight.

Geometric considerations and minimal requirements from phenomenology impose
significant constraints. For example, the deformation of a singularity will
generically contain monodromies in the seven-brane configuration whereby some
of the matter curves will in fact combine to form a single irreducible matter
curve. Group theoretically, the deformations of a singularity are
parameterized by the Cartan modulo the Weyl subgroup of the singularity. \ In
the present context, the Weyl group of $SU(5)_{\bot}$ is $S_{5}$, the
symmetric group on five letters. This discrete group acts as permutations on
the five $t_{i}$'s. The most generic geometry will involve identifications by
the full $S_{5}$ monodromy group. In particular to have any smaller monodromy
we need to assume certain factorization properties of the unfolding
singularity. But this generic choice is already too restrictive: For example
it will force the ${\overline{5}}_{M}$,$H_{d},H_{u}$ to all come from a single
curve, which would be in conflict with the resolution of the doublet-triplet
splitting problem found in \cite{BHVII}, and would also not allow a consistent
identification of matter parity. Thus to fit with phenomenological constraints
we need to assume a less generic monodromy group than $S_{5}$. The bigger the
monodromy group, the more generic the geometry. In this sense, the most
generic monodromies we find are for two Majorana neutrino scenarios discussed
in section \ref{sec:DarkE}, where the monodromy group is the order $8$
symmetry group of the square, $Dih_{4}$. Quite remarkably in these cases, all
the available orbits which are charged under the standard model gauge group
are utilized in the minimal GUT model, including the messenger fields!

Note that the monodromy group cannot be trivial. Indeed, the appearance of
monodromies is quite important for achieving a single massive generation in
the up-type quark sector. \ For example, as found in \cite{Hayashi:2009ge}, a
rank one $5_{H}\times10_{M}\times10_{M}$ is quite natural once monodromies in
the seven-brane configuration are taken into account. In the covering theory,
there are then at least two $10_{M}$ curves, which are exchanged under monodromy.

Consistency with all necessary interaction terms then implies that the
monodromy group may act non-trivially on the other covering theory matter
fields of F-theory GUTs. Each matter field in the covering theory fills out an
orbit under the monodromy group, $G$, so that if $w$ denotes the corresponding
weight, then elements of the form:%
\begin{equation}
Orb(w)=\left\{  \sigma(w)|\sigma\in G\right\}
\end{equation}
are all identified under the action of the monodromy group. We shall often
refer to the \textquotedblleft length\textquotedblright\ of the orbit as the
number of elements so that:%
\begin{equation}
\text{Length}(Orb(w))=\#Orb(w)\text{.}%
\end{equation}
As a final piece of notation, we will often denote the action of a permutation
group element using cycle notation. For example, $(123)$ acts on the weights
$t_{1},...,t_{5}$ as:%
\begin{equation}
(t_{1},t_{2},t_{3},t_{4},t_{5})\overset{(123)}{\longrightarrow}(t_{2}%
,t_{3},t_{1},t_{4},t_{5})\text{.}%
\end{equation}
The element $(12)(34)$ instead acts on the weights as:%
\begin{equation}
(t_{1},t_{2},t_{3},t_{4},t_{5})\overset{(12)(34)}{\longrightarrow}(t_{2}%
,t_{1},t_{4},t_{3},t_{5})\text{.}%
\end{equation}
In our conventions, multiplication of two elements proceeds as in the
composition of two functions, so that $(123)\cdot(12)=(13)$.

The monodromy group will also sometimes identify continuous global symmetries.
The effect of this can lead to the presence of additional discrete symmetries
in the low energy theory.

Returning to the interaction term $5_{H}\times10_{M}\times10_{M}$, the weight
assignments in $SU(5)_{\bot}$ for the $5_{H}$ and $10_{M}$'s are of the form
$-t_{i}-t_{j}$ and $t_{k}$, respectively. If these weight assignments form an
invariant interaction term, then they must satisfy the constraint:%
\begin{equation}
(-t_{i}-t_{j})+(t_{k})+(t_{l})=0\text{,}%
\end{equation}
where we have grouped the weight assignments for each field by parentheses. In
this case, it follows that the weights for the $10_{M}$'s must correspond to
$t_{i}$ and $t_{j}$. The existence of a single orbit for the $10_{M}$'s
implies that there must exist an element of the monodromy group which sends
$t_{i}$ to $t_{j}$. \ Note, however, that a given element in the orbit of the
$5_{H}$ need not form an interaction term with all of the weights in the orbit
of the $10_{M}$'s. Indeed, the only condition is that there is some weight in
the orbit of the $10_{M}$ which can form an appropriate interaction.

\subsection{$E_{7}$ Does Not Suffice}

We now proceed to show that quark and lepton flavor hierarchies are
incompatible with a single point of $E_{7}$ enhancement. To start our
analysis, recall that we require the following interaction terms:%
\begin{equation}
\int d^{2}\theta\text{ }5_{H}\times10_{M}\times10_{M}+\int d^{2}\theta\text{
}\overline{5}_{H}\times\overline{5}_{M}\times10_{M}+\text{Neutrinos.}%
\end{equation}

Suppose to the contrary that an $E_{7}$ point of enhancement did suffice for
all required interaction terms. In this case, the resulting breaking pattern
would fit as $E_{8}\supset E_{7}\times SU(2)_{\bot}$. In particular, of the
five $t_{i}$'s present in $SU(5)_{\bot}$, the monodromy group inside of
$E_{7}$ would have to leave the generators $t_{4}$ and $t_{5}$ invariant.
Indeed, this $SU(2)_{\bot}$ is one factor in the \textquotedblleft Standard
Model subgroup\textquotedblright\ $SU(3)_{\bot}\times SU(2)_{\bot}\times
U(1)_{\bot}\subset SU(5)_{\bot}$. Our strategy will be to show that the orbits
non-trivially involve the weights from the $SU(2)_{\bot}$ factor. Thus, we
will have established that the only available enhancement point must be
$E_{8}$.

A necessary condition in this regard is that all available weights must be
uncharged under the Cartan generator of this $SU(2)_{\bot}$ factor. Viewing
this generator as an element in the dual space, this direction can be written
as:%
\begin{equation}
t_{SU(2)_{\bot}}^{\ast}\equiv t_{4}^{\ast}-t_{5}^{\ast}\text{,}%
\end{equation}
so that:%
\begin{align}
t_{SU(2)_{\bot}}^{\ast}(t_{4})  &  =+1\\
t_{SU(2)_{\bot}}^{\ast}(t_{5})  &  =-1
\end{align}
and for $i=1,2,3$,%
\begin{equation}
t_{SU(2)_{\bot}}^{\ast}(t_{i})=0\text{.}%
\end{equation}

The available monodromy groups which act on $t_{1}$, $t_{2}$ and $t_{3}$ are
given by permutations on these letters. The subgroups of the symmetric group
on three letters are isomorphic to $S_{3}$, $%
\mathbb{Z}
_{3}$, $%
\mathbb{Z}
_{2}$ and the trivial group. Since the $5_{H}\times10_{M}\times10_{M}$
interaction requires at least two weights in the orbit of the $10_{M}$, it
follows that the monodromy group must be non-trivial. We now show that in all
cases, the available monodromy group orbits are inconsistent with a neutrino sector.

\subsubsection{$G_{mono}\simeq%
\mathbb{Z}
_{3}$ or $S_{3}$}

First suppose that the monodromy group is either $S_{3}$ or $%
\mathbb{Z}
_{3}$. In this case, there exists an order three element which acts on the
weights, which without loss of generality we take as:%
\begin{equation}
(123)\in G_{mono}%
\end{equation}
Since the orbit for the $10_{M}$ is non-trivial, it follows that this orbit
must in fact involve $t_{1}$, $t_{2}$ and $t_{3}$. In particular,
compatibility with the interaction term $5_{H}\times10_{M}\times10_{M}$ now
implies that the orbit for the $5_{H}$ must involve $-t_{1}-t_{2}$,
$-t_{2}-t_{3}$, $-t_{1}-t_{3}$. Since index theory considerations require the
$\overline{5}$'s to localize on a curve distinct from $5_{H}$, there are
precisely three orbits available, which we label by subscripts:%
\begin{align}
Orb(\overline{5}_{(1)})  &  =t_{1}+t_{4},t_{2}+t_{4},t_{3}+t_{4}\\
Orb(\overline{5}_{(2)})  &  =t_{1}+t_{5},t_{2}+t_{5},t_{3}+t_{5}\\
Orb(\overline{5}_{(3)})  &  =t_{4}+t_{5}\text{.}%
\end{align}
The last option given by $Orb(\overline{5}_{(3)})=t_{4}+t_{5}$ is incompatible
with the interaction term $\overline{5}_{H}\times\overline{5}_{M}\times10_{M}%
$. Indeed, this interaction term requires the weights for the various fields
to satisfy:%
\begin{equation}
(t_{i}+t_{j})+(t_{k}+t_{l})+t_{m}=0\text{,}%
\end{equation}
Where $t_{m}=t_{1},t_{2}$ or $t_{3}$. In all available cases where
$t_{4}+t_{5}$ appears as a weight, either $t_{4}$ or $t_{5}$ will appear
twice, and the sum will not vanish.

We therefore conclude that the only candidate orbits for the $\overline{5}%
_{H}$ and $\overline{5}_{M}$ involve $t_{i}+t_{4}$ or $t_{i}+t_{5}$ for
$i=1,2,3$. Since nothing distinguishes $t_{4}$ or $t_{5}$ in our analysis so
far, we can now fix the orbits of $\overline{5}_{M}$ to be $t_{i}+t_{4}$. Note
that compatibility with the interaction term $\overline{5}_{H}\times
\overline{5}_{M}\times10_{M}$ now forces the orbit $\overline{5}_{H}$ to be
$t_{i}+t_{5}$. Hence, the available orbits are:%
\begin{align}
Orb(10_{M}) &  =t_{1},t_{2},t_{3}\\
Orb(\overline{5}_{M}) &  =t_{1}+t_{4},t_{2}+t_{4},t_{3}+t_{4}\\
Orb(\overline{5}_{H}) &  =t_{1}+t_{5},t_{2}+t_{5},t_{3}+t_{5}\\
Orb(5_{H}) &  =-t_{1}-t_{2},-t_{1}-t_{3},-t_{2}-t_{3}\text{.}%
\end{align}

It is now straightforward to eliminate both neutrino scenarios. For example,
in the Dirac neutrino scenario, the interaction term $H_{d}^{\dag}%
LN_{R}/\Lambda_{\text{UV}}$ requires the weight assignments to obey the
constraint:%
\begin{equation}
(-t_{i}-t_{5})+(t_{j}+t_{4})+(t_{m}-t_{n})=0\text{.}%
\end{equation}
In particular, this implies that the weight of $N_{R}$ is given by
$t_{4}-t_{5}$. This, however, requires a non-trivial participation from the
roots of $SU(2)$. In other words, the right-handed neutrino will only touch
the $E_{7}$ point of enhancement provided it is actually an $E_{8}$ point of enhancement!

Next consider Majorana neutrino scenarios. In this case, the interaction term
$H_{u}LN_{R}$ leads to the constraint on the weights:%
\begin{equation}
(-t_{i}-t_{j})+(t_{k}+t_{4})+(t_{m}-t_{n})=0\text{.}%
\end{equation}
By inspection of the $5_{H}$ orbit, it now follows that $t_{4}$ can only
cancel out provided $t_{n}=t_{4}$.

On the other hand, the Majorana mass term $N_{R}N_{R}$ requires $t_{n}-t_{m}$
must also be present. Thus, there exists another set of weights such that:%
\begin{equation}
(-t_{i^{\prime}}-t_{j^{\prime}})+(t_{k^{\prime}}+t_{4})+(t_{n}-t_{m}%
)=0\text{.}%
\end{equation}
But this implies that $t_{m}=t_{4}$. In other words, the right-handed neutrino
is given by the weight $t_{n}-t_{m}=t_{4}-t_{4}$, which is a contradiction!

To summarize, in the case where the monodromy group is $%
\mathbb{Z}
_{3}$ or $S_{3}$, the Dirac scenario is consistent with requirements from
flavor, but requires an $E_{8}$ enhancement point, and in the case
of\ Majorana neutrinos, we do not find a consistent scenario.

\subsubsection{$G_{mono}\simeq%
\mathbb{Z}
_{2}$}

Next suppose that the monodromy group is given by $%
\mathbb{Z}
_{2}$, so that it acts by interchanging $t_{1}$ and $t_{2}$:%
\begin{equation}
G_{mono}=\left\langle (12)\right\rangle \simeq%
\mathbb{Z}
_{2}%
\end{equation}
This last conclusion is a consequence of the fact that there are only three
$t_{i}$'s available which can participate in the monodromy group action. In
this case, the orbits for the $10_{M}$ and $5_{H}$ are completely fixed to be:%
\begin{align}
Orb(10_{M})  &  =t_{1},t_{2}\\
Orb(5_{H})  &  =-t_{1}-t_{2}\text{.}%
\end{align}
In this case, the available orbits for the $\overline{5}$'s are:%
\begin{align}
Orb(\overline{5}_{(1)})  &  =t_{1}+t_{3},t_{2}+t_{3}\\
Orb(\overline{5}_{(2)})  &  =t_{1}+t_{4},t_{2}+t_{4}\\
Orb(\overline{5}_{(3)})  &  =t_{1}+t_{5},t_{2}+t_{5}\\
Orb(\overline{5}_{(4)})  &  =t_{3}+t_{4}\\
Orb(\overline{5}_{(5)})  &  =t_{3}+t_{5}\\
Orb(\overline{5}_{(6)})  &  =t_{4}+t_{5}\text{.}%
\end{align}

Fixing the weight of $10_{M}$ as $t_{1}$, this imposes the weight constraint:%
\begin{equation}
(t_{i}+t_{j})+(t_{k}+t_{l})+t_{1}=0\text{.}%
\end{equation}
The available pairs of orbits which can satisfy this constraint are then:%
\begin{align}
\text{Option 1}  &  \text{: }Orb(\overline{5}_{(1)})=t_{1}+t_{3},t_{2}%
+t_{3}\text{ and }Orb(\overline{5}_{(6)})=t_{4}+t_{5}\label{popone}\\
\text{Option 2}  &  \text{: }Orb(\overline{5}_{(2)})=t_{1}+t_{4},t_{2}%
+t_{4}\text{ and }Orb(\overline{5}_{(5)})=t_{3}+t_{5}\\
\text{Option 3}  &  \text{: }Orb(\overline{5}_{(3)})=t_{1}+t_{5},t_{2}%
+t_{5}\text{ and }Orb(\overline{5}_{(4)})=t_{3}+t_{4}\text{.} \label{popthr}%
\end{align}

It is now immediate that all Dirac scenarios are ruled out. Indeed, the
presence of the operator $H_{d}^{\dag}LN_{R}/\Lambda_{\text{UV}}$ requires the
weight constraint:%
\begin{align}
\text{Option 1}\text{: }  &  (-t_{i}-t_{3})+(t_{4}+t_{5})+(t_{m}-t_{n})=0\\
\text{Option 2}\text{: }  &  (-t_{i}-t_{4})+(t_{3}+t_{5})+(t_{m}-t_{n})=0\\
\text{Option 3}\text{: }  &  (-t_{i}-t_{5})+(t_{3}+t_{4})+(t_{m}%
-t_{n})=0\text{.}%
\end{align}
Since $t_{i}=t_{1}$ or $t_{2}$, it follows that there does not exist a
solution for any $t_{m}$ and $t_{n}$. Thus, we conclude that in all Dirac
scenarios, flavor considerations require $E_{8}$.

Next consider Majorana scenarios. In this case, the presence of the operator
$H_{u}LN_{R}$ imposes the weight constraint:%
\begin{equation}
(-t_{1}-t_{2})+(t_{j}+t_{k})+(t_{m}-t_{n})=0\text{.} \label{oop}%
\end{equation}
We therefore conclude that $t_{m}=t_{1}$ or $t_{2}$. In fact, since both
$t_{m}-t_{n}$ and $t_{n}-t_{m}$ must be present in the same orbit, it now
follows that the orbit for $N_{R}$ is precisely given as:%
\begin{equation}
Orb(N_{R})=t_{1}-t_{2},t_{2}-t_{1}\text{.}%
\end{equation}
Returning to equation (\ref{oop}), it now follows that the weight $t_{j}%
+t_{k}$ must satisfy the constraint:%
\begin{equation}
(-t_{1}-t_{2})+(t_{j}+t_{k})+(t_{1}-t_{2})=0\text{,}%
\end{equation}
which is not possible, because $t_{2}$ appears twice in the constraint. Hence,
in all cases we find that an $E_{7}$ point of enhancement cannot accommodate
either a Dirac, or a Majorana scenario.

It is quite remarkable that flavor considerations from quarks and leptons push
the enhancment point all the way up to $E_{8}$. We now turn to a
classification of possible mondromy group orbits compatible with the other
elements of F-theory GUTs.

\section{Matter and Monodromy in $E_{8}$\label{sec:DarkE}}

Taking seriously the idea of unification naturally suggests combining the
various ingredients of F-theory GUTs in the minimal number of geometric
ingredients necessary. Indeed, as we have seen in section \ref{sec:FLAVOR},
the presence of an $E_{8}$ point is not so much an aesthetic criterion, as a
\textit{necessary} one in order to generate hierarchical CKM\ and
PMNS\ matrices. The aim of this section is to classify F-theory GUT\ models
consistent with this $E_{8}$ point of enhancement. The only assumptions we
shall make are that the following interaction terms be present in the low
energy theory:%
\begin{equation}
\int d^{2}\theta\text{ }5_{H}\times10_{M}\times10_{M}+\int d^{2}\theta\text{
}\overline{5}_{H}\times\overline{5}_{M}\times10_{M}+\text{Neutrinos}+\int
d^{4}\theta\frac{X^{\dag}H_{u}H_{d}}{\Lambda_{\text{UV}}}\text{,}\label{okay}%
\end{equation}
where here, $X$ is a GUT\ singlet localized on a curve. In fact, the result we
obtain will not strictly require $X^{\dag}H_{u}H_{d}/\Lambda_{\text{UV}}$, but
will also apply to $\mu$-terms generated through the vev of a GUT singlet $S$
through the F-term\footnote{Note that the field $S$ whose vev gives rise to
the $\mu$-term is related to our field $X$ by $S={\overline{D}}^{2}X^{\dag
}/\Lambda_{\text{UV}}$.}:
\begin{equation}
\int d^{2}\theta\text{ }SH_{u}H_{d}\text{.}%
\end{equation}

In order for a GUT scale $\mu$-term to not be present, we shall also require
the presence of a continuous global Peccei-Quinn symmetry. In the context of
Majorana neutrino scenarios, there is a unique PQ symmetry available, with
charge assignments \cite{BHSV}:%
\begin{equation}%
\begin{tabular}
[c]{|c|c|c|c|c|c|c|}\hline
& $\overline{5}_{M}$ & $10_{M}$ & $5_{H}$ & $\overline{5}_{H}$ & $X^{\dag}$ &
$N_{R}$\\\hline
Majorana $U(1)_{PQ}$ & $+2$ & $+1$ & $-2$ & $-3$ & $+5$ & $0$\\\hline
\end{tabular}
\ \ \ \ \ \ \ \text{.} \label{MAJ}%
\end{equation}
In the case of Dirac neutrino scenarios, there is a certain degree of
flexibility. For simplicity, we shall take the same convention for PQ charge
assignments used in \cite{HVGMSB,BHSV}:%
\begin{equation}%
\begin{tabular}
[c]{|c|c|c|c|c|c|c|}\hline
& $\overline{5}_{M}$ & $10_{M}$ & $5_{H}$ & $\overline{5}_{H}$ & $X^{\dag}$ &
$N_{R}$\\\hline
Dirac $U(1)_{PQ}$ & $+1$ & $+1$ & $-2$ & $-2$ & $+4$ & $-3$\\\hline
\end{tabular}
\ \ \ \ \ \ \ \text{.} \label{DIR}%
\end{equation}
Summarizing, we shall classify all available monodromy group actions in
F-theory GUTs consistent with the assumptions:

\begin{itemize}
\item Hierarchical CKM and PMNS matrices

\item $\mu$-term from the vev of a GUT\ singlet

\item A PQ symmetry of the type given by line (\ref{MAJ}) for Majorana
scenarios and (\ref{DIR}) for Dirac scenarios
\end{itemize}

\begin{figure}
[ptb]
\begin{center}
\includegraphics[
natheight=7.270500in,
natwidth=10.390700in,
height=3.3347in,
width=4.7582in
]%
{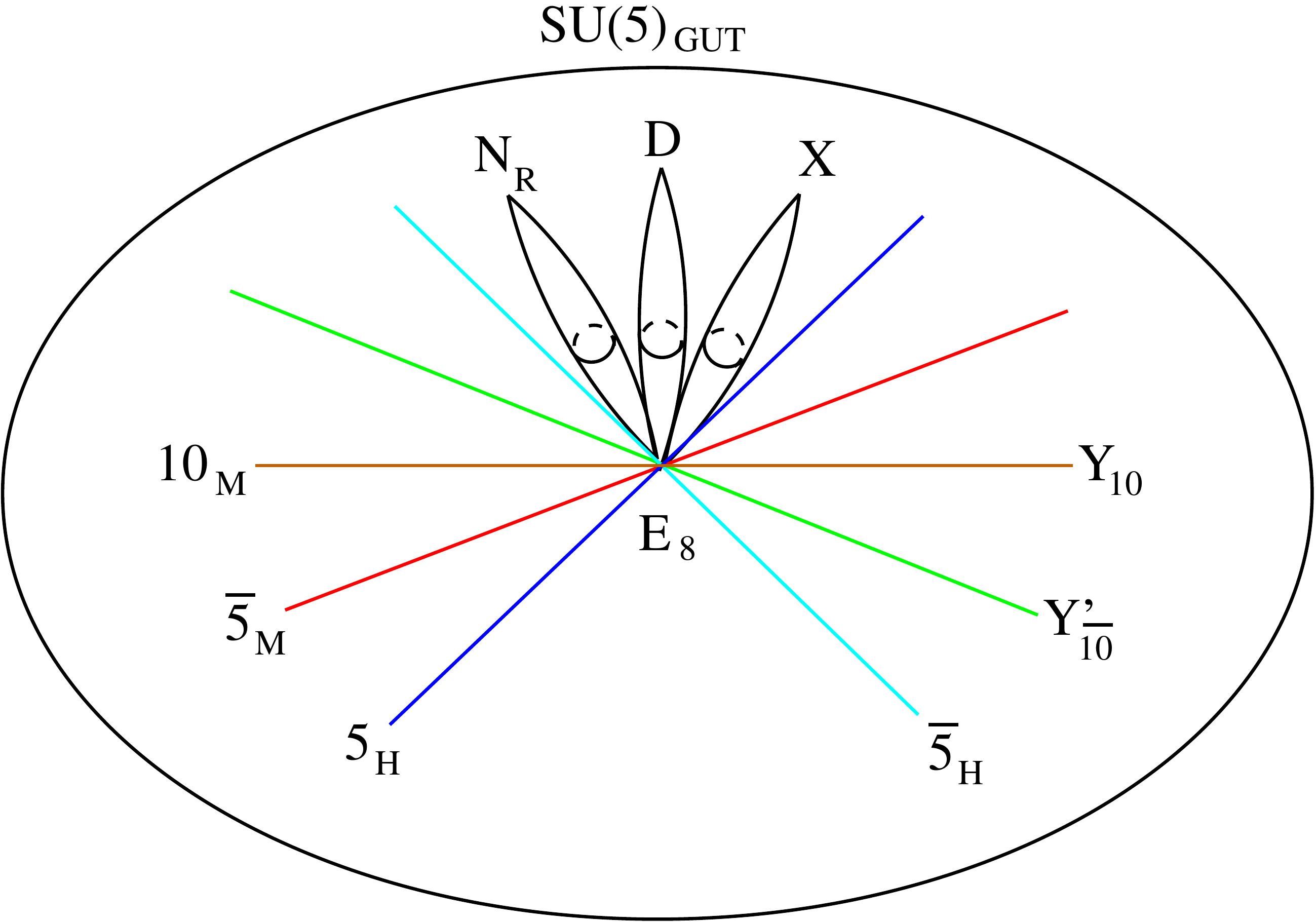}%
\caption{Depiction of an F-theory GUT\ in which all of the necessary
interaction terms descend from a single point of $E_{8}$ enhancement. In all
but one Dirac neutrino scenario, accommodating messenger fields in the gauge
mediated supersymmetry breaking sector turns out to force the messengers
($Y_{10}$ and $Y_{\overline{10}}^{\prime}$) to transform in the $10\oplus
\overline{10}$ of $SU(5)_{GUT}$. GUT\ singlets such as the right-handed
neutrinos $N_{R}$ and $X$ field localize on curves normal to the
GUT\ seven-brane. Here we have also included a dark matter candidate $D$ which
is localized on a curve.}%
\label{egut}%
\end{center}
\end{figure}

A remarkable byproduct of this analysis is that after performing this
classification, there is typically just enough room for a messenger sector of
a minimal gauge mediated supersymmetry breaking scenario where the field $X$
couples to a vector-like pair of messengers $Y$ and $Y^{\prime}$ through the
F-term:%
\begin{equation}
L\supset\int d^{2}\theta\text{ }XYY^{\prime}\text{.}%
\end{equation}
In fact, one of the messengers localizes on the same curve as a matter field.
More surprisingly, in all but one Dirac scenario, the actual representation
content of the messengers will uniquely be fixed to transform in the
$10\oplus\overline{10}$ of $SU(5)_{GUT}$! This has distinctive consequences
for phenomenology, which we shall comment on later in this section. See figure
\ref{egut} for a depiction of the relevant geometry.

Quotienting by a discrete subgroup of $S_{5}$ can also generate highly
non-trivial symmetries in the effective field theory. For example, the
identification of two $U(1)$ factors in $U(1)\times U(1)$ can lead to discrete
$%
\mathbb{Z}
_{2}$ subgroup factors.

The action of the mondromy group in Majorana scenario eliminates essentially
all other gauge degrees of freedom other than $U(1)_{PQ}$. In the case of the
Dirac scenarios, we find that there is one additional gauge boson $U(1)_{\chi
}$ which is a linear combination of $U(1)_{Y}$ and $U(1)_{B-L}$. In addition,
we also classify the available chiral multiplets which can localize on matter
curves normal to the GUT\ seven-brane. These can then constitute potential
dark matter candidates, although we shall return to issues connected with dark
matter later in section \ref{sec:FVS}.

The rest of this section is organized as follows. We first explain in greater
detail the main conditions which we shall demand of the monodromy group action
at the $E_{8}$ enhancement point. Next, we proceed to review the available
monodromy groups in Dirac and Majorana neutrino scenarios. See Appendix A for
further details of this classification. A remarkable feature of the
classification is that in all but one scenario, it forces the messengers to
transform in the $10\oplus\overline{10}$ of $SU(5)_{GUT}$. After commenting on
some of the implications of this for phenomenology, we next discuss potential
\textquotedblleft semi-visible\textquotedblright\ dark matter candidates
corresponding to electrically neutral components of non-trivial $SU(5)_{GUT}$
multiplets.

\subsection{Flux and Monodromy}

Before reviewing the main elements of the classification, we first discuss
some of the necessary conditions on matter curves and fluxes which a monodromy
group action must respect in order to remain consistent with the assumptions
spelled out at the beginning of this section.

Recall that the chiral matter is determined by the choice of background fluxes
through the matter curves of the geometry. In particular, we must require that
if a zero mode in a representation localizes on a curve, then the conjugate
representation cannot appear. In this context, keeping the $5_{H}$,
$\overline{5}_{H}$ and $\overline{5}_{M}$ localized on distinct curves imposes
the following condition on the orbits of these fields:%
\begin{equation}
Orb(\overline{5}_{M}),\text{ }Orb(\overline{5}_{H})\text{, }\overline
{Orb(5_{H})}\text{ all distinct.}%
\end{equation}

It is important to note that the flux passing through a matter curve need not
give precisely three generations. For example, a $10_{M}$ curve could in
principle contain another GUT multiplet, and another curve could contain an
extra $\overline{10}_{M}$. Indeed, as noted in \cite{BHSV}, because both sets
of fields fill out full GUT multiplets, it is in principle possible to allow
some of the messengers and matter fields to localize on the same curve. This
will prove quite important when we discuss conditions necessary for unifying a
messenger sector with the other matter content of F-theory GUTs.

Before closing this subsection, let us also comment on the sense in which we demand an $E_8$ structure.
The primary condition we consider in this paper is that in a sufficiently small patch of the K\"ahler surface wrapped
by the GUT seven-brane, there is a point of enhancement to $E_8$. This does \textit{not} mean
that the geometries we consider must descend from a single globally well-defined $E_8$ singularity over the entire surface,
as in \cite{DonagiWijnholtIII}. Indeed, it is a very strong assumption on the class of compactifications to literally import all of the structures of the
perturbative heterotic string to the F-theory setting. To give just one example of how this can fail, consider the heterotic $E_8 \times E_8$ theory compactified on a $T^2$ with radii chosen so that the full gauge group is $E_8 \times E_8 \times SU(2)$. This latter factor is simply missing
from the spectral cover description, and points to a significant limitation on the class of compactifications covered by such global unfoldings. Along these
lines, let us also note that though there is a Kodaira classification of degenerations of an elliptic fiber for K3 surfaces, no such classification is known
for codimension three singularities, of the type considered here.\footnote{As a related point, let us note that orbifold singularities given by quotienting $\mathbb{C}^2$ by a discrete subgroup of $SU(2)$ admit an ADE classification, though there is no similar ADE classification
for orbifolds of $\mathbb{C}^{3}$ and $\mathbb{C}^{4}$, in part because now collapsing K\"ahler surfaces may be present at these more general orbifold
singularities.} In particular, though we can find a \textit{local patch} which describes the singular fibers in terms
of enhancement to $E_8$, extending this over the surface $S$ may include additional seven-branes which
intersect the GUT stack. These additional six-dimensional matter fields cannot be embedded inside a
single $E_8$ factor, but nevertheless are expected to be present in a general compactification.

The reason that it is important to point out these limitations is that certain statements have appeared in the literature,
which appear to have propagated from \cite{Marsano:2009gv} that there is an obstruction to activating a GUT breaking flux without inducing
charged exotics. Under the strong assumptions that the geometry over $S$ is described by the unfolding of a \textit{single} globally defined $E_8$, and moreover, that no additional factorization occurs in the discriminant locus, this can be established \cite{Marsano:2009gv}. However, this result crucially relies on knowing that everything unfolds from a single global $E_8$. It is through this assumption that one determines more information about the homology classes of the matter curves. Since the index theory is controlled by the cohomology class of the gauge field flux and the homology classes of the matter curves, this can in principle produce a non-trivial constraint on the low energy content.

But in the case of a single $E_8$ unfolding at a point, we in general expect to have some flexibility in the homology
class of the corresponding matter fields. Indeed, it is only after compactifying all curves by specifying the content of the unfolding over all of $S$
that we can hope to read off the homology classes, and thus determine by index theory considerations the chiral matter content on a curve.
To summarize and repeat: In general local models based on a compact K\"ahler surface $S$ with a non-compact normal direction,
\textit{no obstruction to achieving doublet triplet splitting with hyperflux has been proven in general}, and we shall assume
in this paper that this condition can be satisfied.

\subsection{Dirac Scenarios\label{ssec:Dirac}}

We now discuss the various Dirac neutrino scenarios. We refer to Appendix A
for a derivation of the orbit classification. As found in Appendix A, there
are essentially three distinct orbits available for matter fields in Dirac
neutrino scenarios. In all three cases, we find that the monodromy group
preserves two $U(1)$ factors in $SU(5)_{\bot}$. Labeling these $U(1)$ factors
as $U(1)_{PQ}$ and $U(1)_{\chi}$ the charge assignments we find for the
visible matter are:%
\begin{equation}%
\begin{tabular}
[c]{|c|c|c|c|c|c|c|c|c|}\hline
Minimal Matter & $10_{M},Y_{10}$ & $\overline{5}_{M},Y_{\overline{5}}^{\prime
}$ & $Y_{\overline{10}}^{\prime}$ & $Y_{5}$ & $5_{H}$ & $\overline{5}_{H}$ &
$X^{\dag}$ & $N_{R}$\\\hline
$U(1)_{PQ}$ & $+1$ & $+1$ & $+3$ & $+3$ & $-2$ & $-2$ & $+4$ & $-3$\\\hline
$U(1)_{\chi}$ & $-1$ & $+3$ & $+1$ & $-3$ & $+2$ & $-2$ & $0$ & $-5$\\\hline
\end{tabular}
\ \ \ \ \ \ \ \ \ .
\end{equation}
where the $Y$'s denote messenger fields, and the subscript indicates the
representation content. An interesting feature of this analysis is that it
suggests a potentially more general choice for the $U(1)_{PQ}$ charges. The
symmetry $U(1)_{\chi}$ is a linear combination of hypercharge and $U(1)_{B-L}%
$. In an integral normalization of $U(1)_{Y}$ so that the lepton doublet has
$U(1)_{Y}$ charge $-3$, the $B-L$ generator is given as:%
\begin{equation}
U(1)_{B-L}=-\frac{1}{5}U(1)_{\chi}+\frac{2}{15}U(1)_{Y}\text{.}%
\end{equation}
Note that the charge assignments in the messenger sector have non-trivial
$U(1)_{B-L}$ charge. As one final comment, although our conventions for the
$U(1)_{PQ}$ symmetry are chosen so as to coincide with those taken in
\cite{HVGMSB,BHSV}, there is a more general possibility for the
\textquotedblleft PQ deformation\textquotedblright\ given by taking an
arbitrary linear combination of $U(1)_{PQ}$ and $U(1)_{\chi}$.

The actual choice of the $U(1)_{PQ}$ generator naturally splits into two
distinct scenarios, depending on the action of the monodromy group. We find
that the monodromy group is isomorphic to $%
\mathbb{Z}
_{2}$, $%
\mathbb{Z}
_{2}\times%
\mathbb{Z}
_{2}$, $%
\mathbb{Z}
_{3}$ or $S_{3}$. Invariance under the corresponding subgroup largely fixes
the direction of the Cartan. Viewing $U(1)_{PQ}$ and $U(1)_{\chi}$ as elements
in the vector space dual to the weights spanned by the $t_{i}$'s, we find that
when the monodromy group is isomorphic to either $%
\mathbb{Z}
_{2}$ or $%
\mathbb{Z}
_{2}\times%
\mathbb{Z}
_{2}$,%
\begin{align}
G_{mono}^{Dirac}  &  \simeq%
\mathbb{Z}
_{2}\text{ or }%
\mathbb{Z}
_{2}\times%
\mathbb{Z}
_{2}\text{:}\\
t_{PQ}^{\ast}  &  =t_{1}^{\ast}+t_{2}^{\ast}-3t_{3}^{\ast}-3t_{4}^{\ast
}+4t_{5}^{\ast}\\
t_{\chi}^{\ast}  &  =-(t_{1}^{\ast}+t_{2}^{\ast}+t_{3}^{\ast}+t_{4}^{\ast
})+4t_{5}^{\ast}\text{,}%
\end{align}
where the $t_{i}^{\ast}$ correspond to weights in the dual space such that:%
\begin{equation}
t_{i}^{\ast}(t_{j})=\delta_{ij}\text{.}%
\end{equation}

The monodromy group action leaves these generators invariant. With respect to
this convention, $%
\mathbb{Z}
_{2}$ is generated by the permutation group element $(12)(34)$, which acts in
the obvious way on the $t_{i}$'s. The $%
\mathbb{Z}
_{2}\times%
\mathbb{Z}
_{2}$ group is generated by $(12)$ and $(34)$. In the case where the monodromy
group is isomorphic to $%
\mathbb{Z}
_{3}$ or $S_{3}$, the generators for $U(1)_{PQ}$ are instead given as:%
\begin{align}
G_{mono}^{Dirac} &  \simeq%
\mathbb{Z}
_{3}\text{ or }S_{3}\text{:}\\
t_{PQ}^{\ast} &  =t_{1}^{\ast}+t_{2}^{\ast}+t_{3}^{\ast}-3t_{4}^{\ast}\\
t_{\chi}^{\ast} &  =-(t_{1}^{\ast}+t_{2}^{\ast}+t_{3}^{\ast}+t_{4}^{\ast
})+4t_{5}^{\ast}\text{.}%
\end{align}
Without loss of generality, the $%
\mathbb{Z}
_{3}$ group can be taken to be generated by the three cycle $(123)$. Having
specified the underlying symmetries of each Dirac scenario, we now turn to the
explicit orbits in each case.

\subsubsection{$%
\mathbb{Z}
_{2}$ and $%
\mathbb{Z}
_{2}\times%
\mathbb{Z}
_{2}$ Orbits}

In this subsection we summarize the classification of the orbits in the cases
where the monodromy group is isomorphic to either $%
\mathbb{Z}
_{2}$ or $%
\mathbb{Z}
_{2}\times%
\mathbb{Z}
_{2}$. In this case, the generators for $U(1)_{PQ}$ and $U(1)_{\chi}$ lie in
the directions:%
\begin{align}
G  &  \simeq%
\mathbb{Z}
_{2}\text{ or }%
\mathbb{Z}
_{2}\times%
\mathbb{Z}
_{2}\text{:}\\
t_{PQ}^{\ast}  &  =t_{1}^{\ast}+t_{2}^{\ast}-3t_{3}^{\ast}-3t_{4}^{\ast
}+4t_{5}^{\ast}\\
t_{\chi}^{\ast}  &  =-(t_{1}^{\ast}+t_{2}^{\ast}+t_{3}^{\ast}+t_{4}^{\ast
})+4t_{5}^{\ast}\text{.}%
\end{align}
In the case where the monodromy group is generated by the order two element
$(12)(34)$, the orbits of the monodromy group are:%
\begin{align}
G_{mono}^{Dirac}  &  =\left\langle (12)(34)\right\rangle \simeq%
\mathbb{Z}
_{2}\text{:}\\
&  \text{Minimal Matter}\\
Orb(10_{M},Y_{10})  &  =t_{1},t_{2}\\
Orb(Y_{\overline{10}}^{\prime})  &  =-t_{3},-t_{4}\\
Orb(\overline{5}_{M})  &  =t_{4}+t_{5},t_{3}+t_{5}\\
Orb(5_{H})  &  =-t_{1}-t_{2}\\
Orb(\overline{5}_{H})  &  =t_{1}+t_{4},t_{2}+t_{3}\\
Orb(X^{\dag})  &  =t_{2}-t_{4},t_{1}-t_{3}\\
Orb(N_{R})  &  =t_{1}-t_{5},t_{2}-t_{5}%
\end{align}
for minimal matter. There are in principle additional charged matter fields,
which lie in the additional orbits:%
\begin{align}
G_{mono}^{Dirac}  &  =\left\langle (12)(34)\right\rangle \simeq%
\mathbb{Z}
_{2}\text{:}\\
&  \text{Extra Charged Matter}\\
Orb(10_{(1)})  &  =t_{5}\\
Orb(\overline{5}_{(1)})  &  =t_{1}+t_{3},t_{2}+t_{4}\\
Orb(\overline{5}_{(2)})  &  =t_{1}+t_{5},t_{2}+t_{5}\\
Orb(\overline{5}_{(3)})  &  =t_{3}+t_{4}\text{,}%
\end{align}
as well as additional matter curves supporting neutral fields:%
\begin{align}
G_{mono}^{Dirac}  &  =\left\langle (12)(34)\right\rangle \simeq%
\mathbb{Z}
_{2}\text{:}\\
&  \text{Extra Neutral}\\
Orb(D_{(1)})  &  =t_{1}-t_{2},t_{2}-t_{1}\\
Orb(D_{(2)})  &  =t_{1}-t_{4},t_{2}-t_{3}\\
Orb(D_{(3)})  &  =t_{3}-t_{4},t_{4}-t_{3}\\
Orb(D_{(4)})  &  =t_{3}-t_{5},t_{4}-t_{5}\text{.}%
\end{align}
Note that here, and in what follows, by `Extra Neutral' we mean extra fields
which are not charged under the standard model gauge group.

Each such orbit defines a matter curve, so that in addition to the dark chiral
matter listed, we also have the conjugate representations as well. Besides the
chiral matter localized on curves, there are also two zero weights, $Z_{PQ}$
and $Z_{\chi}$ which descend from the adjoint of $SU(5)_{\bot}$. These can be
identified with bulk modes which are roughly the center of mass motion degrees
of freedom for the seven-branes. The $U(1)_{PQ}$ and $U(1)_{\chi}$ charges for
all of the fields listed above are:%
\begin{align}
&
\begin{tabular}
[c]{|c|c|c|c|c|c|c|c|}\hline
Minimal & $10_{M},Y_{10}$ & $\overline{5}_{M}$ & $Y_{\overline{10}}^{\prime}$
& $5_{H}$ & $\overline{5}_{H}$ & $X^{\dag}$ & $N_{R}$\\\hline
$U(1)_{PQ}$ & $+1$ & $+1$ & $+3$ & $-2$ & $-2$ & $+4$ & $-3$\\\hline
$U(1)_{\chi}$ & $-1$ & $+3$ & $+1$ & $+2$ & $-2$ & $0$ & $-5$\\\hline
\end{tabular}
\\
&
\begin{tabular}
[c]{|c|c|c|c|c|}\hline
Extra Charged & $10_{(1)}$ & $\overline{5}_{(1)}$ & $\overline{5}_{(2)}$ &
$\overline{5}_{(3)}$\\\hline
$U(1)_{PQ}$ & $+4$ & $-2$ & $+5$ & $-6$\\\hline
$U(1)_{\chi}$ & $+4$ & $-2$ & $+3$ & $-2$\\\hline
\end{tabular}
\\
&
\begin{tabular}
[c]{|c|c|c|c|c|c|c|}\hline
Extra Neutral & $D_{(1)}$ & $D_{(2)}$ & $D_{(3)}$ & $D_{(4)}$ & $Z_{PQ}$ &
$Z_{\chi}$\\\hline
$U(1)_{PQ}$ & $0$ & $+4$ & $0$ & $-7$ & $0$ & $0$\\\hline
$U(1)_{\chi}$ & $0$ & $0$ & $0$ & $-5$ & $0$ & $0$\\\hline
\end{tabular}
\ \ \ \ \ \ \text{.}\
\end{align}

In the Dirac scenario where the monodromy group is enlarged to $%
\mathbb{Z}
_{2}\times%
\mathbb{Z}
_{2}=\left\langle (12),(34)\right\rangle $, the only difference is that the
size of the orbits are enlarged. The orbits of the monodromy group in this
case are:%
\begin{align}
G_{mono}^{Dirac}  &  =\left\langle (12),(34)\right\rangle \simeq%
\mathbb{Z}
_{2}\times%
\mathbb{Z}
_{2}\text{:}\\
&  \text{Minimal Matter}\\
Orb(10_{M},Y_{10})  &  =t_{1},t_{2}\\
Orb(Y_{\overline{10}}^{\prime})  &  =-t_{3},-t_{4}\\
Orb(\overline{5}_{M})  &  =t_{4}+t_{5},t_{3}+t_{5}\\
Orb(5_{H})  &  =-t_{1}-t_{2}\\
Orb(\overline{5}_{H})  &  =t_{1}+t_{4},t_{2}+t_{3},t_{2}+t_{4},t_{1}+t_{3}\\
Orb(X^{\dag})  &  =t_{2}-t_{4},t_{1}-t_{3},t_{1}-t_{4},t_{2}-t_{3}\\
Orb(N_{R})  &  =t_{1}-t_{5},t_{2}-t_{5},
\end{align}
while the extra charged matter is:%
\begin{align}
G_{mono}^{Dirac}  &  =\left\langle (12),(34)\right\rangle \simeq%
\mathbb{Z}
_{2}\times%
\mathbb{Z}
_{2}\text{:}\\
&  \text{Extra Charged Matter}\\
Orb(10_{(1)})  &  =t_{5}\\
Orb(\overline{5}_{(2)})  &  =t_{1}+t_{5},t_{2}+t_{5}\\
Orb(\overline{5}_{(3)})  &  =t_{3}+t_{4}\text{,}%
\end{align}
and the extra neutral states available are:%
\begin{align}
G_{mono}^{Dirac}  &  =\left\langle (12),(34)\right\rangle \simeq%
\mathbb{Z}
_{2}\times%
\mathbb{Z}
_{2}\text{:}\\
&  \text{Extra Neutral}\\
Orb(D_{(1)})  &  =t_{1}-t_{2},t_{2}-t_{1}\\
Orb(D_{(3)})  &  =t_{3}-t_{4},t_{4}-t_{3}\\
Orb(D_{(4)})  &  =t_{3}-t_{5},t_{4}-t_{5}\text{.}%
\end{align}

\subsubsection{$%
\mathbb{Z}
_{3}$ and $S_{3}$ Orbits}

We now turn to Dirac scenarios where the monodromy group is isomorphic to
either $%
\mathbb{Z}
_{3}$ or $S_{3}$. In this case, the two surviving $U(1)$ directions in the
Cartan of $SU(5)_{\bot}$ are:%
\begin{align}
G_{mono}^{Dirac}  &  \simeq%
\mathbb{Z}
_{3}\text{ or }S_{3}\text{:}\\
t_{PQ}^{\ast}  &  =t_{1}^{\ast}+t_{2}^{\ast}+t_{3}^{\ast}-3t_{4}^{\ast}\\
t_{\chi}^{\ast}  &  =-(t_{1}^{\ast}+t_{2}^{\ast}+t_{3}^{\ast}+t_{4}^{\ast
})+4t_{5}^{\ast}\text{.}%
\end{align}
With respect to this convention, the possible monodromy groups are generated
by either $(123)$, $(132)$, or the entire symmetric group on three letters,
$S_{3}$. In all cases, the orbits for the matter fields are the same. Using
the results of Appendix A, the resulting orbits are:%
\begin{align}
G_{mono}^{Dirac}  &  \simeq%
\mathbb{Z}
_{3}\text{ or }S_{3}\\
&  \text{Minimal Matter}\\
Orb(10_{M},Y_{10})  &  =t_{1},t_{2},t_{3}\\
Orb(Y_{\overline{10}}^{\prime})  &  =-t_{4}\\
Orb(\overline{5}_{M},Y_{\overline{5}}^{\prime})  &  =t_{1}+t_{5},t_{2}%
+t_{5},t_{3}+t_{5}\\
Orb(Y_{5})  &  =-t_{4}-t_{5}\\
Orb(5_{H})  &  =-t_{1}-t_{2},-t_{2}-t_{3},-t_{3}-t_{1}\\
Orb(\overline{5}_{H})  &  =t_{1}+t_{4},t_{2}+t_{4},t_{3}+t_{4}\\
Orb(X^{\dag})  &  =t_{1}-t_{4},t_{2}-t_{4},t_{3}-t_{4}\\
Orb(N_{R})  &  =t_{4}-t_{5}\text{.}%
\end{align}
In this case, the available extra matter fields are more limited, in
comparison to the other Dirac scenarios. The orbits of potentially extra
charged matter are:%
\begin{align}
G_{mono}^{Dirac}  &  \simeq%
\mathbb{Z}
_{3}\text{ or }S_{3}\text{ Orbits}\\
&  \text{Extra Charged Matter }\\
Orb(10_{(1)})  &  =t_{5}\text{,}%
\end{align}

Next consider extra GUT\ singlets which descend from the adjoint of
$SU(5)_{\bot}$. Due to the action of the monodromy group, there are only two
zero weights, which we denote as $Z_{PQ}$ and $Z_{\chi}$. \ The remaining
weights are of the form $t_{m}-t_{n}$ and localize on curves in the geometry.
\ The available orbits can all be obtained by acting with a $%
\mathbb{Z}
_{3}$ subgroup of the monodromy group. Up to complex conjugation, the orbits
are:%
\begin{align}
&  G_{mono}^{Dirac}\simeq%
\mathbb{Z}
_{3}\text{ or }S_{3}\text{ Orbits}\\
&  \text{Extra Singlets}\\
Orb(D_{(1)})  &  =t_{1}-t_{5},t_{2}-t_{5},t_{3}-t_{5}\\
Orb(D_{(2)})  &  =t_{2}-t_{3},t_{3}-t_{1},t_{1}-t_{2}\text{.}%
\end{align}
In addition, there are again two zero weights, $Z_{PQ}$ and $Z_{\chi}$ given
by the two unidentified directions $U(1)_{PQ}$ and $U(1)_{\chi}$. These
correspond to bulk modes which do not localize on curves. This is also the
only scenario in which messengers in the $10\oplus\overline{10}$ and
$5\oplus\overline{5}$ are both in principle possible.

Finally, the charges of the various fields under these $U(1)$ symmetries are:%
\begin{align}
&
\begin{tabular}
[c]{|c|c|c|c|c|c|c|c|c|}\hline
Minimal & $10_{M},Y_{10}$ & $\overline{5}_{M},Y_{\overline{5}}^{\prime}$ &
$Y_{5}$ & $Y_{\overline{10}}^{\prime}$ & $5_{H}$ & $\overline{5}_{H}$ &
$X^{\dag}$ & $N_{R}$\\\hline
$U(1)_{PQ}$ & $+1$ & $+1$ & $+3$ & $+3$ & $-2$ & $-2$ & $+4$ & $-3$\\\hline
$U(1)_{\chi}$ & $-1$ & $+3$ & $-3$ & $+1$ & $+2$ & $-2$ & $0$ & $-5$\\\hline
\end{tabular}
\\
&
\begin{tabular}
[c]{|c|c|}\hline
Extra Charged & $10_{(1)}$\\\hline
$U(1)_{PQ}$ & $0$\\\hline
$U(1)_{\chi}$ & $+4$\\\hline
\end{tabular}
\\
&
\begin{tabular}
[c]{|c|c|c|c|c|}\hline
Extra Neutral & $D_{(1)}$ & $D_{(2)}$ & $Z_{PQ}$ & $Z_{\chi}$\\\hline
$U(1)_{PQ}$ & $+1$ & $0$ & $0$ & $0$\\\hline
$U(1)_{\chi}$ & $-5$ & $0$ & $0$ & $0$\\\hline
\end{tabular}
\ \ \ \text{.}%
\end{align}

\subsection{Majorana Scenarios: $\mathbb{Z}_{2} \times\mathbb{Z}_{2}$ and
$Dih_{4}$ Orbits \label{ssec:Maj}}

In this subsection we discuss minimal Majorana neutrino scenarios. As in
\cite{BHSV}, we shall assume that the right-handed neutrino states localize on
matter curves. In this case, the covering theory must contain interaction
terms which are schematically of the form:%
\begin{equation}
L\supset\int d^{2}\theta\text{ }H_{u}LN_{R}+H_{u}^{\prime}L^{\prime}%
N_{R}^{\prime}+M_{\text{maj}}N_{R}N_{R}^{\prime}\text{,}%
\end{equation}
where the primes denote fields in the same orbit. The actual list of
interaction terms will typically be much larger. This can partially be traced
to the presence of a Majorana mass term, which couples states with conjugate
quantum numbers. This has the effect of also eliminating candidate global
$U(1)$ symmetries in the quotient theory which would be explicitly broken by
the Majorana mass term.

Majorana scenarios admit three consistent choices of orbits. Before proceeding
to explicit examples, we first focus on the common aspects of all such
scenarios, and defer a more complete characterization to Appendix A. In all
cases, the monodromy group action identifies all four $U(1)$ factors of
$SU(5)_{\bot}$. Adopting the same convention used in Appendix A, this leads to
a unique direction in the Cartan for $U(1)_{PQ}$:%
\begin{equation}
t_{PQ}^{\ast}=t_{1}^{\ast}+t_{2}^{\ast}+t_{3}^{\ast}-4t_{4}^{\ast}+t_{5}%
^{\ast}\text{.}%
\end{equation}
By inspection, this choice of $U(1)_{PQ}$ is consistent with permutations
which leave $t_{4}$ invariant. Thus, the monodromy group must be a subgroup of
$S_{4}$, the symmetric group on the remaining $t_{i}$'s. The $U(1)_{PQ}$
charges of all visible matter are:%
\begin{equation}%
\begin{tabular}
[c]{|c|c|c|c|c|c|c|c|}\hline
Visible Matter & $10_{M},Y_{10}$ & $Y_{\overline{10}}^{\prime}$ &
$\overline{5}_{M}$ & $5_{H}$ & $\overline{5}_{H}$ & $X^{\dag}$ & $N_{R}%
$\\\hline
$U(1)_{PQ}$ & $+1$ & $+4$ & $+2$ & $-2$ & $-3$ & $+5$ & $0$\\\hline
\end{tabular}
\ \ \ .
\end{equation}
An interesting feature of all Majorana scenarios is that the only messenger
fields which can be accommodated transform in the $10\oplus\overline{10}$ of
$SU(5)$. In addition, note that in this case, the right-handed neutrinos are
neutral under $U(1)_{PQ}$.

An additional feature of such scenarios is that because the $X$ field carries odd
PQ charge, its vev will not retain a $\mathbb{Z}_{2}$ subgroup, so we cannot embed
matter parity inside of $U(1)_{PQ}$. Rather, we shall assume that there is an
additional source of $\mathbb{Z}_{2}$ matter parity under which $X$ and the Higgs fields have
charge $+1$, while the MSSM chiral matter supefields have charge $-1$. Such symmetries can in principle
originate from the geometry of an F-theory compactification, as discussed for example in \cite{BHVII}.

We now proceed to list the consistent Majorana orbits. In the case where the
mondromy group is $\left\langle (12)(35),(13)(25)\right\rangle \simeq%
\mathbb{Z}
_{2}\times%
\mathbb{Z}
_{2}$, the corresponding orbits are:%
\begin{align}
G_{mono}^{Maj} &  =\left\langle (12)(35),(13)(25)\right\rangle \simeq%
\mathbb{Z}
_{2}\times%
\mathbb{Z}
_{2}\\
&  \text{Minimal Matter}\\
Orb(10_{M},Y_{10}) &  =t_{1},t_{2},t_{3},t_{5}\\
Orb(Y_{\overline{10}}^{\prime}) &  =-t_{4}\\
Orb(\overline{5}_{M}) &  =t_{2}+t_{3},t_{5}+t_{1}\\
Orb(5_{H}) &  =-t_{1}-t_{2},-t_{5}-t_{3}\\
Orb(\overline{5}_{H}) &  =t_{1}+t_{4},t_{2}+t_{4},t_{5}+t_{4},t_{3}+t_{4}\\
Orb(X^{\dag}) &  =t_{2}-t_{4},t_{1}-t_{4},t_{5}-t_{4},t_{3}-t_{4}\\
Orb(N_{R}) &  =\pm\left(  t_{1}-t_{3}\right)  ,\pm\left(  t_{5}-t_{2}\right)
\end{align}
Besides fields with non-trivial weights, there is also a single zero weight
$Z_{PQ}$, which lies in the same direction as the $U(1)_{PQ}$ generator. For
small monodromy group orbits, there is one possible extra charged matter curve
given as:%
\begin{align}
\text{ }G_{mono}^{Maj} &  =\left\langle (12)(35),(13)(25)\right\rangle \simeq%
\mathbb{Z}
_{2}\times%
\mathbb{Z}
_{2}\\
&  \text{Extra Charged}\\
Orb(\overline{5}_{(1)}) &  =t_{1}+t_{3},t_{2}+t_{5}\text{.}%
\end{align}

Next consider extra GUT\ singlets of the theory. Since there is a single
available $U(1)$ in $SU(5)_{\bot}$, only one zero weight denoted as $Z_{PQ}$
can descend from the adjoint of $SU(5)_{\bot}$. The remaining weights of the
adjoint are of the form $t_{m}-t_{n}$. Under the provided monodromy group
action, these separate into the following orbits:%
\begin{align}
G_{mono}^{Maj} &  =\left\langle (12)(35),(13)(25)\right\rangle \simeq%
\mathbb{Z}
_{2}\times%
\mathbb{Z}
_{2}\\
&  \text{Extra Neutral}\\
Orb(D_{(1)}) &  =\pm(t_{1}-t_{2}),\pm(t_{3}-t_{5})\\
Orb(D_{(2)}) &  =\pm(t_{1}-t_{5}),\pm(t_{2}-t_{3})\text{.}%
\end{align}
The actual charge assignments in the case where the monodromy group is
$\mathbb{Z}_{2}\times\mathbb{Z}_{2}$ are given as:%
\begin{align}
&
\begin{tabular}
[c]{|c|c|c|c|c|c|c|c|}\hline
Visible Matter & $10_{M},Y_{10}$ & $Y_{\overline{10}}^{\prime}$ &
$\overline{5}_{M}$ & $5_{H}$ & $\overline{5}_{H}$ & $X^{\dag}$ & $N_{R}%
$\\\hline
$U(1)_{PQ}$ & $+1$ & $+4$ & $+2$ & $-2$ & $-3$ & $+5$ & $0$\\\hline
\end{tabular}
\\
&
\begin{tabular}
[c]{|c|c|}\hline
Extra Charged & $\overline{5}_{(1)}$\\\hline
$U(1)_{PQ}$ & $+2$\\\hline
\end{tabular}
\\
&
\begin{tabular}
[c]{|c|c|c|c|}\hline
Extra Neutral & $D_{(1)}$ & $D_{(2)}$ & $Z_{PQ}$\\\hline
$U(1)_{PQ}$ & $0$ & $0$ & $0$\\\hline
\end{tabular}
\ \ \ \ \text{.}%
\end{align}
Thus, in this case, there is exactly one additional unused orbit,
corresponding to the $\overline{5}_{(1)}$.

The other Majorana scenarios simply correspond to enlarging the monodromy
group by allowing more elements from $S_{4}$ to participate in the various
orbits. There are two remaining monodromy groups of order eight, which are
both isomorphic to the dihedral group $Dih_{4}$ given by rotations and
reflections of the square. The explicit generators for these two monodromy
groups are:%
\begin{align}
G_{8}^{Maj(2)} &  \simeq\left\langle (13)(25),(1253)\right\rangle \simeq
Dih_{4}\simeq%
\mathbb{Z}
_{2}\ltimes%
\mathbb{Z}
_{4}\\
G_{8}^{Maj(3)} &  =\left\langle (13)(25),(1325)\right\rangle \simeq
Dih_{4}\simeq%
\mathbb{Z}
_{2}\ltimes%
\mathbb{Z}
_{4}\text{,}%
\end{align}
where the $%
\mathbb{Z}
_{2}$ acts by inversion on the $%
\mathbb{Z}
_{4}$ factor of the semi-direct product. This also enlarges the orbits of some
of the visible matter field sectors, and identifies some of the orbits for the
dark matter fields. We refer the interested reader to Appendix A for a full
classification of all possible orbits. Quite remarkably, in the two cases of
the maximal monodromy group where $G_{mono}^{Maj}\simeq Dih_{4}$, \textit{all}
of the orbits in the visible sector are utilized except for one $10$ curve.
But this is precisely where the messenger $\overline{10}$ can localize! Thus,
the $E_{8}$ enhancement point is just flexible to accommodate all of the
minimal matter, but nothing more! In this case, the matter content and charge
assignments are:
\begin{align}
&
\begin{tabular}
[c]{|c|c|c|c|c|c|c|c|}\hline
Visible Matter & $10_{M},Y_{10}$ & $Y_{\overline{10}}^{\prime}$ &
$\overline{5}_{M}$ & $5_{H}$ & $\overline{5}_{H}$ & $X^{\dag}$ & $N_{R}%
$\\\hline
$U(1)_{PQ}$ & $+1$ & $+4$ & $+2$ & $-2$ & $-3$ & $+5$ & $0$\\\hline
\end{tabular}
\\
&
\begin{tabular}
[c]{|c|c|c|}\hline
Extra Neutral & $D_{(1)}$ & $Z_{PQ}$\\\hline
$U(1)_{PQ}$ & $0$ & $0$\\\hline
\end{tabular}
\ \ \ \text{.}%
\end{align}

\subsection{$E_{8}$ and the Absence of Exotica}

The presence of a single $E_{8}$ interaction point also addresses a puzzle
encountered in the context of the gauge mediated supersymmetry breaking sector
introduced in \cite{HVGMSB}. While the PQ charge assignments for matter fields
suggestively hinted at higher unification structures such as $E_{6}$, $E_{7}$
and $E_{8}$, in \cite{HVGMSB}, it was assumed that this required full
multiplets such as the $27$ of $E_{6}$ to localize on matter curves in the
$SU(5)$ GUT seven-brane. This would appear to require a local enhancement from
$SU(5)$ to at least $E_{7}$. In analyzing the matter trapped along the
corresponding curve, however, it was often difficult to remove additional
exotic states from the low energy theory. For example, the adjoint $133$ of
$E_{7}$ breaks to $E_{6}\times U(1)$ as:%
\begin{align}
E_{7} &  \supset E_{6}\times U(1)\\
133 &  \rightarrow78_{0}+27_{2}+\overline{27}_{-2}+1_{0}\text{.}%
\end{align}
Breaking further to $SU(5)$, it seemed in \cite{HVGMSB} that in addition to
the states in the $27$, the matter curve would also contain irreducible
components descending from the $78$ of $E_{6}$. Here, we have seen that rank
one enhancements along matter curves are sufficient in realizing these higher
unification structures. Indeed, there are no exotic fields in this case, and
the presence of the E-type singularity is instead concentrated at a point of
the geometry.

\subsection{Monodromy and Messengers}

An intriguing byproduct of the classification of monodromy group orbits is
that while it is indeed possible to accommodate a messenger sector, some of
the messengers must localize on the same curve as the matter fields of the
MSSM. Moreover, we have also seen that aside from the $%
\mathbb{Z}
_{3}$ or $S_{3}$ Dirac scenarios, in all scenarios the only available
messengers transform in the $10\oplus\overline{10}$ of $SU(5)_{GUT}$. It is
important to note that the presence of additional matter fields does not alter
the main features of the flavor hierarchy found in \cite{HVCKM} and
\cite{BHSV} which was obtained by estimating the overlap of wave functions. A
crucial ingredient in the flavor hierarchy story is that the internal wave
function for the massive generation must not vanish near the interaction
point. Indeed, letting $\psi^{(1)},...,\psi^{(N)}$ denote the internal wave
functions for the matter fields localized on a curve, a generic linear
combination of these wave functions corresponds to the matter field which
actually couples to $X$ in the messenger sector. Thus, provided there is
another interaction point where the messengers and $X$ fields meet,
the other linear combinations for the matter fields will not vanish at the interaction point,
so that the analysis of flavor hierarchies found in \cite{HVCKM} and
\cite{BHSV} will carry over to this case as well.

Another consequence of localizing the messenger field $Y_{10}$ on the same
curve as the $10_{M}$ is that these messengers will now couple to MSSM\ fields
through the coupling:%
\begin{equation}
5_{H}Y_{10}10_{M}\text{.}%
\end{equation}
This provides a rapid channel of decay for messengers to MSSM\ fields. Similar
considerations hold for the $%
\mathbb{Z}
_{3}$ or $S_{3}$ Dirac scenarios with messengers in the $5\oplus\overline{5}$.
In particular, this means that the messengers will not persist as thermal
relics. This is a welcome feature of this model, because stable thermal relics
of this type would lead to far greater matter density than is observed. To a
certain extent, this is to be expected because the classification of monodromy
orbits performed earlier eliminates all $U(1)$ symmetries other than
$U(1)_{PQ}$, and in the case of Dirac neutrino scenarios, also $U(1)_{\chi}$.
Since there is no conserved messenger number, it is natural to expect nothing
to protect the messengers from exhibiting such decays.

We have also seen that in all but one scenario, the messenger fields organize
into vector-like pairs in the $10\oplus\overline{10}$ of $SU(5)$, rather than
the $5\oplus\overline{5}$. Recall that in minimal gauge mediation, the soft
mass terms are controlled by $C(R)$ of the messenger fields, where:%
\begin{equation}
T_{ij}^{A}T_{ji}^{B}=C(R)\delta^{AB}%
\end{equation}
for matrices $T_{ij}^{A}$ in the representation $R$. Since $C(10)/C(5)=3$, it
follows that each $10$ effectively counts as three $5$'s.\footnote{More
generally, recall that for the group $SU(N)$, $C(N)=1/2$ and for the two index
anti-symmetric representation of dimension $N(N-1)/2$, $C(N(N-1)/2)=(N-2)/2$.}
The presence of $10$'s also scales the soft scalar masses relative to their
gaugino counterparts. Indeed, letting $m_{(1)}^{scalar}$ and $m_{(1)}%
^{gaugino}$ denote the soft masses in the case of a minimal gauge mediation
scenario with a single vector-like pair in the $5\oplus\overline{5}$, a
minimal gauge mediation model with $N_{5}$ vector-like pairs in the
$5\oplus\overline{5}$ has soft masses:
\begin{align}
m^{scalar}  &  =\sqrt{N_{5}}\cdot m_{(1)}^{scalar}\\
m^{gaugino}  &  =N_{5}\cdot m_{(1)}^{gaugino}\text{.}%
\end{align}
Thus, the soft scalar masses decrease relative to the gaugino masses.

For the generic case of interest where $N_{5}$ is a multiple of three, this
causes the lightest stau to become comparable in mass to the bino. In the
presence of the PQ\ deformation, this can lead to a further decrease in the
lightest stau mass which quite suggestively hints at scenarios with a stau
NLSP. Furthermore, this also lowers the upper bound on $\Delta_{PQ}$. On the
other hand, the size of the PQ deformation is bounded below by the requirement
that the saxion have available a decay channel to MSSM particles
\cite{FGUTSCosmo}. This then leads to an interesting constraint on the viable
window of PQ deformations.

\subsection{Semi-Visible TeV Scale Dark Matter Candidates\label{ssec:SEMIVIS}}

Strictly speaking, a viable dark matter candidate need only be neutral under
$U(1)_{EM}$ and not under all of $SU(5)_{GUT}$. Indeed, bino LSP\ scenarios
provide an explicit example of precisely this type. With this in mind, it is
therefore important to examine potential TeV scale dark matter candidates
which are electrically neutral, but not necessarily fully neutral under
$SU(5)_{GUT}$. The $5$ and $10$ of $SU(5)_{GUT}$ decompose into irreducible
representations of the Standard Model gauge group as:%
\begin{align}
SU(5)_{GUT}  &  \supset SU(3)_{C}\times SU(2)_{L}\times U(1)_{Y}\\
5  &  \rightarrow(3,1)_{-2}+(1,2)_{3}\\
10  &  \rightarrow(1,1)_{6}+(\overline{3},1)_{-4}+(3,2)_{1}\text{.}%
\end{align}
Once $SU(2)_{L}\times U(1)_{Y}$ breaks to $U(1)_{EM}$, it follows that only
the $5$ and $\overline{5}$ possess electrically neutral candidates, which can
be thought of as the \textquotedblleft neutrino component\textquotedblright%
\ of the fundamental and anti-fundamental. Note further that in order to
retain successful gauge coupling unification, a full GUT\ multiplet must
persist below the GUT scale.\footnote{In principle, one could consider
scenarios where a piece of the messenger is retained, and another piece of the
dark matter GUT\ multiplet is also kept, retaining unification. This appears
to be a somewhat artificial means by which to preserve unification in this
context, so we shall not pursue this possibility here. In the context of
F-theory GUTs, more motivated choices for splitting up GUT\ multiplets through
appropriate fluxes can also be arranged, for example as in \cite{Font:2008id}%
.}

The available orbits for additional $5$'s are typically quite limited. As we
now explain, minimal realizations of TeV scale dark matter are problematic in
this context, and can only be accommodated in a single Dirac neutrino
scenario.\footnote{We thank T. Hartman for a related discussion which prompted
this analysis.} \ Since interaction terms with the Higgs fields both involve
couplings to a $10$ of $SU(5)_{GUT}$, we shall assume that the same mechanism
responsible for generating the $\mu$-term is also responsible for generating
mass for the dark matter candidate so that:%
\begin{equation}
\int d^{4}\theta\frac{X^{\dag}D_{5}D_{\overline{5}}^{\prime}}{\Lambda
_{\text{UV}}}\text{.} \label{XDD}%
\end{equation}
Thus, when $X$ develops a supersymmetry breaking vev, it will induce a mass of
$\mu\sim100-1000$ GeV for the bosons and fermions of the dark matter multiplet
$D_{5}$ and $D_{\overline{5}}^{\prime}$.

The interaction term of line (\ref{XDD}) is difficult to incorporate at a
single point of $E_{8}$ unification. The reason for this obstruction is quite
similar to the one already encountered for messengers. In all scenarios
considered in this section, the orbit of $X^{\dag}$ contains a weight of the
form:%
\begin{equation}
Orb(X^{\dag})\ni t_{2}-t_{4}\text{.}%
\end{equation}
Thus, the weights for $D_{5}$ and $D_{\overline{5}}$ must satisfy the
constraint:%
\begin{equation}
(t_{2}-t_{4})+(-t_{i}-t_{j})+(t_{k}+t_{l})=0\text{,}%
\end{equation}
so that the weights for $D_{5}$ and $D_{\overline{5}}$ are respectively of the
form $-t_{2}-t_{i}$ and $t_{4}+t_{i}$. The options for semi-visible dark
matter are therefore quite limited. Indeed, there are essentially three
possible options for all of the dark matter candidates:%
\begin{align}
\text{Option 1} &  \text{: }Orb(D_{5})\ni-t_{2}-t_{1}\text{, }Orb(D_{\overline
{5}})\ni t_{4}+t_{1}\\
\text{Option 2} &  \text{: }Orb(D_{5})\ni-t_{2}-t_{3}\text{, }Orb(D_{\overline
{5}})\ni t_{4}+t_{3}\\
\text{Option 3} &  \text{: }Orb(D_{5})\ni-t_{2}-t_{5}\text{, }Orb(D_{\overline
{5}})\ni t_{4}+t_{5}\text{.}%
\end{align}
Returning to the discussion of viable orbits in Dirac and Majorana scenarios,
the analysis of subsections \ref{ssec:Dirac} and \ref{ssec:Maj} establishes
that typically, the length of the orbits other than $\overline{5}_{M}$ are too
large to admit additional $5$'s in full GUT\ multiplets. Rather, we find that
only in the $%
\mathbb{Z}
_{2}$ and $%
\mathbb{Z}
_{2}\times%
\mathbb{Z}
_{2}$ Dirac neutrino scenarios where the orbit $\overline{5}_{M}$ is:%
\begin{equation}
Orb(\overline{5}_{M})=t_{4}+t_{5},t_{3}+t_{5}%
\end{equation}
is it even possible to accommodate additional dark matter in the
$5\oplus\overline{5}$. The corresponding orbits for these fields are then:%
\begin{align}
Orb(\overline{5}_{M},D_{\overline{5}}) &  =t_{4}+t_{5},t_{3}+t_{5}\\
Orb(D_{5}) &  =-t_{1}-t_{5},-t_{2}-t_{5}\text{,}%
\end{align}
and theire charges under $U(1)_{PQ}$ and $U(1)_{\chi}$ are:%
\begin{equation}%
\begin{tabular}
[c]{|c|c|c|}\hline
Semi-Dark & $\overline{5}_{M},D_{\overline{5}}$ & $D_{5}$\\\hline
$U(1)_{PQ}$ & $+1$ & $-5$\\\hline
$U(1)_{\chi}$ & $+3$ & $-3$\\\hline
\end{tabular}
\ \ \ \ \ \ \ \ \ \text{.}%
\end{equation}
Even so, this by itself is insufficient for specifying a dark matter
candidate. We will encounter some obstructions to realizing such a candidate
in section \ref{sec:FVS}.

\section{{Bifundamentals and E-type Singularities\label{sec:ESHIELD}}}

In the previous section we focussed primarily on the constraints derived from
the condition that all of the matter of the MSSM\ consistently embed inside
the unfolding of a single $E_{8}$ singularity. \ Indeed, the E-type structure
is what leads to an appealing unification structure as well as geometric
rigidity in the construction. In the context of a more complete geometry,
tadpole cancellation considerations generically requires the presence of
additional seven-branes, as well three-branes. \ These additional branes will
introduce additional gauge group factors, which can in principle contribute to
the low energy phenomenology of the theory.

Given the presence of such objects, it is natural to study whether additional
bifundamental matter could localize at a pairwise intersection of an E-type
seven-brane with another seven-brane. \ Even though this would at first appear
to represent a mild generalization of the quiver-type constructions which are
quite common in perturbatively realized intersecting seven-brane
configurations, the presence of an exceptional gauge symmetry significantly
shields the E-type seven-brane from massless bifundamentals charged under
gauge groups of the form $E_{8}\times G$. As we now explain, states localized
at such pairwise intersections correspond to more exotic objects which resist
an interpretation in terms of conventional particles.

To see the source of this restriction, we first review how bifundamental
matter charged under distinct seven-branes comes about in F-theory
constructions.\ Letting $G_{S}$ and $G_{S^{\prime}}$ denote the gauge groups
of the corresponding seven-branes wrapping surfaces $S$ and $S^{\prime}$,
suppose that $S$ and $S^{\prime}$ intersect over a matter curve $\Sigma$.
\ This curve corresponds to the locus of the elliptic fibration where the
singularity $G_{S}\times G_{S^{\prime}}$ enhances to $G_{\Sigma}\supset
G_{S}\times G_{S^{\prime}}$.\ \textquotedblleft
Bifundamental\textquotedblright\ matter trapped on the curve can be analyzed
in terms of a gauge theory with gauge group $G_{\Sigma}$ which locally Higgses
to $G_{S}$ along $S$ and $G_{S^{\prime}}$ along $S^{\prime}$ \cite{KatzVafa}.
\ As an example, note that this covers perturbatively realized intersecting
seven-brane configurations, where $G_{S}=SU(N)$, $G_{S^{\prime}}=SU(M)$ and
$G_{\Sigma}=SU(N+M)$.

While this clearly generalizes (at least in a non-compact model) to arbitrary
$N$ and $M$, exceptional singularities impose far more significant
restrictions. \ Indeed, the gauge theory description just presented requires
that $G_{\Sigma}$ admits an interpretation as a gauge theory which is locally
Higgsed. \ On the other hand, if $G_{S}=E_{8}$, then there is no compact
simple gauge group such that $G\supset E_{8}\times G_{S^{\prime}}%
$.\footnote{\ Although there is no sense in which a compact simple gauge group
contains $E_{8}\times G_{S^{\prime}}$, affine extensions such as $E_{9}\supset
E_{8}$ provide one possible gauge theory interpretation. In fact this seems
compatible with the interpretation of tensionless strings carrying an $E_{8}$
current algebra as the analog of \textquotedblleft bifundamental
matter\textquotedblright.} \ As a consequence, there is little sense in which
\textquotedblleft bifundamental\textquotedblright\ particle states localize
along the intersection locus. \ Note that this does not require the presence
of a single, globally defined $E_{8}$ singularity which unfolds over an
isolated surface. \ Indeed, it is in principle possible to consider geometries
where the singularity enhancements embed in different $E_{8}$ gauge groups.

This begs the question, however, as to what occurs when an $E_{8}$ singularity
collides with another stack of seven-branes with gauge group $G_{S^{\prime}}$.
These types of colliding singularities can occur geometrically and in fact
have been studied in \cite{Bershadsky:1996nu}. In six dimensional theories,
this phenomenon leads to tensionless strings. In the context of F-theory
compactified on a Calabi-Yau threefold, this corresponds to the presence of a
vanishing $\mathbb{P}^{1}$ in the base wrapped by a D3-brane. Compactifying
the six-dimensional theory on a $T^{2}$, a T-dual version of this intersecting
brane configuration maps to zero size $E_{8}$ instantons. For example, a
seven-brane which carries $E_{8}$ gauge symmetry can be Higgsed by turning on
an instanton. A zero size instanton of the seven-brane theory carries non-zero
D3-brane charge. Thus, we can view a zero size $E_{8}$ instanton as an
intersection of a single D3-brane with the $E_{8}$ seven-brane. As argued in
\cite{WittenSmall}, the small instanton limit includes tensionless strings.
The presence of such tensionless strings can be deduced as follows. One can
lift this configuration to an M-theory configuration with an M5-brane near the
boundary of the space which carries an $E_{8}$ symmetry. If the M5-brane is
brought to the boundary, it corresponds to a zero size $E_{8}$ instanton. On
the other hand an M2-brane stretched between the M5-brane and the boundary
leads to a tensionless string as the M5-brane approaches the boundary. This
can be further analyzed in F-theory, as in
\cite{Seiberg:1996vs,MorrisonVafaII,Klemm:1996hh}, and a great deal of
information can be extracted from various duality chains. In particular one
can construct the elliptic genus of such E-strings \cite{Minahan:1998vr}.

Note, however, that although tensionless strings can be BPS objects in 5 and 6
dimensions, this is not the case in four dimensions. The remnants of such
tensionless strings in four dimensions are conformal fixed points. Thus, the
exotic possibility of a brane intersecting an E-brane can be phrased in four
dimensional language as the study of a conformal theory with an E-type global
symmetry, as studied for example in \cite{Minahan:1996fg,Minahan:1996cj}. Some
examples of precisely this type have recently been studied in
\cite{Argyres:2007cn}, and F-theory based constructions involving D3-brane
probes of seven-branes with E-type gauge symmetries have been analyzed in
\cite{Aharony:2007dj}, which can be viewed roughly as T-dual to the theories
we are considering.

In the present context, the $E_{8}$ symmetry need only be restored at a single
point of the GUT\ seven-brane. In other words, one can view our background as
a theory near an $E_{8}$ type seven-brane. Thus, if any other brane intersects
the GUT or PQ seven-branes, we have the remnant of a \textquotedblleft
nearly\textquotedblright\ conformal theory with an approximate E-type global symmetry.

We are thus led to potentially consider the possibility of an additional
sector which is nearly conformal with an approximate E-type symmetry. In other
words we could contemplate having an \textquotedblleft unparticle
sector\textquotedblright, with terminology as in \cite{Georgi:2007ek}. As the
qualification of the previous paragraph suggests, it is first important to
settle how badly the conformal symmetry of the theory is broken. There are at
least two reasons to expect that the conformal symmetry is badly broken at
scales below $M_{GUT}$ or $M_{\ast}$, the string scale. The first reason has
to do with the fact that the scalar vevs in the geometry which break the
$E_{8}$ symmetry to $SU(5)$ are typically of the GUT scale or higher (as they
determine the background geometry), and thus a typical brane will not be
particularly close to the $E_{8}$ point.\ Thus, the would be \textquotedblleft
tensionless string\textquotedblright\ will have a tension on the order of
$T\sim M_{\ast}^{3}\cdot R_{GUT}\sim(10^{15}$ GeV$)^{2}$. Hence, the mass
scale for the operator responsible for breaking conformal invariance is
relatively large.

Another reason why the conformal symmetry should be badly broken is the
assumption of naturalness: If the mass scale for breaking of conformal
symmetry is small, we will have a tower of particles charged under the
Standard Model group which will dramatically accelerate the running of the
coupling constants, and thus make the unification scale much smaller than
$10^{16}$ GeV. This could lead to a certain amount of tension with current
constraints on the lifetime of the proton, since lowering the GUT\ scale will
also decrease the suppression scale of dimension six operators. Moreover,
lowering the GUT\ seven-brane far below $10^{16}$ GeV may be difficult to
arrange in practice without significant fine tuning.

Thus, even if there are such sectors we are naturally led to consider the
corresponding conformal symmetry to be badly broken at energies near the weak
scale. Without knowing more about the nature of such conformal theories and
their breaking, it is difficult to say much about the light spectrum of
particles associated with this conformal system, or whether any such particles
will even survive to low energies. Although perhaps somewhat exotic, this
possibility might produce novel phenomenological signatures at low energies,
and would be interesting to study further.

\section{The Sequestered Sector\label{sec:FULLDARK}}

The progression of ideas so far has been to examine potential dark objects,
first \textquotedblleft inside\textquotedblright\ of the $E_{8}$ singularity.
Next, we examined possible dark objects which may only \textquotedblleft
touch\textquotedblright\ some part of the $E_{8}$ singularity. As the next
logical step, in this section we consider objects which are \textquotedblleft
fully dark\textquotedblright\ in the sense that they do not directly intersect
any seven-brane connected with the $E_{8}$ enhancement point.

Such dark objects can either correspond to degrees of freedom localized on
other branes, or fields which propagate in the threefold base of the
compactification. In the context of a local model, these degrees of freedom
will decouple in the same limit which turns off the effects of gravity. If the
corresponding degrees of freedom do not develop high scale supersymmetric
masses, the corresponding masses of such particles are then cut off from the
PQ seven-brane, and will not lead to TeV scale objects. For example, the
gravitino is of this type, and has a mass in the far lower range of $10-100$
MeV. Other bulk gravitational modes have similar masses, but will also
interact quite weakly with the visible sector.

Next consider modes localized on three-branes and seven-branes, which will
typically be present in order to satisfy tadpole cancellation conditions. Here
there is again a problem with generating a TeV scale dark matter
candidate because the associated degrees of freedom for these branes only indirectly communicate with the PQ seven-brane. Although
somewhat ad hoc, we can still consider scenarios where such objects develop a TeV scale
mass through some other mechanism. Even
though these degrees of freedom are geometrically \textquotedblleft
sequestered\textquotedblright\ from the GUT\ seven-brane, they can still
interact non-trivially with the visible sector. In particular, abelian gauge
fields for these branes can mix non-trivially with the $U(1)_{Y}$ gauge boson
of the Standard Model. As we now explain, this type of effect is difficult to
avoid, and suggests that in fact all of the abelian gauge bosons will mix to a
certain extent. This would suggest that ``dark gauge bosons'' and ``dark gauginos'' cannot
be dark matter candidates, as they will decay too rapidly.

\subsection{Ubiquitous Kinetic Mixing}

In this subsection we show that dark gauge bosons associated with
\textquotedblleft sequestered branes\textquotedblright\ generically mix with
$U(1)_{Y}$ because massive states charged under both groups induce one loop
mixing effects. \ This kinetic mixing then provides a source by which other
aspects of dark physics can potentially communicate with the Standard Model.
\ To illustrate the main features of kinetic mixing, we first review this
phenomenon in the context of a $U(1)_{1}\times U(1)_{2}$ gauge theory with
bifundamental matter. Next, we explain how in the context of string based
constructions, such kinetic mixing will generically arise from massive modes
charged under both group factors.

We now review the primary features of kinetic mixing \cite{Holdom:1985ag} in
the context of a $U(1)\times U(1)$ field theory with Lagrangian:%
\begin{equation}
L\supset\operatorname{Im}\tau_{1}\int d^{2}\theta W_{(1)}^{\alpha}W_{\alpha
}^{(1)}+\operatorname{Im}\tau_{2}\int d^{2}\theta W_{(2)}^{\alpha}W_{\alpha
}^{(2)}+\operatorname{Im}\varepsilon\int d^{2}\theta W_{(1)}^{\alpha}%
W_{\alpha}^{(2)}\text{.}%
\end{equation}
The final term proportional to $\varepsilon$ denotes the effects of kinetic
mixing. For further discussion on kinetic mixing see for example
\cite{Holdom:1985ag,Babu:1996vt,Dienes:1996zr,Babu:1997st,Morrissey:2009ur}.
If matter is charged under both $U(1)$ factors, then loop suppressed
contributions can convert a $U(1)_{1}$ gauge field into a $U(1)_{2}$ gauge
field. \ This introduces a kinetic mixing term which in a holomorphic basis
leads to the coupling:\footnote{We thank D.E. Morrissey for discussion on this
point.}%
\begin{equation}
\varepsilon\left(  \mu\right)  =\varepsilon\left(  \mu_{0}\right)
+\frac{e_{1}e_{2}}{16\pi^{2}}\log\frac{\mu_{0}^{2}}{\mu^{2}},
\end{equation}
where the $e_{i}$ denote the charges of the bifundamental under the two gauge
groups and $\mu_{0}$ denotes the scale of the UV\ boundary conditions for the
field theory. \ The crucial point is that even if the mass of the
bifundamental is quite large, it can still contribute to kinetic mixing
between the $U(1)$ gauge fields. \ In particular, even very heavy states can
also contribute to kinetic mixing.

Returning to more string based constructions, there will generically be
massive modes charged as bifundamentals under the gauge groups of the
compactification. Because kinetic mixing does not decouple, all of these
massive states will induce small effects which mix the various $U(1)$ factors. This
can occur even when considering mixing with $U(1)_{Y}$ embedded inside of $SU(5)$
due to flux breaking of the GUT group. Note, however, that for non-abelian hidden sector gauge groups, gauge invariance forbids
such kinetic mixing terms.\ In particular, this implies that although
sequestered, kinetic mixing between abelian gauge fields will generically
occur in string based models, and is in fact difficult to turn off. Moreover,
this also implies that it is in fact quite difficult to completely sequester
the effects of a \textquotedblleft dark\textquotedblright\ abelian gauge boson
which is light compared to the GUT scale. While a dark matter scenario involving
kinetic mixing is therefore still a possibility,
such scenarios constitute a significant departure from the spirit of minimality
advocated here, and so we shall not pursue this option further in this paper.

\section{Recent Dark Matter Experiments and Theoretical Explanations
\label{sec:Review}}

In the previous sections we have presented a classification of possible dark
matter candidates in the context of F-theory GUTs. In this section we provide
a very brief overview of current experiments which provide tantalizing hints
at the observation of dark matter. After reviewing the primary signals of each
such experiment, we turn to possible theoretical explanations for these
results. Aside from astrophysical sources, such experiments may point towards
the observation of dark matter. In this vein, we discuss possible dark matter
explanations for such experiments. These fall into two main categories,
corresponding to dark matter which annihilates, and dark matter which decays.
As a disclaimer, our aim here is not to provide an exhaustive overview of
experimental constraints and theoretical models, but rather, to provide a
minimal characterization of dark matter explanations of these experiments.

\subsection{Experiments}

There are currently many different astrophysically based experiments which are
potentially capable of detecting dark matter. Recently, the PAMELA satellite
experiment \cite{Adriani:2008zq,Adriani:2008zr} has observed an excess in the
positron fraction $e^{+}/(e^{+}+e^{-})$ in comparison to the cosmic ray
background in the range of $10$ GeV to $100$ GeV. In particular, as opposed to
the expected power law falloff, this experiment instead observes an increase
in the observed positron fraction as a function of energy in this range.
Moreover, PAMELA does not detect an excess in the analogous anti-proton
fraction. If one accepts a dark matter explanation for this experiment, this
suggests the presence of a dark particle with mass at least $100$ GeV which
preferentially produces positrons, rather than anti-protons.

Corroborating evidence for heavy dark matter with mass near the weak scale has
also emerged from various recent experiments which measure the total
$e^{+}+e^{-}$ flux in different energy ranges. Experiments such as ATIC
\cite{ATIC}, PPB-BETS \cite{PPB} and HESS \cite{HESSone} all appear to be
consistent with the PAMELA experiment. Moreover, ATIC appears to detect an
increase in flux up to around $700$ GeV, with a steep falloff at higher
energies. Again assuming a dark matter explanation for such experiments, this
motivates an even heavier source of dark matter.

Most recently, however, the FERMI satellite experiment \cite{FERMI}, which is
also sensitive to the electron energy spectrum from $20$ GeV to $1$ TeV
detects a milder increase in signal at energies above $300$ GeV, when compared
with the results of ATIC. At lower energy scales FERMI appears to be
consistent with ATIC. This leads to a certain amount of tension with the
results of the other experiments. Moreover, the fact that there is no steep
falloff in the signal casts some doubt on dark matter explanations for the
other signals. Nevertheless, FERMI points towards a lower bound on the mass of
the dark matter around $1$ TeV \cite{Meade:2009iu}. Presented with this at
least suggestive evidence for the presence of dark matter, we now turn to some
of the broad features of such scenarios.

\subsection{Theoretical Scenarios}

The suggestive energy range of excess positrons and electrons found in recent
experiments begs for a theoretical explanation of some sort. Either some
astrophysical explanation, or some new production mechanism beyond that found
in the Standard Model must be introduced. In this subsection, we discuss these
two possibilities, focussing first on the exciting prospect that such
experiments may require a novel extension of the Standard Model. After
reviewing some of the features of annihilating and decaying dark matter
scenarios, we next discuss potential astrophysical origins for the observed
experimental results.

\subsubsection{Annihilating Dark Matter}

Assuming that the current experiments possess a particle physics origin, this
suggests an exciting window into the physics of dark objects. Generating an
appropriate excess of electrons can potentially be generated provided dark
matter particles preferentially annihilate into leptonic final states. Modulo
effects related to the propagation of dark matter across the galaxy, the
annihilation cross section for dark matter must obey the relation:%
\begin{equation}
\Phi_{e^{+}}^{\text{ann}}\propto\left(  \frac{\rho_{DM}^{local}}{m_{DM}%
}\right)  ^{2}\cdot\left\langle \sigma_{\text{ann}}v\right\rangle
_{\text{present}}^{DM}\label{fluxann}%
\end{equation}
where in the above, \bigskip$\Phi_{e^{+}}$ denotes the positron flux,
$\rho_{DM}^{local}$ denotes the local energy density of dark matter, and
$\left\langle \sigma_{\text{ann}}v\right\rangle _{\text{present}}^{DM}$
denotes the thermally averaged annihilation cross section for dark matter.
Here, the local density for dark matter is known to be greater than the
overall cold dark matter density as:%
\begin{equation}
\rho_{DM}^{local}=A\cdot\rho_{DM}\text{,}%
\end{equation}
where $A$ is the \textquotedblleft accumulation factor\textquotedblright%
\ which is on the order of $10^{2}$ for weak scale dark matter.

To get a sense of the required cross section necessary to explain the results
of the PAMELA experiment, we note that if the corresponding dark matter
candidate comprises all of the dark matter relic abundance, then as explained
for example in \cite{Grajek:2008pg}:%
\begin{equation}
\left\langle \sigma_{\text{ann}}v\right\rangle _{\text{present}}^{DM}%
\sim10^{-24}-10^{-23}\text{ cm}^{-3}\text{ s}\sim10^{-7}-10^{-6}\text{
GeV}^{-2}\text{,} \label{CROSSSEC}%
\end{equation}
will generate a sufficiently large signal to explain current data available
from PAMELA.

On the other hand, if the dark matter is given by a WIMP thermal relic, then
(ignoring saxion dilution effects), generating the required relic abundance
leads to $\left\langle \sigma_{\text{ann}}v\right\rangle _{\text{WIMP}}%
\sim10^{-9}-10^{-8}$ GeV$^{-2}$. Generating a sufficient signal from this
value of the cross section would then require an enhancement in the overall
size of the cross section, by a factor of $10^{2}$ to $10^{3}$. Indeed, we can
write:%
\begin{equation}
\Phi_{e^{+}}^{\text{PAMELA}}=\Phi_{e^{+}}^{\text{WIMP}}\cdot B_{\text{WIMP}%
}\text{,} \label{PHIWIMP}%
\end{equation}
where
\begin{equation}
B_{\text{WIMP}}\sim10^{2}-10^{3}%
\end{equation}
is some enhancement factor. This enhancement factor can either be argued for
by some astrophysical reason, or through an infrared enhancement in the cross
section. In the latter scenario, an infrared enhancement is in principle
possible if dark matter interacts with a light particle. Summing the
corresponding ladder diagrams associated with this annihilation can lead to a
\textquotedblleft Sommerfeld enhancement\textquotedblright\ in the
annihilation cross section as discussed for example in
\cite{Hisano:2003ec,Hisano:2004ds,Cirelli:2007xd,ArkaniHamed:2008qn}. An
important ingredient for this infrared enhancement is the presence of a light
bosonic mediator between the dark and visible sectors, with a mass around a
few GeV.

It is important to note that because the flux depends on the square of the
number density, this leads to the expectation that a large number of gamma
rays should be generated near the center of the galaxy. This leads to a
certain amount of tension with present observation \cite{VIVIER}. In addition,
there are potential issues with constraints from BBN \cite{Hisano:2009rc}.

\subsubsection{Decaying Dark Matter}

It is also possible to consider scenarios where a sufficiently long-lived dark
matter particle decays preferentially to leptonic final states. In this case,
the corresponding flux is dictated by the local dark matter energy density
as:
\begin{equation}
\Phi_{e^{+}}^{\text{dec}}\propto\left(  \frac{\rho_{DM}^{local}}{m_{DM}%
}\right)  \cdot\Gamma_{DM}\text{,}\label{fluxdec}%
\end{equation}
where $\Gamma_{DM}$ denotes the decay rate to positron final states. In order
to accommodate PAMELA, for example, this requires the corresponding lifetime
to be on the order of \cite{Arvanitaki:2008hq}:%
\begin{equation}
\Gamma_{DM}\sim10^{-51}\text{ GeV}\sim\left(  10^{26}\text{ sec}\right)
^{-1}\text{.}%
\end{equation}

The numerology connected with this lifetime leads to a quite suggestive link
with dimension six operators suppressed by the GUT\ scale
\cite{Arvanitaki:2008hq}. For example, the decay of a TeV scale dark matter
particle via a \textquotedblleft dark analogue\textquotedblright\ of the
operator responsible for proton decay in ordinary GUT\ models naturally lead
to the requisite decay rate of the form:%
\begin{equation}
\Gamma_{DM}\sim\frac{m_{DM}^{5}}{M_{GUT}^{4}}\sim6\times10^{-51}\text{
GeV}\cdot\left(  \frac{m_{DM}}{\text{TeV}}\right)  ^{5}\left(  \frac
{2\times10^{16}\text{ GeV}}{M_{GUT}}\right)  ^{4}\text{.}\label{decrate}%
\end{equation}
Moreover, because the flux scales only linearly with the number density, there
is considerably less tension with current constraints on gamma rays from the
galactic center, in comparison with annihilating dark matter scenarios. It is
important to stress, however, that in order for a decaying dark matter
candidate to generate a sufficiently large signal, it cannot contain any rapid
decay channels.

\subsubsection{Astrophysical Explanations}

Although less exciting from the perspective of particle physics, astrophysical
sources based on pulsars provides a source of distortion in the energy
dependence in the flux measured by various experiments. Pulsar winds appear to
be capable of generating up to $1000$ TeV boosts in the energy of electrons
(see for example \cite{Aharonian:2004yt}). Contributions from both nearby and
distant pulsars can in principle produce signals which can explain the types
of signals found in the current slew of experiments. As shown for example in
\cite{Hooper:2008kg,Yuksel:2008rf,Profumo:2008ms,Blasi:2009bd} pulsars appear
to provide an explanation for the PAMELA\ excess. Most conservatively then,
pulsars appear to provide a promising explanation using established physics.
This is certainly quite economical, and the obstructions we shall encounter
later in realizing dark matter capable of generating a sufficient flux in
F-theory GUTs will point back to pulsars as an attractive option.

Upcoming experiments such as FERMI will study anisotropies in the flux
background \cite{Grasso:2009ma}. This should provide a means by which to
distinguish between astrophysical compact sources and dark matter scenarios
based on a larger halo.

\section{Exceptional Obstructions to Annihilating and Decaying
Scenarios\label{sec:FVS}}

Having discussed potential dark matter candidates in F-theory GUTs, we now
proceed to analyze whether any of these options provide an adequate
explanation for recent dark matter experiments. In all cases we encounter
significant obstructions to the existence of a weak to TeV scale dark matter
candidate which can account for these experiments. \textit{This leads us to
the conclusion that at least with the minimal geometric ingredients necessary
for other aspects of F-theory GUTs, recent dark matter experiments have an
astrophysical explanation, and moreover, that the gravitino remains as the
primary dark matter candidate in F-theory GUTs.}

In principle, there could be other candidate dark matter candidates in the
context of less minimal F-theory GUT models. This would, however, lead to a
less motivated as well as less predictive framework. Given the tight
interrelations between various phenomenological ingredients already found in
previous work on F-theory GUTs, in this paper we shall exclusively focus on
those elements which are compatible with minimal considerations.

Even though one might initially think that global considerations could provide
many dark matter candidates invisible to the local model, such candidates do
not provide weak to TeV scale dark matter candidates. Indeed, recall that the
available dark matter candidates roughly split into those which are
\textquotedblleft inside\textquotedblright\ the minimal $E_{8}$ of the local
GUT\ model, possible candidates \textquotedblleft nearby\textquotedblright%
\ which interact with at least some seven-brane factor of this $E_{8}$, and
candidates which are fully sequestered. Of these possibilities, note that
since the weak scale is controlled by the scale of supersymmetry breaking,
which is in turn controlled by the vev of the chiral superfield $X$, weak or
TeV scale dark matter must interact closely with $X$, or at least with other
degrees of freedom which interact closely with $X$. As shown in section
\ref{sec:massscale}, degrees of freedom which are completely sequestered from
the PQ seven-brane either have very large masses set by high scale
supersymmetric dynamics, or have lower masses on the order of $10-100$ MeV as
for the gravitino, while for some fermions the mass can as be low as $10-100$ keV.

Weak to TeV scale dark matter candidates must interact with degrees of freedom
more closely linked to the supersymmetry breaking sector, and thus, to the PQ
seven-brane. In principle, viable candidates could descend either from the
minimal $E_{8}$ enhancement point, or from \textquotedblleft nearby
branes\textquotedblright\ which intersect the PQ\ seven-brane, but do not
embed in this minimal $E_{8}$. This latter possibility is somewhat exotic, as
well as non-minimal as discussed in section \ref{sec:ESHIELD}, so we shall not
dwell on this possibility here. Rather, in this section we shall instead focus
on those candidates which the presence of the $E_{8}$ enhancement point
allows. Whether or not such candidates are present depends on the specific
choice of fluxes. In principle, there could be many candidates, or none at
all. Some of our discussion will be more general and will also apply to any PQ
charged object.

Restricting further to minimal dark matter which descends from a local
Higgsing of the $E_{8}$ enhancement point, we have seen in section
\ref{sec:DarkE} that there are essentially two types of possible dark matter
candidates, corresponding to electrically neutral matter which can descend
either from the $5$ or $\overline{5}$ of $SU(5)$, or as a GUT\ singlet. This
first possibility is quite limited, and meets with significant constraints.

Having dispensed with this option, we are then left with dark matter
candidates which transform as singlets under the GUT\ group. In the covering
theory, such matter fields descend from the adjoint representation of
$SU(5)_{\bot}$.\ Our aim in this section will be to show that these singlets
also meet with little success because all available candidates decay faster
than cosmological timescales. This again points to the natural role of
gravitino dark matter in F-theory GUTs.

The rest of this section is organized as follows. We first illustrate that
already at the level of cosmological considerations, the decay of the saxion
presents significant obstructions to realizing a dark matter scenario capable
of generating large positron fluxes, as required to provide a dark matter
explanation of experiments such as PAMELA. Indeed, this decay dilutes the
abundance of thermally produced relics, and, in certain instances can
overproduce relics through non-thermal processes. After explaining some of the
issues with such scenarios, we next turn to a list of potential dark matter
candidates. First, we explain in greater detail why current experimental
limits strongly disfavor electrically neutral components in extra
$5\oplus\overline{5}$'s. Next, we study the available GUT singlets which can
descend from the local Higgsing of an $E_{8}$ point of enhancement. We find
that in all cases, dark matter candidates which develop a suitable mass are
also unstable against rapid decay to MSSM\ fields, and briefly comment on some
non-minimal possibilities. Finally, we discuss other aspects of annihilating
dark matter scenarios, such as the absence of light GeV scale gauge bosons
inside of $E_{8}$, and constraints on light scalars.

\subsection{Saxion Decay and Dark Matter Production in F-theory}

In order to generate a sufficiently large signal capable of providing an
explanation for recent dark matter experiments, the relic abundance of the
corresponding dark matter particles must comprise at least a portion of the
total dark matter relic abundance. Note that as the fraction of dark matter
given by a candidate decreases, the resulting strength of the signal generated
by this candidate must increase. For example, the flux generated in
annihilating and decaying dark matter scenarios respectively scale as:%
\begin{align}
\Phi_{e^{+}}^{\text{ann}}  &  \propto\left(  \frac{\rho_{DM}^{local}}{m_{DM}%
}\right)  ^{2}\cdot\left\langle \sigma_{\text{ann}}v\right\rangle
_{\text{present}}^{DM}\\
\Phi_{e^{+}}^{\text{dec}}  &  \propto\left(  \frac{\rho_{DM}^{local}}{m_{DM}%
}\right)  \cdot\Gamma_{DM}\text{,}%
\end{align}
with notation as in equations (\ref{fluxann}) and (\ref{fluxdec}). Thus,
decreasing the fraction of the dark matter relic abundance requires a
corresponding increasing in either the annihilation cross section, or decay rate.

Generally speaking, the relic abundance for dark matter can be generated
either from the thermal bath of the early Universe, or at a later stage due to
a late decaying particle. This distinction is especially important in F-theory
GUTs, because in the most common cosmological scenario, oscillations of the
saxion will dominate the energy density of the Universe prior to its
decay.\footnote{It is in principle possible to decrease the initial reheating
temperature $T_{RH}^{0}$ to such a low value that the oscillation of the
saxion never comes to dominate the energy density of the Universe. In keeping
with the broad outlines of the cosmological scenario found in
\cite{FGUTSCosmo}, we shall typically take the initial reheating temperature
higher than this low value.} Indeed, once the saxion decays, it will release a
significant amount of entropy, diluting the relic abundance of all previously
generated thermal relics through the relation:%
\begin{equation}
\Omega_{\text{after}}=D_{\text{sax}}\Omega_{\text{before}}\text{,}%
\end{equation}
where as reviewed in section \ref{sec:gravDM}, the typical size of the
dilution factor is:%
\begin{equation}
D_{\text{sax}}\sim10^{-4}\text{.}%
\end{equation}
On the other hand, the decay of the saxion can also generate new sources of
dark matter relics. In the following subsections we analyze both possibilities.

\subsubsection{Thermal Production\label{thermprod}}

We now consider the expected flux from annihilating and decaying scenarios,
under the assumption that a species of dark matter is produced thermally from
the primordial bath, in which it freezes out at a temperature $T_{(i)}^{f}$.
The decay of the saxion imposes quite severe conditions on annihilating
scenarios, but somewhat milder conditions on decaying dark matter scenarios.

First consider annihilating scenarios. In the case of an annihilating
scenario, the flux is:%
\begin{equation}
\Phi_{e^{+}}^{(i)}\propto\left(  \frac{r_{i}\rho_{DM}^{local}}{m_{i}}\right)
^{2}\cdot\left\langle \sigma_{(i)}v\right\rangle _{\text{present}}\text{,}%
\end{equation}
where $r_{i}$ denotes the actual fraction of dark matter of the $i^{th}$
species participating in annihilation processes so that:%
\begin{equation}
r_{i}\equiv\frac{\Omega_{i}}{\Omega_{DM}^{\text{total}}}\leq1\text{.}
\label{ratio}%
\end{equation}
Since we are assuming the relic abundance is generated thermally, we have the
relation:%
\begin{equation}
\Omega_{i}h^{2}\propto D_{\text{sax}}\cdot\frac{1}{\left\langle \sigma
_{(i)}v\right\rangle _{T=T_{(i)}^{f}}}, \label{ispec}%
\end{equation}
where here, we have taken into account the effects of saxion dilution.

To compare this with the expected signal from PAMELA, we first compare the
expected signal of this scenario with dilution, to a WIMP scenario with no
dilution. Recall that (in the absence of saxion dilution), WIMPs naturally
produce the correct relic abundance to account for all of the dark matter. In
particular, we have:%
\begin{equation}
\Omega_{DM}^{\text{total}}h^{2}=\Omega_{\text{WIMP}}h^{2}\propto\frac
{1}{\left\langle \sigma v\right\rangle _{\text{WIMP}}}\text{,} \label{WIMP}%
\end{equation}
where $\left\langle \sigma v\right\rangle _{\text{WIMP}}$ is the annihilation
cross section for WIMPs and the constant of proportionality is roughly the
same in lines (\ref{ispec}) and (\ref{WIMP}). Taking the ratio of $\Omega
_{i}h^{2}$ to $\Omega_{DM}^{\text{total}}h^{2}$, we therefore obtain:%
\begin{equation}
\frac{\Omega_{i}h^{2}}{\Omega_{DM}^{\text{total}}h^{2}}=D_{\text{sax}}%
\cdot\frac{\left\langle \sigma v\right\rangle _{\text{WIMP}}}{\left\langle
\sigma_{(i)}v\right\rangle _{T=T_{(i)}^{f}}}\text{,}%
\end{equation}
In other words, returning to equation (\ref{ratio}), we obtain:%
\begin{equation}
\left\langle \sigma_{(i)}v\right\rangle _{T=T_{(i)}^{f}}=D_{\text{sax}}%
\cdot\frac{\left\langle \sigma v\right\rangle _{\text{WIMP}}}{r_{i}}\text{.}%
\end{equation}
Assuming that the cross section at freeze out is roughly the same as at
present, we can now express the flux as:%
\begin{equation}
\Phi_{e^{+}}^{(i)}\propto\left(  \frac{r_{i}\rho_{DM}^{local}}{m_{\text{WIMP}%
}}\right)  ^{2}\cdot D_{\text{sax}}\cdot\frac{\left\langle \sigma
v\right\rangle _{\text{WIMP}}}{r_{i}}\text{.}%
\end{equation}
In other words, in comparison with a WIMP scenario, the expected flux from the
annihilating scenario is:%
\begin{equation}
\Phi_{e^{+}}^{(i)}=r_{i}D_{\text{sax}}\Phi_{e^{+}}^{\text{WIMP}}=\frac
{r_{i}D_{\text{sax}}}{B_{\text{WIMP}}}\Phi_{e^{+}}^{\text{PAMELA}}%
\lesssim10^{-7}\cdot\Phi_{e^{+}}^{\text{PAMELA}}\text{,}%
\end{equation}
where in the second relation we have used equation (\ref{PHIWIMP}), and in the
last inequality, we have used the fact that $r_{i}$, $D_{\text{sax}}%
\sim10^{-4}$ and $B_{\text{WIMP}}^{-1}\sim10^{-2}-10^{-3}$ are all factors
less than one. Thus, saxion dilution exacerbates the problems with generating
a sufficiently large signal already present in WIMP\ scenarios! It is
therefore necessary to boost the cross section by a factor of at least
$10^{7}$, which is unappealing. To bypass this constraint, it seems necessary
to consider cosmological scenarios where the saxion never comes to dominate
the energy density of the Universe.\ This requires a significant decrease in
the initial reheating temperature $T_{RH}^{0}$, to values in the range of
$10^{4}-10^{6}$ GeV \cite{FGUTSCosmo}. In principle, this is a logical
possibility, but runs counter to the idea that gravity should in principle
decouple from gauge theory considerations. Indeed, lowering the initial
reheating temperature so much is in conflict with this parameter range.

The consequences of saxion dilution in decaying dark matter scenarios are
somewhat milder. Assuming that the decay rate for weak scale decaying dark
matter is controlled by a dimension six operator with suppression scale
$\Lambda_{\text{UV}}$, the relevant decay rate scales much as in equation
(\ref{decrate}):%
\begin{equation}
\Gamma_{DM}\sim\frac{m_{DM}^{5}}{\Lambda_{\text{UV}}^{4}}\sim6\times
10^{-51}\text{ GeV}\cdot\left(  \frac{m_{DM}}{\text{TeV}}\right)  ^{5}\left(
\frac{2\times10^{16}\text{ GeV}}{\Lambda_{\text{UV}}}\right)  ^{4}\text{.}%
\end{equation}
Thus, while dilution introduces a decrease in the expected flux by a factor of
$10^{-4}$, note that a factor of $10$ decrease in $\Lambda_{\text{UV}}$ below
the GUT\ scale easily makes up this difference. In the context of F-theory
GUTs, such suppression scales are quite commonplace, either as the mass scale
of a normal curve, or from enhancements in the effective scale due to the
local behavior of Green's functions. In other words, the effects of saxion
dilution are far milder in decaying scenarios.

\subsubsection{Non-Thermal Production}

Even though the decay of the saxion dilutes the thermally produced relics, its
decays can generate a new source for such particles. This type of production
mechanism requires the mass of the saxion to be greater than the decay
products. When this is not the case, the relevant decay channel is closed, and
some other means must be found to generate a sufficient relic abundance.
Indeed, we find that if saxion decays can generate the relevant dark matter,
then they will be overproduced.

Kinematic considerations require that the saxion mass be at least as heavy as
the dark matter candidate. Otherwise, the needed decay channel will be
unavailable. Assuming a TeV scale dark matter candidate, this is already quite
problematic, because the mass of the saxion is more typically in the range of
$500$ GeV, than a TeV. The reason for this upper bound comes about because of
the interplay between the mass of the saxion and the PQ\ deformation through
the relation \cite{FGUTSCosmo}:%
\begin{equation}
m_{sax}\propto\Delta_{PQ},
\end{equation}
where the constant of proportionality depends on the embedding of $SU(5) \times U(1)_{PQ}$ in $E_8$,
and details of the saxion potential. For illustrative purposes,
we use the mass scale induced by the PQ deformation for $X$ as a rough estimate that this constant is on the order of four.
On the other hand, the PQ deformation also induces a tachyonic contribution to the
mass squared of many of the scalars, including the stau \cite{HVGMSB}:%
\begin{equation}
m^{2}=m_{0}^{2}-q\Delta_{PQ}^{2}\text{,}%
\end{equation}
where $m_{0}$ denotes the mass in the absence of the PQ deformation. This
leads to an upper bound on the size of the PQ\ deformation. To overcome this
obstacle, it is necessary to increase $m_{0}$, which will increase the amount
of fine-tuning present in the gauge mediation sector.

To take an explicit example, in a Dirac neutrino scenario with zero
PQ\ deformation, the mass of the lightest stau with minimal fine tuning is on
the order of $200$ GeV. With a single pair of $10\oplus\overline{10}$
messengers with minimal fine tuning, the maximum allowed $\Delta_{PQ}$ is on
the order of $150$ GeV. This would lead to a saxion mass on the order of $600$ GeV, which
is significantly less than $1$ TeV. To generate a TeV scale mass would require
a factor of two increase in the soft scalar masses generated by gauge
mediation effects. This would also increase the mass of the squarks, and would
exacerbate the fine-tuning already present.

Leaving aside this potential worry, we now show that there are further issues
with generating the required relic abundance from saxion decays. It is
important to note that the decay to a dark matter candidate produced in this
way need not be the dominant decay channel. For example, in F-theory GUTs, the
dominant decay channel for the saxion is to a pair of Higgs fields. The
branching fraction to gravitinos is significantly smaller, as the relevant
decay amplitude derives from a Planck suppressed higher-dimension operator.
Even so, because the gravitino is stable, the resulting relic abundance of
gravitinos produced in this way can still comprise up to ten percent of the
relic abundance \cite{FGUTSCosmo}. Thus, even if a dark matter candidate
couples only very weakly to the saxion through a higher dimension operator,
the branching fraction can still generate a sizable dark matter relic abundance.

We now compute the relic abundance generated by the decay of the saxion:%
\begin{equation}
\Omega_{\phi}^{NT}h^{2}=\left(  \frac{s_{0}}{\rho_{c,0}}\right)  m_{\phi
}Y_{\phi}^{NT}\text{,}%
\end{equation}
where $s_{0}$ and $\rho_{c,0}$ respectively denote the present entropy and
critical density, and $Y_{\phi}^{NT}$ denotes the yield non-thermally
generated by the decay of the saxion:%
\begin{equation}
Y_{\phi}^{NT}\sim\frac{n_{\phi,after}^{NT}}{s_{after}}\text{,}%
\end{equation}
where $n_{\phi,after}^{NT}$ denotes the number density of the dark matter
generated non-thermally, and $s_{after}$ denotes the entropy density after the
decay of the saxion. As reviewed for example in \cite{FGUTSCosmo}, this can be
related to the reheating temperature for the saxion, and the branching
fraction of the saxion to dark matter so that:%
\begin{equation}
Y_{\phi}^{NT}\sim\frac{3}{2}B_{\phi}\frac{T_{RH}^{sax}}{m_{sax}}\text{,}%
\end{equation}
where $B_{\phi}$ denotes the saxion branching fraction to the dark matter
species $\phi$. The resulting relic abundance is then given by:%
\begin{equation}
\Omega_{\phi}^{NT}h^{2}=\left(  \frac{s_{0}}{\rho_{c,0}}h^{2}\right)
\cdot\frac{3}{2}m_{\phi}B_{\phi}\frac{T_{RH}^{sax}}{m_{sax}}\text{.}%
\end{equation}
Using $s_{0}/\rho_{c,0}\sim3\times10^{8}$ $(h^{2}$ GeV$)^{-1}$, and the rough
estimate $m_{\phi}\sim m_{sax}/2$, we therefore find:%
\begin{equation}
\Omega_{\phi}^{NT}h^{2}\sim2\times10^{8}\text{ GeV}^{-1}\cdot B_{\phi} \cdot
T_{RH}^{sax}\text{.} \label{NTPROD}%
\end{equation}

To estimate the resulting relic abundance, we now determine the branching
fraction to the dark matter candidate $\phi$. Assuming the mass of dark matter
chiral superfields $\Phi_{1}$ and $\Phi_{2}$ is generated through the higher
dimension operator:%
\begin{equation}
L\supset\gamma\int d^{4}\theta\frac{X^{\dag}\Phi_{1}\Phi_{2}}{\Lambda
_{\text{UV}}},
\end{equation}
we now show that the coupling to the saxion leads to an overproduction of dark
matter particles.

For the purposes of this discussion we may take the bosonic component of the
chiral superfield to be the quasi-stable dark matter candidate. If the
fermionic component is lighter, then the bosonic component will eventually
decay to its fermionic partner, also emitting a gravitino. Letting $x$,
$\phi_{1}$ and $\phi_{2}$ denote the scalar components of these chiral
superfields, taking derivatives of the K\"{a}hler potential shows that the
bosonic component fields couple as:%
\begin{equation}
L\supset\gamma\frac{\phi_{2}}{\Lambda_{\text{UV}}}\partial^{\mu}\overline
{x}\partial_{\mu}\phi_{1}+\gamma\frac{\phi_{1}}{\Lambda_{\text{UV}}}%
\partial^{\mu}\overline{x}\partial_{\mu}\phi_{2}+h.c.
\end{equation}
Here we shall assume that the bosons are stable against further decays, and
then compute the branching fraction. Writing $x=\langle x \rangle\exp(ia+s)$,
with $s$ the saxion, it follows that the saxion indeed couples to the
$\phi_{i}$'s. Provided the mass of the saxion is larger than the combined mass
of the two decay products, we obtain a rough estimate for the decay rate:%
\begin{equation}
\Gamma_{x\rightarrow\phi_{1}\phi_{2}}\sim\frac{\gamma^{2}}{32\pi}\frac
{m_{sax}^{3}}{\Lambda_{\text{UV}}^{2}}\text{.}%
\end{equation}
With the decay rate in hand, we now compute the expected branching fraction.
Returning to equation (\ref{TSAX}), the total saxion decay rate is related to
the axion branching fraction as:%
\begin{equation}
\Gamma_{sax}\sim\frac{1}{B_{a}}\Gamma_{a}\sim\frac{1}{B_{a}}\frac{1}{64\pi
}\frac{m_{sax}^{3}}{f_{a}^{2}}\text{,}%
\end{equation}
where $B_{a}$ is the branching fraction to axions and $f_{a}$ is the axion
decay constant. Hence, the branching fraction to dark matter is:%
\begin{equation}
B_{\phi}=\frac{\Gamma_{x\rightarrow\phi_{1}\phi_{2}}}{\Gamma_{sax}}\sim
B_{a}\frac{m_{\phi}^{2}}{\Lambda_{GMSB}^{2}}\sim1\times10^{-4}\cdot
B_{a}\left(  \frac{m_{\phi}}{1\text{ TeV}}\right)  ^{2}\left(  \frac
{10^{5}\text{ GeV}}{\Lambda_{GMSB}}\right)  ^{2}\text{,}%
\end{equation}
where we have used the fact that in F-theory GUTs, the scale of supersymmetry
breaking and axion decay constant are related to the characteristic scale of
gauge mediation as $F_{X}/f_{a}=\Lambda_{GMSB}\sim10^{5}$ GeV \cite{HVGMSB}.
Further, we have taken the natural value $m_{\phi} \sim$ TeV for the mass of
the dark matter candidate. The saxion reheating temperature is:%
\begin{equation}
T_{RH}^{sax}\sim0.5\sqrt{\Gamma_{sax}M_{PL}}\sim10\text{ GeV}\cdot\frac
{1}{B_{a}^{1/2}}\left(  \frac{m_{\phi}}{1\text{ TeV}}\right)  ^{3/2}\text{.}%
\end{equation}
Returning to equation (\ref{NTPROD}), the resulting thermal relic abundance is
then:%
\begin{equation}
\Omega_{\phi}^{NT}h^{2}\sim2\times10^{5}\cdot B_{a}^{1/2}\left(  \frac
{m_{\phi}}{1\text{ TeV}}\right)  ^{7/2}\left(  \frac{10^{5}\text{ GeV}%
}{\Lambda_{GMSB}}\right)  ^{2}\text{.}%
\end{equation}
The actual branching fraction to axions depends on the mass of the saxion.
Indeed, as the saxion becomes more massive than the corresponding states in
the MSSM, additional channels will begin to open up. As a consequence, the
value of $B_{a}$ can decrease, even though the decay rate to axions scales
with $m_{sax}^{3}$. Note, though, that to generate a suitable relic abundance
would require $B_{a}\sim10^{-10}$, which is far smaller than the value of
$B_{a}\sim10^{-3}-10^{-1}$ found in \cite{FGUTSCosmo}. Hence, we deduce that
the decay of the saxion actually overproduces the corresponding relic. In the
above we also assumed that the bosonic component was the dark matter. If this
is not the case, note that it will decay, generating a further source of
gravitino relics, which appears to be far greater than can otherwise be allowed.

Summarizing, the above analysis demonstrates that the decay of the saxion
overproduces TeV scale dark matter candidates, when such a decay channel is
available. Turning this discussion around, we can view this either as a
constraint on how the saxion couples to the dark matter, or as an upper bound
on the mass of the saxion, so that kinematics can forbid such decays.\ In the
former case, this leaves open exactly how the dark matter develops a weak
scale mass. This can in principle be accommodated if another GUT\ singlet
develops a weak scale, but this is somewhat less minimal. Assuming that the
saxion can decay to the dark matter candidate, this imposes an intriguing
cosmological upper bound on the mass, of around $1$ TeV. Combined with the
condition that the saxion must be allowed to decay to an MSSM\ scalar such as
the Higgs, this leads to the mass range:%
\begin{equation}
230\text{ GeV}\lesssim m_{sax}\lesssim1\text{ TeV.}%
\end{equation}

We have also seen that in the case of thermally produced annihilating dark
matter, saxion dilution tends to eliminate any candidate signatures. On the
other hand, thermally produced decaying dark matter can still generate a
sizable signal capable of explaining excess fluxes. Even so, it is possible
that including additional non-minimal ingredients could produce an appropriate
relic abundance. Assuming this is the case, we now turn to other obstructions
with realizing such scenarios in minimal F-theory GUT\ models.

\subsection{Eliminating Semi-Visible Candidates}

In this section we discuss obstructions in F-theory associated with realizing
dark matter from the electrically neutral component of the $5\oplus
\overline{5}$ of $SU(5)$. In subsection \ref{ssec:SEMIVIS} we found that in
the case of a Dirac scenario with $%
\mathbb{Z}
_{2}$ monodromy group, it is indeed possible to consider vector-like pairs in
the $5\oplus\overline{5}$ which develop a mass through the Giudice-Masiero
operator:%
\begin{equation}
\int d^{4}\theta\frac{X^{\dag}D_{5}D_{\overline{5}}^{\prime}}{\Lambda
_{\text{UV}}}\rightarrow\int d^{2}\theta\text{ }\mu_{D}D_{5}D_{\overline{5}%
}^{\prime}.
\end{equation}
This induces a mass $\mu_{D}$ on the order of the $\mu$ parameter. The
electrically neutral components of the $5$ and $\overline{5}$ then provide
potential dark matter candidates. A quite attractive feature of such
semi-visible dark matter candidates is that although electrically neutral,
this component can still couple to the Standard Model through $SU(2)$ interactions.

Current experimental limits impose strong restrictions on the existence of
such electrically neutral objects. First note that in order not to spoil the
unification of the coupling constants, all of the components of the
GUT\ multiplet must have comparable mass. Thus, we can expect extra states
with non-trivial $SU(3)_{C}$ and $U(1)_{EM}$ charges. Direct searches at the
Tevatron only rule out extra colored states with mass less than about $300$
GeV \cite{PDG}, which is well below the expected mass of $D_{5}$ and
$D_{\overline{5}}^{\prime}$.

A more stringent constraint comes from the $SU(2)$ doublet components of the
dark matter candidate.\footnote{We thank D.E. Morrissey for discussion on this
point.} Experimental bounds on dark matter scattering off of nuclei in
experiments such as CDMS are sensitive to a convolution between the total flux
from dark matter and the effective cross section for scattering with nuclei
from $Z^{0}$ gauge boson exchange. In this context, the corresponding
effective cross section for matter far above the GeV scale is
\cite{Jungman:1995df}:\footnote{See for example \cite{Cui:2009xq} for a brief
discussion of how to extract the value of this cross section from
\cite{Jungman:1995df}.}%
\begin{equation}
\sigma_{DM,nuc}\sim7.5\times10^{-39}\text{ cm}^{2}\text{.}%
\end{equation}
This is to be contrasted with current experimental constraints from CDMS.
Assuming that all of the dark matter relic abundance participates in nuclei
scattering, and moreover, that the local dark matter density is roughly $0.3$
GeV/cm$^{3}$, this leads to a lower bound on the size of the allowed cross
section \cite{Akerib:2003px}:%
\begin{equation}
\sigma_{CDMS}\lesssim10^{-43}\text{ cm}^{2}\text{.}%
\end{equation}
It is important to note that this result scales linearly with the overall dark
matter number density, so in principle, lowering the local density by roughly
a factor of $10^{-4}$ can provide one way for a dark matter candidate to evade
such a bound. Note, however, that this will also lower the expected positron
flux which experiments such as PAMELA can in principle produce.

An alternative way to evade such a bound is in principle possible in inelastic
dark matter (iDM) scenarios, in which a small Majorana mass is also included
in the effective Lagrangian \cite{TuckerSmith:2001hy}. This induces a mass
splitting between the two Dirac mass states. In particular, when the lighter
dark matter state interacts with the nuclei, it can \textquotedblleft
up-scatter\textquotedblright\ to the heavier state. Kinematic considerations
now imply that the resulting phase space available for the convolution between
dark matter flux and the total cross section is somewhat lower. In order for
this mechanism to lead to a sufficient suppression in the reaction rate
expected from CDMS, this requires a Majorana mass term of at least $100$ keV.

Generating such a large Majorana mass term is quite problematic in minimal
F-theory GUT setups. Indeed, this requires the presence of a higher dimension
operator such as:%
\begin{equation}
\int d^{2}\theta\frac{\left(  H_{u}L_{D_{\overline{5}}^{\prime}}\right)  ^{2}%
}{\Lambda_{\text{UV}}}\text{,}%
\end{equation}
where $L_{D_{\overline{5}}}$ denotes the ``lepton doublet'' part of
$D_{\overline{5}}$. On the other hand, similar operators in the Majorana
neutrino scenario of \cite{BHSV} generate masses closer to $10$ meV. Without
adding a significantly more elaborate sector with which $D_{5}$ and
$D_{\overline{5}}^{\prime}$ interact, this effectively rules out such
semi-visible dark matter candidates.

\subsection{Unstable GUT\ Singlets\label{UNSTAB}}

In the previous subsections we found that the decay of the saxion in F-theory
GUTs dilutes thermal relics, and can also overproduce dark matter candidates.
Leaving aside these potential concerns, in this section we show that all of
the available dark matter candidates decay far too rapidly. This then
eliminates all annihilating and decaying dark matter scenarios based on
particles embedded in $E_{8}$.

Any dark matter candidate must be sufficiently long-lived. Dimension five
decay operators pose significant problems for the lifetime of a dark matter
candidate. Indeed, the expected decay rate in this case is of the form:%
\begin{align}
\Gamma_{DM} &  \sim\frac{m_{DM}^{3}}{M_{GUT}^{2}}\sim10^{-24}\text{ GeV}%
\cdot\left(  \frac{m_{DM}}{1\text{ TeV}}\right)  ^{3}\left(  \frac
{3\times10^{16}\text{ GeV}}{M_{GUT}}\right)  ^{2}\\
&  \sim10\text{ sec}^{-1}\cdot\left(  \frac{m_{DM}}{1\text{ TeV}}\right)
^{3}\left(  \frac{3\times10^{16}\text{ GeV}}{M_{GUT}}\right)  ^{2}\text{,}%
\end{align}
which is far shorter than any cosmological timescale. In other words, if a
dark matter candidate can couple to MSSM\ fields through a dimension five
operator, and if the decay is kinematically allowed, the corresponding field
will be unstable on cosmological timescales.

The aim of this section is to show that all of the GUT singlets which descend
from the local Higgsing of $E_{8}$ decay far too rapidly to constitute dark
matter candidates. This is a general statement, and does not require delving
into the particular details of a specific annihilating or decaying scenario.
The relevant higher dimension operators can all be generated much as in
\cite{BHVII,HVGMSB,BHSV} by integrating out heavy modes localized on curves.

Available GUT singlets which can constitute dark matter descend from the
adjoint of $SU(5)_{\bot}$, so in the languague of section \ref{sec:DarkE} are
either given as non-trivial weights of the form $t_{m}-t_{n}$, or as zero
weights. Of the available states which descend from the adjoint of
$SU(5)_{\bot}$, note that the zero weights correspond to bulk modes of other
seven-branes. Since no symmetry forbids F-terms involving one zero weight and
three MSSM\ chiral superfields, we conclude that such modes cannot constitute
dark matter candidates. This leaves only the weights of the form $t_{m}-t_{n}$
as possibilities.

With notation as in section \ref{sec:DarkE}, the available GUT\ singlets are
given as either $D_{(i)},$ the conjugates $D_{(i)}^{c}$ which localize on the
same curve, or as GUT\ singlets which localize on either the $X$ curve or the
$N_{R}$ curve. Index theory considerations exclude the possibility of fields
conjugate to $X$ or $N_{R}$, but in principle, extra \textquotedblleft
generations\textquotedblright\ on these curves could be present, which we
denote as $\widehat{X}$ and $\widehat{N}_{R}$. Using the classification of
section \ref{sec:DarkE}, the available dark matter candidates, and their
respective charges under the available symmetries in the various neutrino
scenarios are:%
\begin{align}
&
\begin{tabular}
[c]{|c|c|c|c|c|c|c|c|c|}\hline
$G_{mono}^{Dirac}\simeq%
\mathbb{Z}
_{2}$ or $%
\mathbb{Z}
_{2}\times%
\mathbb{Z}
_{2}$ & $\widehat{X}$ & $\widehat{N}_{R}$ & $D_{(1)},D_{(1)}^{c}$ & $D_{(2)}$
& $D_{(2)}^{c}$ & $D_{(3)},D_{(3)}^{c}$ & $D_{(4)}$ & $D_{(4)}^{c}$\\\hline
$U(1)_{PQ}$ & $-4$ & $-3$ & $0$ & $+4$ & $-4$ & $0$ & $-7$ & $+7$\\\hline
$U(1)_{\chi}$ & $0$ & $-5$ & $0$ & $0$ & $0$ & $0$ & $-5$ & $+5$\\\hline
\end{tabular}
\\
&
\begin{tabular}
[c]{|c|c|c|c|c|c|}\hline
$G_{mono}^{Dirac}\simeq%
\mathbb{Z}
_{3}$ or $S_{3}$ & $\widehat{X}$ & $\widehat{N}_{R}$ & $D_{(1)}$ &
$D_{(1)}^{c}$ & $D_{(2)},D_{(2)}^{c}$\\\hline
$U(1)_{PQ}$ & $-4$ & $-3$ & $+1$ & $-1$ & $0$\\\hline
$U(1)_{\chi}$ & $0$ & $-5$ & $-5$ & $+5$ & $0$\\\hline
\end{tabular}
\\
&
\begin{tabular}
[c]{|c|c|c|c|c|}\hline
$G_{mono}^{Maj}$ & $\widehat{X}$ & $\widehat{N}_{R}$ & $D_{(1)},D_{(1)}^{c}$ &
$D_{(2)},D_{(2)}^{c}$\\\hline
$U(1)_{PQ}$ & $-5$ & $0$ & $0$ & $0$\\\hline
\end{tabular}
\ . \label{MAJO}%
\end{align}

Of these candidates, it is necessary to state how each available dark matter
candidate can develop a weak to TeV scale mass. The most minimal option is to
correlate the mass through the same type of mechanism which generates the
$\mu$ term, through a Giudice-Masiero operator. It is in principle possible to
also consider cubic superpotential couplings between three GUT\ singlets where
one field develops a weak scale vev. Although less minimal, this possibility
can occur in F-theory GUTs provided the field which develops a vev has a
suitable PQ charge. Indeed, the PQ deformation can induce a tachyonic
contribution to the mass squared for the bosonic component of this field.
Letting $\Phi_{(i)}$ denote candidate dark matter chiral multiplets, this
leaves the following possibilities for how the dark fields can develop a
suitable mass:%
\begin{equation}
\int d^{4}\theta\frac{X^{\dag}\Phi_{(1)}\Phi_{(2)}}{\Lambda_{\text{UV}}}\text{
or }\int d^{2}\theta\text{ }\Phi_{(1)}\Phi_{(2)}\Phi_{(3)}, \label{massterm}%
\end{equation}
where in the second possibility, at least one of the $\Phi_{(i)}$'s develops a
weak scale vev.

Our strategy to deduce the absence of any quasi-stable dark matter candidates
is as follows. Given a pair of chiral superfields $\Phi_{(i)}$ and $\Phi
_{(j)}$which develop a weak to TeV scale mass through the coupling:,%
\begin{equation}
\mu_{ij}\int d^{2}\theta\text{ }\Phi_{(i)}\Phi_{(j)},
\end{equation}
then either both fields are quasi-stable, or neither is. Indeed, if the
components of $\Phi_{(i)}$ decay rapidly, then mixing through the mass term
will also induce a decay for $\Phi_{(j)}$.\ It is therefore enough to show
that for any available pair of chiral superfields, at least one member of the
pair decays rapidly to MSSM\ fields.

We now demonstrate that in all cases, a TeV scale GUT singlet possesses a
rapid decay channel. Since the actual weight assignments are somewhat specific
to the details of a given monodromy group, we discuss these possibilities separately.

\subsubsection{Dirac Scenarios: $G_{mono}^{Dirac}\simeq%
\mathbb{Z}
_{2}$ or $%
\mathbb{Z}
_{2}\times%
\mathbb{Z}
_{2}$\label{firstdec}}

First consider monodromy scenarios where $G_{mono}^{Dirac}\simeq%
\mathbb{Z}
_{2}$ or $%
\mathbb{Z}
_{2}\times%
\mathbb{Z}
_{2}$. Consistency with $U(1)_{PQ}$ and $U(1)_{\chi}$ leads to the following
options for generating weak to TeV scale masses in this Dirac scenario:%
\begin{align}
G_{mono}^{Dirac}  &  \simeq%
\mathbb{Z}
_{2}\text{ or }%
\mathbb{Z}
_{2}\times%
\mathbb{Z}
_{2}\label{GMON}\\
\int d^{4}\theta\frac{X^{\dag}\Phi_{(1)}\Phi_{(2)}}{\Lambda_{\text{UV}}}  &
\Longrightarrow\Phi_{(1)}=D_{(2)}^{c}\text{, }\Phi_{(2)}=D_{(1)},D_{(1)}%
^{c},D_{(3)},D_{(3)}^{c}\label{GMSEC}\\
\int d^{2}\theta\text{ }\Phi_{(1)}\Phi_{(2)}\Phi_{(3)}  &  \Longrightarrow
\Phi_{(1)}=\widehat{X}\text{, }\Phi_{(2)}=D_{(2)}\text{, }\Phi_{(3)}%
=D_{(1)},D_{(1)}^{c},D_{(3)},D_{(3)}^{c}\label{pah}\\
\int d^{2}\theta\text{ }\Phi_{(1)}\Phi_{(2)}\Phi_{(3)}  &  \Longrightarrow
\Phi_{(1)}=\widehat{X}\text{, }\Phi_{(2)}=\widehat{N}_{R}\text{, }\Phi
_{(3)}=D_{(4)}^{c}\label{blah}\\
\int d^{2}\theta\text{ }\Phi_{(1)}\Phi_{(2)}\Phi_{(3)}  &  \Longrightarrow
\Phi_{(1)}=\widehat{N}_{R}\text{, }\Phi_{(2)}=D_{(2)}^{c}\text{, }\Phi
_{(3)}=D_{(4)}^{c}. \label{cubiccoup}%
\end{align}

As we now explain, all available dark matter candidates are unstable on
cosmological timescales. To establish this, recall that the available orbits
for minimal matter and possible dark matter candidates are:%
\begin{align}
G_{mono}^{Dirac}  &  =\left\langle (12)(34)\right\rangle \simeq%
\mathbb{Z}
_{2}\text{:}\\
Orb(10_{M},Y_{10})  &  =t_{1},t_{2}\\
Orb(Y_{\overline{10}}^{\prime})  &  =-t_{3},-t_{4}\\
Orb(\overline{5}_{M})  &  =t_{4}+t_{5},t_{3}+t_{5}\\
Orb(5_{H})  &  =-t_{1}-t_{2}\\
Orb(\overline{5}_{H})  &  =t_{1}+t_{4},t_{2}+t_{3}\\
Orb(X^{\dag})  &  =t_{2}-t_{4},t_{1}-t_{3}\\
Orb(\widehat{X})  &  =t_{4}-t_{2},t_{3}-t_{1}\\
Orb(N_{R},\widehat{N}_{R})  &  =t_{1}-t_{5},t_{2}-t_{5}\\
Orb(D_{(1)},D_{(1)}^{c})  &  =t_{1}-t_{2},t_{2}-t_{1}\\
Orb(D_{(2)})  &  =t_{1}-t_{4},t_{2}-t_{3}\\
Orb(D_{(2)}^{c})  &  =t_{4}-t_{1},t_{3}-t_{2}\\
Orb(D_{(3)},D_{(3)}^{c})  &  =t_{3}-t_{4},t_{4}-t_{3}\\
Orb(D_{(4)}^{c})  &  =t_{5}-t_{3},t_{5}-t_{4}\text{.}%
\end{align}
By inspection, $D_{(1)}$ and $D_{(1)}^{c}$ are in the same orbit, and similar
considerations apply for $D_{(3)}$ and $D_{(3)}^{c}$. Hence, nothing forbids
the mass terms:%
\begin{equation}
\int d^{2}\theta\text{ }D_{(1)}D_{(1)}+\int d^{2}\theta\text{ }D_{(3)}%
D_{(3)}\text{,}%
\end{equation}
which will generically be at the GUT\ scale. Thus, these fields are unsuitable
as dark matter candidates.\footnote{It is also possible to show that these
candidates also couple to MSSM\ fields through dimension five operators, so
that they will also decay rapidly.} In particular, this means that potential
couplings involving these operators are now excluded as dark matter
candidates, since as in the usual seesaw mechanism, the mass of the heavy
state is near the GUT scale, and the light state will have mass closer to
$\mu^{2}/M_{GUT}$. This eliminates lines (\ref{GMSEC}) and (\ref{pah}) as
viable options.

The remaining dark matter candidate fields are then given by either the
bosonic or fermionic component of the chiral superfields $\widehat{N}_{R}$,
$\widehat{X}$, $D_{(2)}^{c}$, or $D_{(4)}^{c}$. As we now argue, $\widehat
{N}_{R}$ must constitute a TeV scale dark matter candidate. Note that of the
two remaining options given by lines (\ref{blah}) and (\ref{cubiccoup}), both
involve the field $\widehat{N}_{R}$. Either $\widehat{N}_{R}$ corresponds to a
dark matter candidate, or it generates a weak to TeV scale vev, inducing a
mass term for the other dark matter candidates.

In the context of F-theory GUTs, however, $\widehat{N}_{R}$ will typically not
acquire a vev due to the PQ\ deformation. Indeed, since it has the same charge
as the $X$ field, the PQ\ deformation will induce a positive mass squared
contribution. Moreover, if $\widehat{N}_{R}$ did develop a vev, it would
induce matter parity couplings in the low energy theory:%
\begin{equation}
\int d^{4}\theta\frac{H_{d}^{\dag}LN_{R}}{\Lambda_{\text{UV}}}\rightarrow\int
d^{4}\theta\frac{H_{d}^{\dag}L\left\langle N_{R}\right\rangle }{\Lambda
_{\text{UV}}}\text{.}%
\end{equation}
This is a somewhat more exotic possibility in the context of F-theory GUTs,
and so we will not dwell on this in any great detail. Another fine-tuned
possibility would be to require that the mass of $\widehat{N}_{R}$ is quite
low. In this case, kinematic considerations could exclude a potential decay
channel. This is again a somewhat exotic possibility in the context of
F-theory GUTs. Indeed, if anything, the PQ deformation tends to further
increase the mass of $\widehat{N}_{R}$. Thus, in all cases we conclude that in
any candidate dark matter scenario, there must exist a quasi-stable
$\widehat{N}_{R}$ field of TeV scale mass which is localized on the same curve
as the other right-handed neutrinos.

As we now explain, the candidate field $\widehat{N}_{R}$ always couples to
operators which induce rapid decays of the field. Since $\widehat{N}_{R}$
couples through a mass term to the remaining dark matter candidates, it
follows that all remaining states will be unstable against rapid decay. In
this regard, it is tempting to simply end the discussion here by invoking the
presence of operators of the form:%
\begin{equation}
\int d^{4}\theta\frac{H_{d}^{\dag}L\widehat{N}_{R}}{\Lambda_{\text{UV}}%
}\text{.}%
\end{equation}
Note, however, that this is the same type of term which induces a Dirac
neutrino mass \cite{BHSV}. Since there are only three generations of $L$'s,
one linear combination of the modes localized on the $N_{R}$ curve will not
couple to $H_{d}^{\dag}L$. Thus, this coupling is by itself not enough to
guarantee that $\widehat{N}_{R}$ decays rapidly.

Besides $H_{d}^{\dag}L\widehat{N}_{R}/\Lambda_{\text{UV}}$, $\widehat{N}_{R}$
also couples to MSSM\ fields through the dimension five operator:%
\begin{equation}
\int d^{2}\theta\frac{\widehat{N}_{R}\overline{5}_{M}\overline{5}_{M}10_{M}%
}{\Lambda_{\text{UV}}}:\widehat{N}_{R}=t_{1}-t_{5}\text{, }\overline{5}%
_{M}=t_{4}+t_{5}\text{, }\overline{5}_{M}=t_{3}+t_{5}\text{, }10_{M}%
=t_{2}\text{,}%
\end{equation}
where we have also listed the corresponding weight assignments. Both the
bosonic and fermionic components of the $\widehat{N}_{R}$ chiral multiplet can
now decay rapidly.

Some examples of the available decay channels for the bosonic and fermionic
components of $\widehat{N}_{R}$ are respectively:%
\begin{align}
\phi_{\widehat{N}_{R}}  &  \rightarrow\widetilde{\tau}_{1}^{\pm}\tau^{\mp}%
\nu_{\tau}\text{.}\\
\psi_{\widehat{N}_{R}}  &  \rightarrow\widetilde{\tau}_{1}^{\pm}%
\widetilde{\tau}_{1}^{\mp}\nu_{\tau}\text{,}%
\end{align}
where $\widetilde{\tau}_{1}^{\pm}$ denotes the lightest stau. In F-theory
GUTs, the mass of the lightest stau is typically in the range of $100-300$
GeV. Since the components of $\widehat{N}_{R}$ are required to be in the TeV
range just to provide a possible dark matter candidate, it follows that this
dark matter candidate is unstable against rapid decay. It now follows that
since all remaining candidates mix by a mass term with the components of
$\widehat{N}_{R}$, all available dark matter candidates are unstable against
rapid decay.

\subsubsection{Dirac Scenarios: $G_{mono}^{Dirac}\simeq%
\mathbb{Z}
_{3}$ or $S_{3}$}

Next consider monodromy scenarios where $G_{mono}^{Dirac}\simeq%
\mathbb{Z}
_{3}$ or $S_{3}$. Returning to our general discussion of available GUT
singlets, compatibility with $U(1)_{PQ}$ and $U(1)_{\chi}$ leads to the
following options:%
\begin{align}
G_{mono}^{Dirac}  &  \simeq%
\mathbb{Z}
_{3}\text{ or }S_{3}\\
\int d^{4}\theta\frac{X^{\dag}\Phi_{(1)}\Phi_{(2)}}{\Lambda_{\text{UV}}}  &
\Longrightarrow\Phi_{(1)}=\widehat{N}_{R}\text{, }\Phi_{(2)}=D_{(1)}^{c}\\
\int d^{4}\theta\frac{X^{\dag}\Phi_{(1)}\Phi_{(2)}}{\Lambda_{\text{UV}}}  &
\Longrightarrow\Phi_{(1)}=\widehat{X}\text{, }\Phi_{(2)}=D_{(2)},D_{(2)}^{c}\\
\int d^{2}\theta\text{ }\Phi_{(1)}\Phi_{(2)}\Phi_{(3)}  &  \Longrightarrow
\text{\ }\Phi_{(1)}=D_{(1)}\text{, }\Phi_{(2)}=D_{(1)}^{c}\text{, }\Phi
_{(3)}=D_{(2)},D_{(2)}^{c}\text{.}%
\end{align}
Of these options, we can immediately eliminate the last possibility, because
it involves a vector-like pair of fields localized on the same curve. Such
modes will generically lift from the low energy theory, and so this coupling
is ruled out. This leaves the first two options, and $\widehat{N}_{R}$,
$D_{(1)}^{c}$, $\widehat{X}$, $D_{(2)}$ and $D_{(2)}^{c}$ as potential dark
matter candidates. To establish the absence of a stable dark matter candidate,
it is enough to show that the candidates $\widehat{N}_{R}$, $D_{(2)}$ and
$D_{(2)}^{c}$ are all unstable on cosmological timescales. To this end, recall
that the orbits for the minimal matter and relevant dark matter candidates are
as obtained in section \ref{sec:DarkE}:%
\begin{align}
G_{mono}^{Dirac}  &  \simeq%
\mathbb{Z}
_{3}\text{ or }S_{3}\\
Orb(10_{M},Y_{10})  &  =t_{1},t_{2},t_{3}\\
Orb(Y_{\overline{10}}^{\prime})  &  =-t_{4}\\
Orb(\overline{5}_{M})  &  =t_{1}+t_{5},t_{2}+t_{5},t_{3}+t_{5}\\
Orb(5_{H})  &  =-t_{1}-t_{2},-t_{2}-t_{3},-t_{3}-t_{1}\\
Orb(\overline{5}_{H})  &  =t_{1}+t_{4},t_{2}+t_{4},t_{3}+t_{4}\\
Orb(X^{\dag})  &  =t_{1}-t_{4},t_{2}-t_{4},t_{3}-t_{4}\\
Orb(N_{R},\widehat{N}_{R})  &  =t_{4}-t_{5}\text{.}\\
Orb(D_{(1)}^{c})  &  =t_{5}-t_{1},t_{5}-t_{2},t_{5}-t_{3}\\
Orb(D_{(2)})  &  =t_{2}-t_{3},t_{3}-t_{1},t_{1}-t_{2}\\
Orb(D_{(2)}^{c})  &  =t_{3}-t_{2},t_{1}-t_{3},t_{2}-t_{1}\text{.}%
\end{align}

We first show that the candidate dark matter field $\widehat{N}_{R}$ again
decays too rapidly. Indeed, this field participates in the dimension five
operator:%
\begin{equation}
\int d^{2}\theta\frac{\widehat{N}_{R}\overline{5}_{M}\overline{5}_{M}10_{M}%
}{\Lambda_{\text{UV}}}:\widehat{N}_{R}=t_{4}-t_{5}\text{, }\overline{5}%
_{M}=t_{1}+t_{5}\text{, }\overline{5}_{M}=t_{2}+t_{5}\text{, }10_{M}%
=t_{3}\text{,}%
\end{equation}
with the weights of all fields indicated. Thus, just as in subsection
\ref{firstdec}, we conclude that both the bosonic and fermionic components of
$\widehat{N}_{R}$ decay too rapidly to constitute viable dark matter candidates.

Next consider the dark matter candidates $D_{(2)}$ and $D_{(2)}^{c}$. These
candidates also decay quite rapidly due to the higher dimension operators:%
\begin{align}
\int d^{2}\theta\frac{D_{(2)}5_{H}10_{M}10_{M}}{\Lambda_{\text{UV}}}  &
:D_{(2)}=t_{1}-t_{2},5_{H}=-t_{3}-t_{1},10_{M}=t_{2},10_{M}=t_{3}%
\label{alph}\\
\int d^{2}\theta\frac{D_{(2)}^{c}\overline{5}_{H}\overline{5}_{M}10_{M}%
}{\Lambda_{\text{UV}}}  &  :D_{(2)}^{c}=t_{2}-t_{1},\overline{5}_{H}%
=t_{1}+t_{4},\overline{5}_{M}=t_{1}+t_{5},10_{M}=t_{3}, \label{bet}%
\end{align}
where in the above, we have also indicated the corresponding weights of the fields.

Since the Higgs fields develop a weak scale vev anyway, it is enough to
consider some of the induced cubic couplings of lines (\ref{alph}) and
(\ref{bet}) with non-zero Higgs vevs. These operators have the form:%
\begin{equation}
\frac{M_{\text{weak}}}{\Lambda_{\text{UV}}}\int d^{2}\theta\text{ }DE_{L}%
E_{R},
\end{equation}
where $E_{L}$ is the charged part of the lepton doublet. Thus, both the scalar
and fermionic components of $D$ can decay to leptons quite rapidly. For
example, the fermionic component of the $D$'s can decay via:%
\begin{equation}
\psi_{D}\rightarrow\widetilde{\tau}_{1}^{\pm}\tau^{\mp}\text{.}
\label{fermdecay}%
\end{equation}
Since the lightest stau typically has mass in the range of $100-300$ GeV,
nothing kinematically forbids the decay of a TeV scale $D$.

Thus, we conclude that in all Dirac scenarios, generating a TeV scale mass for
the dark matter candidate is incompatible with it being quasi-stable.

\subsubsection{Majorana Scenarios}

The Majorana scenarios are significantly more constrained than their Dirac
counterparts because the sizes of the orbits are that much longer. This also
means that the number of available GUT\ singlets will also be significantly
reduced. Returning to line (\ref{MAJO}), the available GUT\ singlets are
$\widehat{X}$, $\widehat{N}_{R}$, $D_{(1)}$, $D_{(1)}^{c}$, $D_{(2)}$ and
$D_{(2)}^{c}$. Of these possibilities, note that only $\widehat{X}$ has
non-trivial PQ\ charge. Indeed, the analysis of Appendix A establishes that in
all Majorana scenarios, the weights for the orbit of $D_{(i)}$ and
$\widehat{N}_{R}$ contains both $t_{m}-t_{n}$ as well as $t_{n}-t_{m}$. Hence,
there is no symmetry in the covering theory which forbids the dark matter
candidates $\Phi_{(i)}$ from developing a large mass through the terms:
\begin{equation}
M_{GUT}\int d^{2}\theta\text{ }\Phi_{(i)}\Phi_{(i)}:\Phi_{(i)}=t_{m}%
-t_{n},\Phi_{(i)}=t_{n}-t_{m}.
\end{equation}
In other words, the dark matter candidate lifts from the low energy theory.\footnote{Note 
that such heavy fields are not cosmologically problematic because they
can also decay due to couplings of the form
$\Phi_{(i)}\overline{5}_{H}\overline{5}_{M}10_{M}/\Lambda_{\text{UV}}$.} This also
means that no light states are available with which $\widehat{X}$ can pair up
to develop a weak to TeV scale mass. Thus, in all Majorana neutrino scenarios,
there are simply no candidates available.

Combining this with the analysis of the Dirac scenarios, we conclude that in
no case do we have a TeV scale quasi-stable dark matter candidate.

\subsubsection{Non-Minimal Scenarios}

Although beyond the scope of this paper, we note that less minimal models in
which nearby seven-branes intersecting the PQ seven-brane also participate can
potentially lead to additional possibilities, such as when an $E_{8}$-type
singularity nearly collides with other singularities.

At least at the level of effective field theory considerations, it is
conceivable that a dark matter candidate could have PQ\ charges and additional
discrete symmetries as necessary to exclude operators which induce rapid
decay. For example, in the context of a Dirac neutrino scenario, we can
introduce dark matter chiral multiplets $\Phi_{(1)}$ and $\Phi_{(2)}$, with
PQ\ charges:%
\begin{equation}%
\begin{tabular}
[c]{|c|c|c|c|c|c|c|c|c|}\hline
& $\overline{5}_{M}$ & $10_{M}$ & $5_{H}$ & $\overline{5}_{H}$ & $X^{\dag}$ &
$N_{R}$ & $\Phi_{(1)}$ & $\Phi_{(2)}$\\\hline
Dirac $U(1)_{PQ}$ & $+1$ & $+1$ & $-2$ & $-2$ & $+4$ & $-3$ & $q_{1}$ &
$q_{2}=-q_{1}-4$\\\hline
\end{tabular}
\ \ \ \ \ \ \ \ \ \text{.}%
\end{equation}
Provided $q_{1}$ is at least an order ten number, it is now immediate that an
odd number of $\Phi_{(i)}$'s cannot appear in an operator. It is therefore
enough to consider operators which are quadratic in the $\Phi_{(i)}$'s
appearing in the combinations $\Phi_{(1)}\Phi_{(2)}$ or $\Phi_{(i)}^{\dag}%
\Phi_{(i)}$. To induce a decay, one of the $\Phi_{(i)}$'s must then develop a
weak scale vev. Note that although very large PQ\ charges constitute a
significant departure from the minimal scenarios considered in this paper, PQ
charges as large as $\pm7$ naturally appeared in the monodromy orbit
classification reviewed in section \ref{sec:DarkE}.

We now rule out potentially problematic dimension five or lower operators
which could destabilize the dark matter candidate. The available operators are
of the form $\Phi_{(1)}\Phi_{(2)}\mathcal{O}_{12}$ or $\Phi_{(i)}^{\dag}%
\Phi_{(i)}\mathcal{O}_{\overline{i}i}$, where $\mathcal{O}_{12}$ and
$\mathcal{O}_{\overline{i}i}$ are operators constructed from the minimal
matter of F-theory GUTs with respective PQ\ charges $+4$ and $0$. In the
former case, note that by inspection of the available minimal matter and their
charges, no dimension four or five F-terms can be generated, and as for
D-terms, only $\mathcal{O}_{12}=X^{\dag}$ is available. By gauge invariance,
the combination $\Phi_{(i)}^{\dag}\Phi_{(i)}$ must involve at least two
minimal fields. We therefore conclude that in all cases, no dangerous operator
is available.\footnote{Since we have already excluded all combinations
involving $X$ and $X^{\dag}$, this also rules out instanton generated
operators. Indeed, an instanton suppressed contribution can effectively be
viewed as multiplication by a field with the same charge as $X^{\dag}$ since
instantons also induce the Polonyi term $q_{\text{inst}}X$, as in the gauge
mediated supersymmetry breaking scenario in \cite{HVGMSB} (see also
\cite{MarsanoGMSB}).}

On the other hand, symmetry considerations are compatible with operators of
the form:%
\begin{equation}
L\supset\int d^{4}\theta\frac{\Phi_{(i)}^{\dag}\Phi_{(j)}\Sigma_{MSSM}^{\dag
}\Sigma_{MSSM}}{\Lambda_{\text{UV}}^{2}}\text{,}%
\end{equation}
where $\Sigma_{MSSM}$ denotes a chiral superfield of the MSSM. See
\cite{Arvanitaki:2008hq} for further discussion on this and other operators
which are suitable for decaying dark matter scenarios. Letting the same
variables denote the bosonic components of $\Sigma$, and $\phi$ and $\psi$ the
respective bosonic and fermionic components of $\Phi$, the Lagrangian
therefore contains the terms:%
\begin{equation}
L\supset\frac{\phi_{(j)}\Sigma_{MSSM}^{\dag}}{\Lambda_{\text{UV}}^{2}}%
\partial^{\mu}\phi_{(i)}^{\dag}\partial_{\mu}\Sigma_{MSSM}+\frac{\phi
_{(j)}\Sigma_{MSSM}^{\dag}}{\Lambda_{\text{UV}}^{2}}\overline{\psi}%
_{(i)}i\overline{\sigma}^{\mu}\partial_{\mu}\psi_{\Sigma}\text{.}%
\end{equation}
In other words, provided one of the $\phi_{(i)}$'s develops a weak scale vev,
this leads to the required decay rate associated with a dimension six operator.

As a cautionary note, this is by itself still insufficient to demonstrate that
such possibilities can be realized in F-theory. A complete realization would
require introducing a somewhat exotic sector of the type discussed in section
\ref{sec:ESHIELD}.

\subsection{Other Aspects of Annihilating Scenarios}

In the previous subsections we have encountered significant obstructions to
realizing other dark matter candidates besides the gravitino in F-theory GUTs.
These obstructions are especially severe in the case of annihilating
scenarios, where we have seen that the dilution of thermally produced dark
matter requires an even bigger boost in the annihilation cross section, as
compared with WIMP\ scenarios. Moreover, when the saxion decay generates such
candidates non-thermally, we have also found that the relic abundance is too
large by several orders of magnitude! In addition, all of the available
GUT\ singlets from $E_{8}$ which could in principle develop a weak to TeV
scale mass are unstable on cosmological timescales, and are therefore
unsuitable as dark matter candidates.

In this subsection we comment on some additional aspects of annihilating
scenarios where the candidate dark sector embeds inside of $E_{8}$. The first
point is that a common ingredient in many annihilating scenarios, namely the
existence of an additional light \textquotedblleft dark\textquotedblright%
\ gauge boson is simply unavailable within $E_{8}$. On the other hand, we show
that with sufficient fine tuning of the parameters of F-theory GUTs, a light
scalar could potentially communicate between the dark and visible sectors.
This by itself is interesting, although it is not completely clear even in
this case how to overcome all of the other obstacles found earlier in this section.

As reviewed in section \ref{sec:Review}, there is already some tension in
annihilating dark matter scenarios because the expected annihilation cross
section from thermally produced WIMPs leads to a signal which is too small to
explain the results of PAMELA by a factor of $10^{-2}-10^{-3}$. One way that
the annihilation cross section can be boosted (as required by PAMELA), is
through an infrared enhancement due to the exchange of a light boson with mass
far below the weak scale, and closer to a few GeV. See for example
\cite{Hisano:2003ec,Hisano:2004ds,Cirelli:2007xd,ArkaniHamed:2008qn} for
discussion on this possibility. It is therefore natural to ask whether such
degrees of freedom are available in F-theory GUTs.

The classification of monodromy orbits reviewed in section \ref{sec:DarkE}
already shows that no spin one mediators are available inside of a minimal
$E_{8}$ F-theory GUT. Indeed, in the Dirac neutrino scenarios, there are only
two $U(1)$'s, $U(1)_{PQ}$ and $U(1)_{\chi}$, and in the\ Majorana neutrino
scenarios, only $U(1)_{PQ}$ is available. Since $U(1)_{PQ}$ is anomalous, the
generalized Green-Schwarz mechanism cancels the corresponding anomaly, but
also cause this gauge boson to develop a large mass via the St\"{u}ckelberg
mechanism. This leaves only $U(1)_{\chi}$ in the Dirac neutrino scenario as
even a candidate for consideration. As reviewed in section \ref{sec:DarkE},
$U(1)_{\chi}$ is a very specific linear combination of $U(1)_{Y}$ and
$U(1)_{B-L}$ dictated by the embedding $SO(10)\supset SU(5)_{GUT}\times
U(1)_{\chi}$. The experimental bounds on the mass of this type of gauge boson
are quite stringent, and well above $100$ GeV.\footnote{See for example
\cite{Langacker:2008yv} for further discussion on the physics of extra $U(1)$
gauge bosons.} We therefore conclude that there are no available gauge bosons
in an $E_{8}$ F-theory GUT which can provide the required Sommerfeld enhancement.

This leaves open the possibility, however, that a light scalar could play the
role of the required mediator. Indeed, in the case of Dirac neutrino scenarios
there are typically several extra GUT\ singlet curves available. Rather than
present an exhaustive list of possible scenarios involving such models, here
we simply comment on the fact that at least some of the ingredients of such
scenarios can indeed be accommodated within $E_{8}$. Even so, this by itself
does not overcome all of the obstacles reviewed at the beginning of this
subsection, and we will encounter additional problems in the specific example considered.

To illustrate some of the issues involved, consider the model of
\cite{Allahverdi:2008jm}. This model includes a $U(1)_{B-L}$ gauge boson, and
a chiral superfield $S$ which couples to another GUT\ singlet $S^{\prime}$
through a \textquotedblleft dark $\mu$-term\textquotedblright, so that the
following terms are present in the effective Lagrangian:%
\begin{equation}
L\supset\int d^{4}\theta S^{\dag}e^{V_{B-L}}S+\int d^{4}\theta S^{\prime\dag
}e^{-V_{B-L}}S^{\prime}+\mu_{\text{dark}}\int d^{2}\theta\text{ }SS^{\prime}.
\label{LDARK}%
\end{equation}
When the $S$'s develop TeV scale vevs, this will Higgs the $U(1)_{B-L}$
symmetry. Letting $\tan\beta_{\text{dark}}$ denote the ratio:%
\begin{equation}
\tan\beta_{\text{dark}}=\frac{\left\langle S\right\rangle }{\left\langle
S^{\prime}\right\rangle },
\end{equation}
when $\tan\beta_{\text{dark}}$ is sufficiently close to one (which requires
fine tuning in the parameters of the F-theory GUT), the physical mass spectrum
will consist of a triplet of TeV scale \textquotedblleft
neutralinos\textquotedblright, given by the $U(1)_{B-L}$ gaugino and the two
$S$ and $S^{\prime}$ \textquotedblleft Higgsinos\textquotedblright, and a
scalar with mass on the order of a few GeV \cite{Allahverdi:2008jm}. Our aim
in this section will not be to realize every detail of the construction in
\cite{Allahverdi:2008jm}, but rather to show that the elements in the
Lagrangian of line (\ref{LDARK}) can indeed be accommodated in some Dirac
neutrino scenarios.

To present one example of this type, recall that when the monodromy group of a
Dirac scenario is either $%
\mathbb{Z}
_{3}$ or $S_{3}$, we have available the following GUT\ singlets as dark matter
candidates:%
\begin{equation}%
\begin{tabular}
[c]{|c|c|c|c|c|c|c|c|c|}\hline
$G_{mono}^{Dirac}\simeq%
\mathbb{Z}
_{3}$ or $S_{3}$ & $\widehat{X}$ & $\widehat{N}_{R}$ & $D_{(1)}$ &
$D_{(1)}^{c}$ & $D_{(2)}$ & $D_{(2)}^{c}$ & $Z_{PQ}$ & $Z_{\chi}$\\\hline
$U(1)_{PQ}$ & $-4$ & $-3$ & $+1$ & $-1$ & $0$ & $0$ & $0$ & $0$\\\hline
$U(1)_{\chi}$ & $0$ & $-5$ & $-5$ & $+5$ & $0$ & $0$ & $0$ & $0$\\\hline
\end{tabular}
\ \ .
\end{equation}
At the level of symmetries then, we can consider an extra right-handed
neutrino \textquotedblleft generation\textquotedblright\ $\widehat{N}_{R}$,
and pair this with $D_{(1)}^{c}$. Making the identification:%
\begin{align}
S  &  =\widehat{N}_{R}\\
S^{\prime}  &  =D_{(1)}^{c}\text{,}%
\end{align}
note that the effective field theory does not forbid the operator $X^{\dag
}SS^{\prime}/\Lambda_{\text{UV}}$, which induces a \textquotedblleft%
$\mu_{\text{dark}}$-term\textquotedblright\ once $X$ develops a supersymmetry
breaking vev.

It is important to check that $U(1)_{\chi}$ (and hence $U(1)_{B-L}$) is
non-anomalous with respect to this choice of matter assignments. Note that in
addition to the MSSM\ matter fields, as in \cite{BHSV}, there are three
generations of right-handed neutrinos, and, by inspection of the $U(1)_{\chi}$
charge assignments of the messengers, $\widehat{N}_{R}$ and $D_{(1)}^{c}$, all
of these additional contributions appear in vector-like pairs with respect to
$U(1)_{\chi}$. Thus, there is at least the possibility that the mass of the
$U(1)_{B-L}$ gauge boson could remain close to the TeV scale.

In fact, it is also straightforward to check that the operator $X^{\dag
}SS^{\prime}/\Lambda_{\text{UV}}$ is compatible with the orbits of the
monodromy group. Indeed, using the orbits reviewed in section \ref{sec:DarkE},
we can take the following weights for $X^{\dag}$, $S=\widehat{N}_{R}$, and
$S^{\prime}=D_{(1)}^{c}$:%
\begin{align}
Orb(X^{\dag})  &  \ni t_{1}-t_{4}\\
Orb(\widehat{N}_{R})  &  \ni t_{4}-t_{5}\\
Orb(D_{(1)}^{c})  &  \ni t_{5}-t_{1}.
\end{align}
Thus, it is therefore in principle possible to reproduce the Lagrangian of
line (\ref{LDARK}).

Note, however, that as argued in subsection \ref{UNSTAB}, it is unappealing to
allow $\widehat{N}_{R}$ to develop a vev, as this breaks matter parity in the
model. Moreover, it is unclear in the present context how to arrange for the
vevs of $\widehat{N}_{R}$ and $D_{(1)}^{c}$ to be very close to each other so
that $\tan\beta_{\text{dark}}\simeq1$. Indeed, the PQ charges of these two
fields are different, and will therefore experience different
PQ\ deformations. Thus, without significant fine tuning of the model, it
appears difficult to generate the required scalar with mass on the order of a
few GeV.

Although all of the ingredients of the construction presented above fall
inside of $E_{8}$, making this model fully realistic still appears to be quite
problematic. Indeed, the required elements include a TeV scale $U(1)_{\chi}$
gauge boson, an extra generation on the right-handed neutrino curve, and an
additional GUT\ singlet matter curve. This constitutes a significant departure
from the spirit of minimality adopted in this paper. Coupled with the fact
that we have also encountered significant phenomenological problems with
saxion dilution, and the instability of all available dark matter candidates,
we again end with the same conclusion that TeV scale dark matter in both
annihilating and decaying scenarios is unfeasible in minimal F-theory GUTs.

\section{Conclusions\label{sec:CONC}}

F-theory GUTs provide a quite predictive framework for making contact between
string theory and phenomenology. In particular, the rich interplay between
geometric and group theoretic ingredients leads to significant constraints
which appear unmotivated from the perspective of the four-dimensional
effective field theory. In this paper we have exploited the rigid structure
present in this unified approach to study whether additional charged and
uncharged matter of such models can provide potential candidates for dark
matter beyond the gravitino. Assuming that all of the interaction terms of the
F-theory GUT descend from a single point of $E_{8}$ enhancement, we have found
a remarkably rigid structure which does not admit many options for additional
matter. A surprising outcome of this analysis is that in all but one Dirac
neutrino scenario, incorporating a minimal gauge mediated supersymmetry
breaking sector forces the messengers to transform in vector-like pairs in the
$10\oplus\overline{10}$ of $SU(5)$. Moreover, embedding in $E_{8}$ also
significantly limits the available dark matter candidates. Indeed, since
supersymmetry breaking is correlated with the dynamics of a seven-brane which
is present in the local model, TeV scale dark matter candidates are forced to
also interact closely with the GUT\ seven-brane. Expanding out from the
$E_{8}$ point of enhancement, we have studied other potential dark objects
which either intersect some locus connected with the $E_{8}$ point of
enhancement, or which are fully cut off from the visible sector. In all cases,
we have encountered obstructions which render all available candidates as
unsuitable explanations for recent experimental results connected with the
detection of dark matter. \textit{This again reinforces the point that in
F-theory GUTs, gravitinos remain as the prime candidate for dark matter}. In
the remainder of this section we speculate on possible directions of future investigation.

While aesthetically quite pleasing, the existence of a single point of $E_{8}$
enhancement leads to surprisingly strong restrictions on the form of both the
visible and dark sectors of F-theory GUTs. In particular, contrary to a
perhaps naive notion that a single vector-like pair of messengers in the
$5\oplus\overline{5}$ is \textquotedblleft minimal\textquotedblright, here we
have seen that geometric minimality instead selects as the more natural option
messengers organized into multiples of $10\oplus\overline{10}$ (where each
such pair effectively plays the role of \textit{three} $5\oplus\overline{5}$
pairs). This leads to a class of signatures which are distinct from the single
$5\oplus\overline{5}$ messenger case, and it would be interesting to study the
phenomenology of such models in greater detail. In this vein, we have seen
that in the Dirac scenario, there is naturally another candidate $U(1)$
symmetry which can combine with the $U(1)_{PQ}$ symmetry. This defines a whole
family of PQ deformations, with different consequences for phenomenology.

More broadly speaking, however, less minimal implementations of F-theory GUTs
need not contain a single point of $E_{8}$ enhancement. In this case, some of
the constraints will nevertheless survive provided there is still a sense in
which all matter descends from a single well-defined $E_{8}$ singularity. On
the other hand, more general F-theory compactifications provide another
element of flexibility. Our goal here has to been to study a particularly
motivated choice which is compatible with the requirements of unification. It
would be interesting to determine what can be said about dark matter in this
more general case.

The primary restrictions we have encountered have centered on obstructions
encountered in trying to fit a TeV scale dark matter candidate with recent
experimental results. We have also seen that F-theory GUTs can quite naturally
accommodate additional fields with masses in the range of $10$ MeV to a few
GeV. It would be interesting to study the consequences of such objects for
other dark matter experiments, such as DAMA.

\section*{Acknowledgements}

We especially thank D.E. Morrissey for many helpful and informative
discussions. We also thank N. Arkani-Hamed, D.P. Finkbeiner, T. Hartman, J.D.
Mason, D. Poland, T. Slatyer and S. Watson for helpful discussions. The work
of the authors is supported in part by NSF grant PHY-0244821.

\appendix

\section*{Appendices}

\section{Monodromy Orbit Classification \label{app:DarkE}}

In this Appendix we present a full classification of possible monodromy groups
such that all of the interaction terms of an F-theory GUT consistently embed
at a single unified $E_{8}$ interaction point. With notation as in section
\ref{sec:DarkE}, we consider the decomposition of the adjoint representation
of $E_{8}$ into irreducible representations of $SU(5)_{GUT}\times SU(5)_{\bot
}$:%
\begin{align}
E_{8}  &  \supset SU(5)_{GUT}\times SU(5)_{\bot}\\
248  &  \rightarrow(1,24)+(24,1)+(5,\overline{10})+(\overline{5}%
,10)+(10,5)+(\overline{10},\overline{5})\text{.}%
\end{align}
The dark objects transform as singlets under $SU(5)_{GUT}$ and must therefore
descend from the $24$ of $SU(5)_{\bot}$. The Cartan of $SU(5)_{\bot}$ can be
parameterized by the coordinates $t_{1},...,t_{5}$ subject to the constraint:%
\begin{equation}
t_{1}+...+t_{5}=0\text{.}%
\end{equation}
The weights of the $5_{\bot}$, $10_{\bot}$ and $24_{\bot}$ of $SU(5)_{\bot}$
are then given as:%
\begin{align}
5_{\bot}  &  :t_{i}\\
10_{\bot}  &  :t_{i}+t_{j}\\
24_{\bot}  &  :\pm\left(  t_{i}-t_{j}\right)  +4\times(0\text{ weights})
\end{align}
for $1\leq i,j\leq5$ such that $i\neq j$. \ It thus follows that matter fields
with appropriate $SU(5)_{GUT}$ representation content localize along the
following curves:%
\begin{align}
5_{GUT}  &  :-t_{i}-t_{j}=0\\
10_{GUT}  &  :t_{i}=0\\
1_{GUT}  &  :t_{i}-t_{j}=0\text{,}%
\end{align}
so that the vanishing loci then correspond to local enhancements in the
singularity type of the compactification. A matter field with a given weight
will necessarily also be charged under a $U(1)$ subgroup of $SU(5)_{\bot}$,
dictated by its weight.

In this Appendix we classify possible monodromy orbits consistent with the
interaction terms:%
\begin{equation}
\int d^{4}\theta\frac{X^{\dag}H_{u}H_{d}}{\Lambda_{\text{UV}}}+\int
d^{2}\theta\text{ }5_{H}\times10_{M}\times10_{M}+\int d^{2}\theta\text{
}\overline{5}_{H}\times\overline{5}_{M}\times10_{M}+\text{Neutrinos,}
\label{neccinteraction}%
\end{equation}
with the neutrino sector as specified in section \ref{sec:DarkE}. More
precisely, we shall also demand that that all three generations in the
$10_{M}$ localize on one curve, and further all three generations in the
$\overline{5}_{M}$ localize on one curve.

The rest of this Appendix is organized as follows. First, we discuss
constraints on the orbit imposed by the interaction terms of the MSSM. Next,
we separate our classification of orbits under the monodromy group into Dirac
and Majorana neutrino scenarios.

\subsection{Constraints From the MSSM}

We now proceed to classify the possible orbits of the monodromy group
inherited from $SU(5)_{\bot}$ which are compatible with the MSSM\ interactions
of line (\ref{neccinteraction}). We will return to constraints imposed by the
neutrino sector in later subsections.

Although we do not know the full orbit of the monodromy group $G$, we do know
that it must be compatible with the presence of the interaction term
$5_{H}\times10_{M}\times10_{M}$. \ In particular, without loss of generality,
we may fix the weight assignment of the $10_{M}$ orbit to include the
$SU(5)_{\bot}$ weights $t_{1}$ and $t_{2}$. \ Hence, we also conclude that the
orbit of the $5_{H}$ matter field must also contain a weight of the form
$-t_{1}-t_{2}$. \ Note that there can in principle be additional weights in
each orbit. \ Nevertheless, we can now conclude that the orbits of $5_{H}$ and
$10_{M}$ minimally contain:%
\begin{align}
Orb(10_{M})  &  =t_{1},t_{2},...\label{10mono}\\
Orb(5_{H})  &  =-t_{1}-t_{2},....
\end{align}

Next consider the presence of the interaction term $X^{\dag}H_{u}H_{d}%
/\Lambda_{\text{UV}}$. \ In principle, various components of each orbit in the
cover can participate in such an interaction term. Thus, the $H_{u}$ field
which participates in a given covering theory interaction term need not
correspond to the weight $-t_{1}-t_{2}$ but might instead correspond to a
weight of the form $-t_{i}-t_{j}$.\ Note, however, that since $-t_{i}-t_{j}$
and $-t_{1}-t_{2}$ both lie in the same orbit, there exists an element of the
monodromy group which maps $-t_{i}-t_{j}$ to $-t_{1}-t_{2}$. This group action
will also map the weights for $X^{\dag}$ and $H_{d}$ to other elements of
their respective orbits. \ Since $X^{\dag}$ is a singlet under $SU(5)_{GUT}$
which is charged under $U(1)_{PQ}$, it must correspond to a weight of the form
$t_{k}-t_{l}$. \ In addition, $H_{d}$ must correspond to a weight of the form
$t_{m}+t_{n}$. \ The presence of the interaction term $X^{\dag}H_{u}%
H_{d}/\Lambda_{\text{UV}}$ then imposes the constraint:%
\begin{equation}
(t_{k}-t_{l})+(-t_{1}-t_{2})+(t_{m}+t_{n})=0\text{,}%
\end{equation}
which implies one of the two sets of conditions must be met:%
\begin{align}
t_{m}+t_{n}  &  =t_{1}+t_{l}\text{ and }t_{k}-t_{l}=t_{2}-t_{l}\\
t_{m}+t_{n}  &  =t_{2}+t_{l}\text{ and }t_{k}-t_{l}=t_{1}-t_{l}\text{.}%
\end{align}
In fact, since line (\ref{10mono}) already requires the monodromy group to
identify $t_{1}$ and $t_{2}$ in the same orbit, the distinction between these
two options is ambiguous. Since nothing in our discussion fixes $t_{l}$,
without loss of generality, we may take one element of the orbit for $X^{\dag
}$ to be $t_{2}-t_{4}$, with one weight in the orbit of $H_{d}$ then fixed to
be $t_{1}+t_{4}$. We therefore conclude that in addition to the $10_{M}$ and
$5_{H}$ orbits, we must minimally also include the following terms in the
orbit for $H_{d}$ and $X^{\dag}$:%
\begin{align}
Orb(10_{M})  &  =t_{1},t_{2},...\\
Orb(5_{H})  &  =-t_{1}-t_{2},...\\
Orb(\overline{5}_{H})  &  =t_{1}+t_{4},...\\
Orb(X^{\dag})  &  =t_{2}-t_{4},....
\end{align}

The final non-neutrino constraint which we can obtain derives from the
presence of the interaction term $\overline{5}_{H}\times\overline{5}_{M}%
\times10_{M}$. \ Again, by a suitable action of the monodromy group, we know
that there will be an interaction in the cover which involves the
$\overline{5}_{H}$ with weight $t_{1}+t_{4}$. \ The weight of $\overline
{5}_{M}$ is then of the form $t_{i}+t_{j}$, and the weight for $10_{M}$ is
$t_{k}$. \ The presence of the interaction term then requires:%
\begin{equation}
(t_{1}+t_{4})+(t_{i}+t_{j})+t_{k}=0. \label{finalgencon}%
\end{equation}
Hence, $t_{i}$, $t_{j}$ and $t_{k}$ must all be distinct, and all different
from $t_{1}$ and $t_{4}$.

We now further fix the weight assignments for the $\overline{5}_{M}$ field.
Recall that the orbit of this field contains a weight of the form $t_{i}%
+t_{j}$ with $t_{i}$ and $t_{j}$ distinct from $t_{1}$ and $t_{2}$. There are
in principle three possibilities available, once we fix the weight of the
$10_{M}$:
\begin{align}
\text{Option 1}  &  \text{: }Orb(\overline{5}_{M})\ni t_{3}+t_{5}\text{,
}Orb(10_{M})\ni t_{2}\label{opone}\\
\text{Option 2}  &  \text{: }Orb(\overline{5}_{M})\ni t_{2}+t_{5}\text{,
}Orb(10_{M})\ni t_{3}\label{optwo}\\
\text{Option 3}  &  \text{: }Orb(\overline{5}_{M})\ni t_{2}+t_{3}\text{,
}Orb(10_{M})\ni t_{5}\text{.} \label{opthr}%
\end{align}
Options two and three are in some sense the same because up to this point our
discussion has not distinguished between $t_{3}$ and $t_{5}$. \ We may
therefore focus without loss of generality on options one and two.

To proceed further, we next turn to constraints derived from the specific
content of the Dirac and Majorana neutrino scenarios. After fixing a
convention for the direction of the $U(1)_{PQ}$ symmetry, we return to the
three options listed in lines (\ref{opone})-(\ref{opthr}).

\subsection{Dirac Neutrino Scenarios}

First consider Dirac neutrino scenarios. Our strategy will be to first
constrain the form of $U(1)_{PQ}$, and to then use this information and
compatibility with the remaining interaction terms to fix further properties
of the weights.

Our first constraint stems from a proper identification of the $U(1)_{PQ}$
symmetry. \ This can be viewed as vector in the space dual to the weight
space, such that for each weight $t_{i}$, we assign a definite charge. \ The
general form of $U(1)_{PQ}$ is:%
\begin{equation}
t_{PQ}^{\ast}=a_{1}t_{1}^{\ast}+a_{2}t_{2}^{\ast}+a_{3}t_{3}^{\ast}+a_{4}%
t_{4}^{\ast}+a_{5}t_{5}^{\ast}\text{,}%
\end{equation}
where:%
\begin{equation}
t_{i}^{\ast}(t_{j})=\delta_{ij}\text{.}%
\end{equation}
In order for an element of the Cartan to remain intact under monodromy, it
must be invariant under the action of the monodromy group.

To fix conventions, as in \cite{HVGMSB,BHSV} we use the same PQ\ charge
assignments associated with the embedding $E_{6}\supset SO(10)\times
U(1)_{PQ}$:\footnote{We will find below that there is in fact another $U(1)$
in Dirac scenarios, so that the actual \textquotedblleft
PQ\ deformation\textquotedblright\ can correspond to a linear combination of
this $U(1)$ and the other $U(1)$ we find. Note, however, that since all of the
interaction terms we find will be invariant under both $U(1)$'s, there is no
loss of generality in remaining with this convention.}
\begin{equation}%
\begin{tabular}
[c]{|c|c|c|c|c|c|c|}\hline
& $\overline{5}_{M}$ & $10_{M}$ & $5_{H}$ & $\overline{5}_{H}$ & $X^{\dag}$ &
$N_{R}$\\\hline
Dirac $U(1)_{PQ}$ & $+1$ & $+1$ & $-2$ & $-2$ & $+4$ & $-3$\\\hline
\end{tabular}
.
\end{equation}
With this convention, we now fix some of the values of the $a_{i}$. The orbit
of the $10_{M}$ contains the weights $t_{1}$ and $t_{2}$.\ Since the $10_{M}$
has PQ\ charge $+1$, we conclude:%
\begin{equation}
a_{1}=a_{2}=1\text{.}%
\end{equation}
Note that this also fixes the PQ\ charge of the $5_{H}$ weight $-t_{1}-t_{2}$.
\ Next consider the PQ\ charge for the $\overline{5}_{H}$, which is $-2$.
\ The orbit of this field contains the weight $t_{1}+t_{4}$, which implies:%
\begin{equation}
a_{1}+a_{4}=-2\text{.}%
\end{equation}
Hence, we also deduce that $a_{4}=-3$.

\subsubsection{First Dirac Scenario\label{FIRSTDIR}}

Returning to lines (\ref{opone})-(\ref{opthr}), we now show that both
physically distinct options yield an acceptable monodromy group action. First
consider the possibility that option 1 of line (\ref{opone}) is realized so
that:%
\begin{equation}
\text{Option 1}\text{: }Orb(\overline{5}_{M})\ni t_{3}+t_{5}\text{,
}Orb(10_{M})\ni t_{2}\text{.}%
\end{equation}
Since the PQ\ charges of $\overline{5}_{M}$ and $10_{M}$ are both $+1$, this
imposes the constraint:%
\begin{equation}
a_{3}+a_{5}=1\text{.}%
\end{equation}
Combining all of the constraints derived previously, the form of the
$U(1)_{PQ}$ generator in the dual to the weight space is fixed to be of the
form:%
\begin{equation}
t_{PQ}^{\ast}=t_{1}^{\ast}+t_{2}^{\ast}+a_{3}t_{3}^{\ast}-3t_{4}^{\ast
}+(1-a_{3})t_{5}^{\ast}\text{.}%
\end{equation}

We now determine the weight assignments for the neutrino sector by requiring
the presence of the interaction term $H_{d}^{\dag}LN_{R}/\Lambda_{\text{UV}}$.
Fixing the weight assignment for $H_{d}^{\dag}$, $L$ and $N_{R}$ as
$-t_{1}-t_{4}$, $t_{i}+t_{j}$ and $t_{m}-t_{n}$, it follows that these weights
obey the constraint:
\begin{equation}
(-t_{1}-t_{4})+(t_{i}+t_{j})+(t_{m}-t_{n})=0\text{.}%
\end{equation}
There are two possible solutions, corresponding to the weight assignments:%
\begin{align}
\text{Option 1a}  &  \text{: }Orb(\overline{5}_{M})\ni t_{1}+t_{n}\text{,
}Orb(N_{R})\ni t_{4}-t_{n}\\
\text{Option 1b}  &  \text{: }Orb(\overline{5}_{M})\ni t_{4}+t_{n}\text{,
}Orb(N_{R})\ni t_{1}-t_{n}\text{.}%
\end{align}
Since the PQ\ charge of $\overline{5}_{M}$ and $N_{R}$ are respectively $+1$
and $-3$, this leads to the constraints:%
\begin{align}
\text{Option 1a}  &  \text{:}\text{ }1+a_{n}=1\text{, }-3-a_{n}=0\\
\text{Option 1b}  &  \text{:}-3+a_{n}=1\text{, }1-a_{n}=-3\text{.}%
\end{align}
Option 1a does not possess a consistent solution. We therefore conclude that
$a_{n}=4$ is the unique possibility. \ This implies that either $(a_{3}%
,a_{5})=(4,-3)$ or $(a_{3},a_{5})=(-3,4)$. \ Since none of our analysis has so
far distinguished $t_{3}$ or $t_{5}$, without loss of generality, we may take
$a_{3}=-3$. \ Hence, the $U(1)_{PQ}$ direction is fixed to be:%
\begin{equation}
t_{PQ}^{\ast}=t_{1}^{\ast}+t_{2}^{\ast}-3t_{3}^{\ast}-3t_{4}^{\ast}%
+4t_{5}^{\ast}\text{.}%
\end{equation}

Returning to the weight assignment for the right-handed neutrino, this also
fixes the orbit of $N_{R}$ to contain the weight $t_{1}-t_{5}$, and the orbit
$\overline{5}_{M}$ to contain the weight $t_{4}+t_{5}$. \ We therefore
conclude that the orbits for each field are of the form:%
\begin{align}
Orb(10_{M})  &  =t_{1},t_{2},...\label{first}\\
Orb(\overline{5}_{M})  &  =t_{4}+t_{5},...\\
Orb(5_{H})  &  =-t_{1}-t_{2},...\\
Orb(\overline{5}_{H})  &  =t_{1}+t_{4},...\\
Orb(X^{\dag})  &  =t_{2}-t_{4},...\\
Orb(N_{R})  &  =t_{1}-t_{5},.... \label{last}%
\end{align}

The form of the $U(1)_{PQ}$ generator $t_{PQ}$ also allows us to deduce the
allowed monodromy groups. \ The essential point is that a permutation on the
$t_{i}$'s must leave $t_{PQ}^{\ast}$ fixed. In particular, this means that the
only possible generators are, in terms of cycles $(12)$, $(34)$, and
$(12)(34)$. \ Thus, the only available monodromy groups are generated by such
elements, and are therefore isomorphic to either $%
\mathbb{Z}
_{2}$ or $%
\mathbb{Z}
_{2}\times%
\mathbb{Z}
_{2}$.

In fact, it is possible to go further and deduce that the monodromy group must
act non-trivially on $t_{3}$ and $t_{4}$. Suppose to the contrary that $t_{4}$
is invariant under the monodromy group. Returning to the weight assignments of
lines (\ref{first})-(\ref{last}), the interaction $\overline{5}_{H}%
\times\overline{5}_{M}\times10_{M}$ would then be of the form $2t_{4}+...$.
Since $t_{4}$ appears twice, this interaction term would then be forbidden.
This in particular implies that either the orbit of $\overline{5}_{M}$, or
$\overline{5}_{H}$ must include a contribution from $t_{3}$. \ Hence, the
monodromy group acts non-trivially on $t_{3}$ and $t_{4}$. \ In this case,
there are therefore precisely two monodromy groups:%
\begin{align}
G_{(2)}^{Dir\left(  1\right)  }  &  =\left\langle (12)(34)\right\rangle \simeq%
\mathbb{Z}
_{2}\\
G_{(4)}^{Dir\left(  1\right)  }  &  =\left\langle (12),(34)\right\rangle
\simeq%
\mathbb{Z}
_{2}\times%
\mathbb{Z}
_{2}\text{,}%
\end{align}
where the subscript indicates the order of the group. \ It is therefore
possible to write out the full orbits in this case. \ In the case where the
monodromy group is $G_{(2)}^{Dir\left(  1\right)  }$, we find:%
\begin{align}
G_{(2)}^{Dir\left(  1\right)  }  &  =\left\langle (12)(34)\right\rangle \simeq%
\mathbb{Z}
_{2}\text{ Orbits}\text{:}\\
Orb(10_{M})  &  =t_{1},t_{2}\\
Orb(\overline{5}_{M})  &  =t_{4}+t_{5},t_{3}+t_{5}\\
Orb(5_{H})  &  =-t_{1}-t_{2}\\
Orb(\overline{5}_{H})  &  =t_{1}+t_{4},t_{2}+t_{3}\\
Orb(X^{\dag})  &  =t_{2}-t_{4},t_{1}-t_{3}\\
Orb(N_{R})  &  =t_{1}-t_{5},t_{2}-t_{5}\text{.}%
\end{align}
Note that the orbit for the $10_{M}$, for example truncates at exactly two
weights. Adding additional weights would lead to distinct orbits, which is
counter the condition that there is a single $10_{M}$ curve. Similar
considerations apply for the other length two orbits. In the case of
$G_{(4)}^{Dir\left(  1\right)  }$ we have:%
\begin{align}
G_{(4)}^{Dir\left(  1\right)  }  &  =\left\langle (12),(34)\right\rangle
\simeq%
\mathbb{Z}
_{2}\times%
\mathbb{Z}
_{2}\text{ Orbits}\text{:}\\
Orb(10_{M})  &  =t_{1},t_{2}\\
Orb(\overline{5}_{M})  &  =t_{4}+t_{5},t_{3}+t_{5}\\
Orb(5_{H})  &  =-t_{1}-t_{2}\\
Orb(\overline{5}_{H})  &  =t_{1}+t_{4},t_{2}+t_{3},t_{2}+t_{4},t_{1}+t_{3}\\
Orb(X^{\dag})  &  =t_{2}-t_{4},t_{1}-t_{3},t_{1}-t_{4},t_{2}-t_{3}\\
Orb(N_{R})  &  =t_{1}-t_{5},t_{2}-t_{5}\text{.}%
\end{align}

It is very tempting to include in this list the messenger fields $Y$ and
$Y^{\prime}$ reviewed in section \ref{sec:DarkE} required for gauge mediated
supersymmetry breaking \cite{HVGMSB}. A priori, such messenger fields can
either localize on curves distinct from the matter curves, or as noted in
\cite{BHSV} can potentially reside on the same curve as other matter fields.
For example, the $\overline{5}_{M}$ curve could in principle support another
generation in the $\overline{5}$ provided another curve is free to support an
entire GUT\ multiplet in the $5$.

To classify possible matter curve assignments for messenger fields, we demand
that the interaction terms~$XYY^{\prime}$ be present. First consider the case
of messengers in the $10\oplus\overline{10}$. Fixing the weight of the $X$
field as $t_{4}-t_{2}$, the constraint on the weights now reads:%
\begin{equation}
(t_{4}-t_{2})+t_{i}-t_{j}=0\text{.}%
\end{equation}
Hence, one of the messenger fields must localize on $t_{2}$, as an additional
$10$ in the orbit of the $10_{M}$, and the other messenger in the
$\overline{10}$ has weight assignment $-t_{4}$:%
\begin{align}
Orb(10_{M},Y_{10})  &  =t_{1},t_{2}\\
Orb(Y_{\overline{10}}^{\prime})  &  =-t_{3},-t_{4}\text{.}%
\end{align}
We therefore conclude that messengers in the $10\oplus\overline{10}$ are
indeed possible, but that $Y_{10}$ must localize on the same curve as the
$10_{M}$.

Next consider messengers in the $5\oplus\overline{5}$ of $SU(5)$. \ Again
fixing the weight of $X$ as $t_{4}-t_{2}$, this imposes the constraint:%
\begin{equation}
(t_{4}-t_{2})-(t_{i}+t_{j})+(t_{k}+t_{l})=0\text{.}%
\end{equation}
Hence, the $Y_{\overline{5}}^{\prime}$ must be of the form $t_{2}+t_{i}$, and
the weight of the $Y_{5}$ is given by $-t_{4}-t_{i}$. There are in principle
three possible options for the weight assignments of the messengers:%
\begin{align}
\text{Option 1}  &  \text{: }Orb(Y_{5})\ni-t_{4}-t_{1},Orb(Y_{\overline{5}%
}^{\prime})\ni t_{2}+t_{1}\\
\text{Option 2}  &  \text{: }Orb(Y_{5})\ni-t_{4}-t_{3},Orb(Y_{\overline{5}%
}^{\prime})\ni t_{2}+t_{3}\\
\text{Option 3}  &  \text{: }Orb(Y_{5})\ni-t_{4}-t_{5},Orb(Y_{\overline{5}%
}^{\prime})\ni t_{2}+t_{5}\text{.}%
\end{align}
Returning to the list of already specified orbits, we see that in the case of
option 1, $Y_{\overline{5}}^{\prime}$ would localize on the $5_{H}$ curve.
Similarly, for option 2 $Y_{\overline{5}}^{\prime}$ would localize on the
$\overline{5}_{H}$ curve. Since a non-trivial hyperflux pierces both curves, a
full GUT\ multiplet of messengers cannot localize on these curves. This leaves
only option 3. While this is in principle possible, note that this requires
$Y_{5}$ to localize on the same matter curve as the $\overline{5}_{M}$. Index
theory considerations then imply that while it is perhaps possible to arrange
for three $\overline{5}_{M}$'s and one $Y_{5}$ to localize on the
\textit{same} curve, this is quite unnatural. Hence, the messengers cannot
transform in the $5\oplus\overline{5}$.

The orbits for all of the visible matter fields are then:%
\begin{align}
G_{(2)}^{Dir\left(  1\right)  }  &  =\left\langle (12)(34)\right\rangle \simeq%
\mathbb{Z}
_{2}\text{ }\text{Orbits}\text{:}\\
Orb(10_{M},Y_{10})  &  =t_{1},t_{2}\\
Orb(Y_{\overline{10}}^{\prime})  &  =-t_{3},-t_{4}\\
Orb(\overline{5}_{M})  &  =t_{4}+t_{5},t_{3}+t_{5}\\
Orb(5_{H})  &  =-t_{1}-t_{2}\\
Orb(\overline{5}_{H})  &  =t_{1}+t_{4},t_{2}+t_{3}\\
Orb(X^{\dag})  &  =t_{2}-t_{4},t_{1}-t_{3}\\
Orb(N_{R})  &  =t_{1}-t_{5},t_{2}-t_{5}\text{.}%
\end{align}

Similarly, the existence of messengers in the $\overline{5}$ is not possible
in the case of the larger monodromy group $G_{(4)}^{Dir\left(  1\right)
}\simeq%
\mathbb{Z}
_{2}\times$ $%
\mathbb{Z}
_{2}$. It thus follows that in the case of $G_{(4)}^{Dir\left(  1\right)  }$,
only $10$'s can correspond to messenger fields. \ In this case, the full
orbits including the messengers are:%
\begin{align}
G_{(4)}^{Dir\left(  1\right)  }  &  =\left\langle (12),(34)\right\rangle
\simeq%
\mathbb{Z}
_{2}\times%
\mathbb{Z}
_{2}\text{ Orbits}\text{:}\\
&  \text{Minimal Matter}\\
Orb(10_{M},Y_{10})  &  =t_{1},t_{2}\\
Orb(Y_{\overline{10}}^{\prime})  &  =-t_{3},-t_{4}\\
Orb(\overline{5}_{M})  &  =t_{4}+t_{5},t_{3}+t_{5}\\
Orb(5_{H})  &  =-t_{1}-t_{2}\\
Orb(\overline{5}_{H})  &  =t_{1}+t_{4},t_{2}+t_{3},t_{2}+t_{4},t_{1}+t_{3}\\
Orb(X^{\dag})  &  =t_{2}-t_{4},t_{1}-t_{3},t_{1}-t_{4},t_{2}-t_{3}\\
Orb(N_{R})  &  =t_{1}-t_{5},t_{2}-t_{5}\text{.}%
\end{align}

Having specified the content of the visible matter, we now turn to extra
charged and neutral fields which can fit inside of $E_{8}$. \ An interesting
consequence of the classification already performed is that it eliminates half
of the possible $U(1)$ dark gauge bosons. Indeed, since the $1$ and $2$
directions, and the $3$ and $4$ directions are also identified, it follows
that of the original $U(1)^{4}\subset SU(5)_{\bot}$ abelian gauge bosons, only
two gauge bosons remain. \ We have identified one such gauge boson with
$U(1)_{PQ}$. \ The other invariant can be identified as:%
\begin{equation}
t_{PQ^{\prime}}^{\ast}=a\left(  t_{1}^{\ast}+t_{2}^{\ast}\right)
+b(t_{3}^{\ast}+t_{4}^{\ast})+ct_{5}^{\ast}\text{.}%
\end{equation}
In keeping with conventions associated with the embedding $E_{6}\supset
SO(10)\times U(1)_{PQ}\supset SU(5)\times U(1)_{\chi}\times U(1)_{PQ}$ used in
\cite{HVGMSB,BHSV}, we demand that the pattern of charges consistent with the
decomposition of the spinor $16$ of $SO(10)$ appear as:%
\begin{align}
SO(10)  &  \supset SU(5)\times U(1)_{\chi}\\
16  &  \rightarrow1_{-5}+\overline{5}_{+3}+10_{-1}\text{,}%
\end{align}
which will occur provided $a=-1$, $b=+3$ and $c=0$. \ Summarizing, the two
$U(1)$ generators are given as:%
\begin{align}
t_{PQ}^{\ast}  &  =\left(  t_{1}^{\ast}+t_{2}^{\ast}\right)  -3(t_{3}^{\ast
}+t_{4}^{\ast})+4t_{5}^{\ast}\\
t_{\chi}^{\ast}  &  =-(t_{1}^{\ast}+t_{2}^{\ast}+t_{3}^{\ast}+t_{4}^{\ast
})+4t_{5}^{\ast}\text{.}%
\end{align}
Under these two $U(1)$'s the minimal matter fields of the F-theory GUT have
charges:%
\begin{equation}%
\begin{tabular}
[c]{|c|c|c|c|c|c|c|c|}\hline
Minimal & $10_{M},Y_{10}$ & $\overline{5}_{M}$ & $Y_{\overline{10}}^{\prime}$
& $5_{H}$ & $\overline{5}_{H}$ & $X^{\dag}$ & $N_{R}$\\\hline
$U(1)_{PQ}$ & $+1$ & $+1$ & $+3$ & $-2$ & $-2$ & $+4$ & $-3$\\\hline
$U(1)_{\chi}$ & $-1$ & $+3$ & $+1$ & $+2$ & $-2$ & $0$ & $-5$\\\hline
\end{tabular}
\ \ \ \ \ .
\end{equation}

Besides the minimal matter required for realizing an F-theory GUT, there could
in principle be additional matter either charged or uncharged under the GUT
group. Under the monodromy group actions just encountered, there are a few
additional matter curves on which a $5$ or a $10$ could localize. The
corresponding orbits for extra $5$'s and $10$'s not already used are:%
\begin{align}
&  \text{ }G_{(2)}^{Dir\left(  1\right)  }=\left\langle (12)(34)\right\rangle
\simeq%
\mathbb{Z}
_{2}\text{ }\text{Orbits}\text{:}\\
&  \text{Extra Charged}\\
Orb(10_{(1)})  &  =t_{5}\\
Orb(\overline{5}_{(1)})  &  =t_{1}+t_{3},t_{2}+t_{4}\\
Orb(\overline{5}_{(2)})  &  =t_{1}+t_{5},t_{2}+t_{5}\\
Orb(\overline{5}_{(3)})  &  =t_{3}+t_{4}\text{,}%
\end{align}
while in the case of the larger monodromy group $G_{(4)}^{Dir\left(  1\right)
}$, $5_{(1)}$ combines to form a larger orbit in the minimal matter sector, so
that we now have:%
\begin{align}
G_{(4)}^{Dir\left(  1\right)  }  &  =\left\langle (12),(34)\right\rangle
\simeq%
\mathbb{Z}
_{2}\times%
\mathbb{Z}
_{2}\text{ Orbits}\text{:}\\
&  \text{Extra Charged}\\
Orb(10_{(1)})  &  =t_{5}\\
Orb(\overline{5}_{(2)})  &  =t_{1}+t_{5},t_{2}+t_{5}\\
Orb(\overline{5}_{(3)})  &  =t_{3}+t_{4}\text{.}%
\end{align}

Besides these possibilities, there can also be extra matter fields neutral
under $SU(5)_{GUT}$ which transform in the adjoint of $SU(5)_{\bot}$. \ These
correspond either to the zero weights, which we can identify with directions
in the Cartan, as well as weights of the form $t_{m}-t_{n}$. \ Of the zero
weights, only two survive, because of the action of the monodromy group, which
we denote as $Z_{PQ}$ and $Z_{\chi}$. \ Next consider the non-zero weights.
\ Some of these have already been encountered in terms of the $X$ \ and
$N_{R}$ fields. \ Besides these possibilities, the full list of singlets
(omitting conjugate weights unless in the same orbit) then fill out the
following distinct orbits:%
\begin{align}
G_{(2)}^{Dir\left(  1\right)  }  &  =\left\langle (12)(34)\right\rangle \simeq%
\mathbb{Z}
_{2}\text{ }\text{Orbits}\text{:}\\
&  \text{Extra Singlets}\\
Orb(D_{(1)})  &  =t_{1}-t_{2},t_{2}-t_{1}\\
Orb(D_{(2)})  &  =t_{1}-t_{4},t_{2}-t_{3}\\
Orb(D_{(3)})  &  =t_{3}-t_{4},t_{4}-t_{3}\\
Orb(D_{(4)})  &  =t_{3}-t_{5},t_{4}-t_{5}\text{.}%
\end{align}
In the case of the larger monodromy group, $D_{(2)}$ joins the orbit of
$X^{\dag}$, and the orbits are instead:%
\begin{align}
G_{(4)}^{Dir\left(  1\right)  }  &  =\left\langle (12),(34)\right\rangle
\simeq%
\mathbb{Z}
_{2}\times%
\mathbb{Z}
_{2}\text{ Orbits}\text{:}\\
&  \text{Extra Singlets}\\
Orb(D_{(1)})  &  =t_{1}-t_{2},t_{2}-t_{1}\\
Orb(D_{(3)})  &  =t_{3}-t_{4},t_{4}-t_{3}\\
Orb(D_{(4)})  &  =t_{3}-t_{5},t_{4}-t_{5}\text{.}%
\end{align}
Under the two surviving $U(1)$'s, the charges of the various extra objects
(omitting complex conjugates) are then:%
\begin{align}
&
\begin{tabular}
[c]{|c|c|c|c|c|}\hline
Extra Charged & $10_{(1)}$ & $\overline{5}_{(1)}$ & $\overline{5}_{(2)}$ &
$\overline{5}_{(3)}$\\\hline
$U(1)_{PQ}$ & $+4$ & $-2$ & $+5$ & $-6$\\\hline
$U(1)_{\chi}$ & $+4$ & $-2$ & $+3$ & $-2$\\\hline
\end{tabular}
\\
&
\begin{tabular}
[c]{|c|c|c|c|c|c|c|}\hline
Extra Neutral & $D_{(1)}$ & $D_{(2)}$ & $D_{(3)}$ & $D_{(4)}$ & $Z_{PQ}$ &
$Z_{\chi}$\\\hline
$U(1)_{PQ}$ & $0$ & $+4$ & $0$ & $-7$ & $0$ & $0$\\\hline
$U(1)_{\chi}$ & $0$ & $0$ & $0$ & $-5$ & $0$ & $0$\\\hline
\end{tabular}
\ \ \
\end{align}

\subsubsection{Second Dirac Scenario}

In the previous section we presented a complete classification of possible
orbits and monodromy groups under the assumption that option 1 of line
(\ref{opone}) is realized in a Dirac scenario. \ Next consider the possibility
that option 2 of line (\ref{opone}) is realized so that:%
\begin{equation}
\text{Option 2}\text{: }Orb(\overline{5}_{M})\ni t_{2}+t_{5}\text{,
}Orb(10_{M})\ni t_{3}%
\end{equation}
Since the PQ\ charges of $\overline{5}_{M}$ and $10_{M}$ are both $+1$, this
imposes the constraint:%
\begin{align}
a_{2}+a_{5}  &  =1\\
a_{3}  &  =1\text{.}%
\end{align}
Combining all of the constraints derived previously, the form of the
$U(1)_{PQ}$ generator in the dual to the weight space is fixed to be of the
form:%
\begin{equation}
t_{PQ}^{\ast}=t_{1}^{\ast}+t_{2}^{\ast}+t_{3}^{\ast}-3t_{4}^{\ast}\text{.}
\label{secondPQ}%
\end{equation}
In order for this choice to be invariant under the action of the monodromy
group, it follows that the monodromy group can only act non-trivially on the
directions $t_{1}^{\ast},t_{2}^{\ast}$ and $t_{3}^{\ast}$. \ Hence, the
monodromy group must be a subgroup of the permutation group $S_{3}$ acting on
these three letters. We now deduce the remaining orbits and weight assignments
by demanding consistency with the interaction term $\overline{5}_{H}%
\times\overline{5}_{M}\times10_{M}$ and the presence of the Dirac interaction
term $H_{d}^{\dag}LN_{R}/\Lambda_{\text{UV}}$.

Returning to the interaction term $\overline{5}_{H}\times\overline{5}%
_{M}\times10_{M}$, fix the weight appearing in the orbit of $\overline{5}_{H}$
as $t_{1}+t_{4}$. \ The existence of this interaction term then requires:%

\begin{equation}
(t_{1}+t_{4})+(t_{i}+t_{j})+t_{k}=0\text{.}%
\end{equation}
Since $10_{M}$ already involves $t_{1}$ and $t_{2}$ it cannot involve $t_{5}$,
as this would lead to two distinct orbits for $10_{M}$. \ Thus, $\overline
{5}_{M}$ must be of the form $t_{i}+t_{5}$ for some $i=2,3$. \ Now, although
we do not know the full action of the monodromy group, we do know that $t_{4}$
is also invariant. \ As a consequence, since $10_{M}$ has in its orbit $t_{1}$
and $t_{2}$, it follows that the orbit of $\overline{5}_{H}$ must include
$t_{1}+t_{4}$, as well as its image, $t_{2}+t_{4}$. \ Repeating the same
argument as before, it now follows that $\overline{5}_{M}$ must be of the form
$t_{i}+t_{5}$ for $i=1$ or $3$. Summarizing, we now learn that the orbits of
the various matter fields contains the following weights:%
\begin{align}
Orb(10_{M})  &  =t_{1},t_{2},...\\
Orb(\overline{5}_{M})  &  =t_{i}+t_{5},...\\
Orb(5_{H})  &  =-t_{1}-t_{2},...\\
Orb(\overline{5}_{H})  &  =t_{1}+t_{4},t_{2}+t_{4},...\\
Orb(X^{\dag})  &  =t_{1}-t_{4},t_{2}-t_{4},...
\end{align}
for some $t_{i}$ given by either $t_{1},t_{2}$ or $t_{3}$.

To fully specify the weight assignment for the $\overline{5}_{M}$, next
consider the interaction term $H_{d}^{\dag}LN_{R}/\Lambda_{\text{UV}}$. Fixing
the weight of $H_{d}^{\dag}$ as $-t_{1}-t_{4}$, this interaction term
requires:%
\begin{equation}
(-t_{1}-t_{4})+(t_{i}+t_{j})+(t_{m}-t_{n})=0\text{.}%
\end{equation}
Thus, we deduce that the orbits for $\overline{5}_{M}$ and $N_{R}$ must
satisfy one of two options:%
\begin{align}
\text{Option 1}  &  \text{: }Orb(\overline{5}_{M})\ni t_{1}+t_{j}\text{,
}Orb(N_{R})\ni t_{4}-t_{j}\\
\text{Option 2}  &  \text{: }Orb(\overline{5}_{M})\ni t_{4}+t_{j}\text{,
}Orb(N_{R})\ni t_{1}-t_{j}\text{.}%
\end{align}
In the latter case, it is note possible for $\overline{5}_{M}$ to attain
PQ\ charge $+1$ with the PQ direction of equation (\ref{secondPQ}). Hence,
only option 1 is available. Since $N_{R}$ must have PQ\ charge $-3$, $t_{j}$
must correspond to $t_{5}$.\ Indeed, for all other choices of $t_{j}$,
$t_{4}-t_{j}$ has PQ charge $-4$. \ We can therefore further fix the orbit of
$\overline{5}_{M}$ and $N_{R}$ to include:%
\begin{align}
Orb(10_{M})  &  =t_{1},t_{2},...\\
Orb(\overline{5}_{M})  &  =t_{1}+t_{5},t_{2}+t_{5},...\\
Orb(5_{H})  &  =-t_{1}-t_{2},...\\
Orb(\overline{5}_{H})  &  =t_{1}+t_{4},t_{2}+t_{4},...\\
Orb(X^{\dag})  &  =t_{1}-t_{4},t_{2}-t_{4},...\\
Orb(N_{R})  &  =t_{4}-t_{5}\text{.}%
\end{align}
In fact, the listed weights do not fill out a complete orbit. \ Indeed,
returning to the $\overline{5}_{H}\times\overline{5}_{M}\times10_{M}$
interaction, using the weights listed above, $t_{1}$ or $t_{2}$ always appears twice.

Fixing the weight assignment for the $\overline{5}_{M}$ as $t_{1}+t_{5}$, note
that the remaining weights for the $10_{M}$ and $\overline{5}_{H}$ must
satisfy the relation:%
\begin{equation}
(t_{1}+t_{5})+(t_{i}+t_{j})+t_{k}=0\text{.}%
\end{equation}
Thus, either $t_{i}$, $t_{j}$ or $t_{k}$ must involve $t_{3}$. Since $t_{4}$
and $t_{5}$ are fixed under monodromy, the orbits for $10_{M}$ and
$\overline{5}_{H}$ respectively include $t_{1}$ and $t_{1}+t_{5}$, this
implies that the monodromy group contains an element which sends $t_{1}$ to
$t_{3}$. Hence, the orbit of $10_{M}$ must include $t_{3}$, the orbit of
$\overline{5}_{M}$ must include $t_{3}+t_{5}$, and the orbit of $\overline
{5}_{H}$ must include $t_{3}+t_{4}$. Similar considerations apply to the orbit
of $X^{\dag}$ so that the orbits include $t_{1}-t_{4},t_{2}-t_{4},t_{3}-t_{4}%
$. \ Hence, the orbits include:%
\begin{align}
Orb(10_{M})  &  =t_{1},t_{2},t_{3}\\
Orb(\overline{5}_{M})  &  =t_{1}+t_{5},t_{2}+t_{5},t_{3}+t_{5}\\
Orb(5_{H})  &  =-t_{1}-t_{2},...\\
Orb(\overline{5}_{H})  &  =t_{1}+t_{4},t_{2}+t_{4},t_{3}+t_{4}\\
Orb(X^{\dag})  &  =t_{1}-t_{4},t_{2}-t_{4},t_{3}-t_{4}\\
Orb(N_{R})  &  =t_{4}-t_{5}\text{.}%
\end{align}

To fix the remaining orbits, we now appeal to some facts from group theory. As
we have already noted, the monodromy group is a subgroup of $S_{3}$. Recall
that the elements of $S_{3}$ are given by the following list of cycles:%
\begin{equation}
S_{3}=\left\{  id,\text{ }(12),\text{ }(13)\text{, }(23)\text{, }(123)\text{,
}(132)\right\}  .
\end{equation}
On the other hand, since the orbit for $10_{M}$ contains three distinct
weights, the monodromy group must have order at least three. Now, the order of
any subgroup of $S_{3}$ must divide that of the subgroup, so $G_{mono}\simeq%
\mathbb{Z}
_{3}$ or $S_{3}$. In either case, it follows that the monodromy group contains
a $%
\mathbb{Z}
_{3}$ subgroup, and thus a three cycle. There are only two available three
cycles, namely $(123)$ and $(132)$, and both generate the same orbits. We thus
conclude that the full set of orbits is:
\begin{align}
G_{mono}^{Dir(2)}  &  \simeq%
\mathbb{Z}
_{3}\text{ or }S_{3}\\
Orb(10_{M})  &  =t_{1},t_{2},t_{3}\\
Orb(\overline{5}_{M})  &  =t_{1}+t_{5},t_{2}+t_{5},t_{3}+t_{5}\\
Orb(5_{H})  &  =-t_{1}-t_{2},-t_{2}-t_{3},-t_{3}-t_{1}\\
Orb(\overline{5}_{H})  &  =t_{1}+t_{4},t_{2}+t_{4},t_{3}+t_{4}\\
Orb(X^{\dag})  &  =t_{1}-t_{4},t_{2}-t_{4},t_{3}-t_{4}\\
Orb(N_{R})  &  =t_{4}-t_{5}\text{.}%
\end{align}

As in subsection \ref{FIRSTDIR}, we now address whether it is possible to
combine this monodromy group action with a messenger sector. \ First consider
messenger fields in the $5\oplus\overline{5}$ of $SU(5)$. Fixing the weight of
$X$ as $t_{4}-t_{1}$, the interaction term $XY_{5}Y_{\overline{5}}^{\prime}$
imposes the weight constraint:%
\begin{equation}
(t_{4}-t_{1})+(-t_{i}-t_{j})+(t_{k}+t_{l})=0\text{.}%
\end{equation}
Thus, the orbit of the $Y_{5}$ contains $-t_{4}-t_{i}$ and the orbit of
$Y_{\overline{5}}^{\prime}$ contains $t_{1}+t_{i}$. Compatibility with the
other orbits then fixes this weight to be $t_{1}+t_{5}$, which is in the same
orbit as the $\overline{5}_{M}$ field.

Next consider messengers in the $10\oplus\overline{10}$ of $SU(5)$. Fixing the
weight of $X$ as $t_{4}-t_{1}$, the interaction term $XY_{10}Y_{\overline{10}%
}^{\prime}$ imposes the constraint:%
\begin{equation}
(t_{4}-t_{1})+t_{i}-t_{j}=0\text{.}%
\end{equation}
Thus, the orbit of $Y_{10}$ contains the weight $t_{1}$, and the orbit of
$Y_{\overline{10}}^{\prime}$ contains the weight $-t_{4}$. Acting with the
three cycle present in the monodromy group it follows that $Y_{10}$ lies in
the same orbit as the $10_{M}$, while the orbit of $Y_{\overline{10}}^{\prime
}$ is given by $t_{4}$. Summarizing, the weight assignments for all of the
visible matter are:%
\begin{align}
G_{mono}^{Dir(2)}  &  \simeq%
\mathbb{Z}
_{3}\text{ or }S_{3}\\
Orb(10_{M},Y_{10})  &  =t_{1},t_{2},t_{3}\\
Orb(Y_{\overline{10}}^{\prime})  &  =-t_{4}\\
Orb(\overline{5}_{M},Y_{\overline{5}}^{\prime})  &  =t_{1}+t_{5},t_{2}%
+t_{5},t_{3}+t_{5}\\
Orb(Y_{5})  &  =-t_{4}-t_{5}\\
Orb(5_{H})  &  =-t_{1}-t_{2},-t_{2}-t_{3},-t_{3}-t_{1}\\
Orb(\overline{5}_{H})  &  =t_{1}+t_{4},t_{2}+t_{4},t_{3}+t_{4}\\
Orb(X^{\dag})  &  =t_{1}-t_{4},t_{2}-t_{4},t_{3}-t_{4}\\
Orb(N_{R})  &  =t_{4}-t_{5}\text{.}%
\end{align}

The action of the monodromy group manifestly identifies three directions in
the Cartan. \ Hence, of the four possible $U(1)$ gauge bosons in $SU(5)_{\bot
}$, only two remain. \ We have already identified one such direction with
$U(1)_{PQ}$. With notation as in subsection \ref{FIRSTDIR}, the other
remaining direction is given as $U(1)_{\chi}$ which is now specified by the
direction:%
\begin{equation}
t_{\chi}^{\ast}=-(t_{1}^{\ast}+t_{2}^{\ast}+t_{3}^{\ast}+t_{4}^{\ast}%
)+4t_{5}^{\ast}\text{.}%
\end{equation}
The charge assignments for each matter field are then:%
\begin{equation}%
\begin{tabular}
[c]{|c|c|c|c|c|c|c|c|c|}\hline
Minimal Matter & $10_{M},Y_{10}$ & $Y_{\overline{10}}^{\prime}$ &
$\overline{5}_{M},Y_{\overline{5}}^{\prime}$ & $Y_{5}$ & $5_{H}$ &
$\overline{5}_{H}$ & $X^{\dag}$ & $N_{R}$\\\hline
$U(1)_{PQ}$ & $+1$ & $+3$ & $+1$ & $+3$ & $-2$ & $-2$ & $+4$ & $-3$\\\hline
$U(1)_{\chi}$ & $-1$ & $+1$ & $+3$ & $-3$ & $+2$ & $-2$ & $0$ & $-5$\\\hline
\end{tabular}
\ \ \ \ .
\end{equation}

As in the case of the first Dirac scenario, it is of interest to classify
extra matter fields which live in unidentified orbits. In this case, the
available orbits for extra $10$ curves are:%
\begin{align}
G_{mono}^{Dir(2)}  &  \simeq%
\mathbb{Z}
_{3}\text{ or }S_{3}\text{ Orbits}\\
&  \text{Extra Charged }\\
Orb(10_{(1)})  &  =t_{5}%
\end{align}

Next consider extra GUT\ singlets which descend from the adjoint of
$SU(5)_{\bot}$. In this case, we have only two possible gauge bosons
corresponding to $U(1)_{PQ}$ and $U(1)_{\chi}$. \ Moreover, due to the action
of the monodromy group, there are only two zero weights, which we denote as
$Z_{PQ}$ and $Z_{\chi}$. \ The remaining weights are of the form $t_{m}-t_{n}$
and localize on curves in the geometry. \ The available orbits can all be
obtained by acting with a $%
\mathbb{Z}
_{3}$ subgroup of the monodromy group. Thus, (omitting complex conjugate
weights) we have the orbits:%
\begin{align}
G_{mono}^{Dir(2)}  &  \simeq%
\mathbb{Z}
_{3}\text{ or }S_{3}\text{ Orbits}\\
&  \text{Extra Singlets}\\
Orb(D_{(1)})  &  =t_{1}-t_{5},t_{2}-t_{5},t_{3}-t_{5}\\
Orb(D_{(2)})  &  =t_{2}-t_{3},t_{3}-t_{1},t_{1}-t_{2}\text{.}%
\end{align}
Up to complex conjugates, the corresponding charges of the extra matter are
then:%
\begin{align}
&
\begin{tabular}
[c]{|c|c|}\hline
Extra Charged & $10_{(1)}$\\\hline
$U(1)_{PQ}$ & $0$\\\hline
$U(1)_{\chi}$ & $+4$\\\hline
\end{tabular}
\\
&
\begin{tabular}
[c]{|c|c|c|c|c|}\hline
Extra Neutral & $D_{(1)}$ & $D_{(2)}$ & $Z_{PQ}$ & $Z_{\chi}$\\\hline
$U(1)_{PQ}$ & $+1$ & $0$ & $0$ & $0$\\\hline
$U(1)_{\chi}$ & $-5$ & $0$ & $0$ & $0$\\\hline
\end{tabular}
\ \ \ \ \text{.}%
\end{align}

\subsection{Majorana Neutrino Scenarios}

In the previous subsections we classified all possible monodromy group
interactions consistent with a Dirac neutrino scenario. \ In this subsection
we return to the case of a minimal Majorana neutrino scenario. \ Returning to
our general discussion, the appropriate MSSM\ interaction terms required the
following weights to be present in the orbit of each field:%
\begin{align}
Orb(10_{M})  &  =t_{1},t_{2},...\\
Orb(5_{H})  &  =-t_{1}-t_{2},...\\
Orb(\overline{5}_{H})  &  =t_{1}+t_{4},...\\
Orb(X^{\dag})  &  =t_{2}-t_{4},....
\end{align}

In the case of a minimal Majorana neutrino scenario, the action of the
monodromy group is more involved, as it involves a more non-trivial
identification of matter fields. We will be interested in covering theories
which contain the F-term:%
\begin{equation}
\int d^{2}\theta\text{ }H_{u}LN_{R}\text{,}%
\end{equation}
where the $N_{R}$'s develop a suitable Majorana mass. \ In fact, as explained
in \cite{BHSV}, the existence of the higher-dimension operator:%
\begin{equation}
\int d^{2}\theta\text{ }\frac{\left(  H_{u}L\right)  ^{2}}{\Lambda_{\text{UV}%
}}%
\end{equation}
requires the PQ$\ $charge assignments:%
\begin{equation}%
\begin{tabular}
[c]{|c|c|c|c|c|c|c|}\hline
& $\overline{5}_{M}$ & $10_{M}$ & $5_{H}$ & $\overline{5}_{H}$ & $X^{\dag}$ &
$N_{R}$\\\hline
Majorana $U(1)_{PQ}$ & $+2$ & $+1$ & $-2$ & $-3$ & $+5$ & $0$\\\hline
\end{tabular}
\ \ \ \ \ \text{.}%
\end{equation}
Letting $t_{PQ}^{\ast}$ correspond to the direction corresponding to the
$U(1)_{PQ}$ generator in the Cartan, the general form of $t_{PQ}^{\ast}$ is:%
\begin{equation}
t_{PQ}^{\ast}=a_{1}t_{1}^{\ast}+a_{2}t_{2}^{\ast}+a_{3}t_{3}^{\ast}+a_{4}%
t_{4}^{\ast}+a_{5}t_{5}^{\ast}\text{.}%
\end{equation}
Since the charge of $10_{M}$ is $+1$, we conclude that $a_{1}=a_{2}=1$.
Moreover, since the PQ charge of $\overline{5}_{H}$ is $-3$, we further
conclude that $a_{4}=-4$. \ Thus, we can already fix part of $t_{PQ}^{\ast}$
to be of the form:%
\begin{equation}
t_{PQ}^{\ast}=t_{1}^{\ast}+t_{2}^{\ast}+a_{3}t_{3}^{\ast}-4t_{4}^{\ast}%
+a_{5}t_{5}^{\ast}\text{.}%
\end{equation}

We now use the presence of the F-term $H_{u}LN_{R}$ to further constrain the
orbits of the various fields. Since $N_{R}$ is assumed to localize on a curve,
it has weight $t_{m}-t_{n}$. Fixing the weight of $H_{u}$ as $-t_{1}-t_{2}$,
it follows that since the weight for $L$ must be of the form $t_{i}+t_{j}$, we
obtain the constraint:%
\begin{equation}
(-t_{1}-t_{2})+(t_{i}+t_{j})+(t_{m}-t_{n})\text{.}%
\end{equation}
Thus, $N_{R}$ must involve a weight of the form $t_{1}-t_{n}$ as well as
(under the action of the monodromy group sending $t_{1}$ to $t_{2}$)
$t_{2}-t_{n^{\prime}}$.\ The weight $t_{n}$ must be distinct from $t_{1}$ and
$t_{2}$ since otherwise $L$ would localize on the same curve as $H_{u}$.
Nothing in our discussion so far has distinguished $t_{3}$ and $t_{5}$, so
without loss of generality, we may further take the $N_{R}$ orbit to contain
the weight $t_{1}-t_{3}$. This also implies that the orbit of the
$\overline{5}_{M}$ contains the weight\ $t_{2}+t_{3}$. Since $N_{R}$ has
PQ\ charge $0$ in the minimal Majorana scenario, it follows that we can also
fix $a_{3}=1$. \ Thus, the form of $t_{PQ}^{\ast}$ is constrained to be of the
form:%
\begin{equation}
t_{PQ}^{\ast}=t_{1}^{\ast}+t_{2}^{\ast}+t_{3}^{\ast}-4t_{4}^{\ast}+a_{5}%
t_{5}^{\ast}\text{.}%
\end{equation}
Note that invariance of this generator under the monodromy group implies that
$t_{4}$ cannot be mapped to any of $t_{1}$, $t_{2}$ or $t_{3}$. To summarize,
the orbits then must contain the terms:%
\begin{align}
Orb(10_{M})  &  =t_{1},t_{2},...\\
Orb(\overline{5}_{M})  &  =t_{2}+t_{3},...\\
Orb(5_{H})  &  =-t_{1}-t_{2},...\\
Orb(\overline{5}_{H})  &  =t_{1}+t_{4},...\\
Orb(X^{\dag})  &  =t_{2}-t_{4},...\\
Orb(N_{R})  &  =t_{1}-t_{3},....
\end{align}

To further fix the form of the monodromy group, note that the listed weights
are incompatible with the interaction term $\overline{5}_{H}\times\overline
{5}_{M}\times10_{M}$. Returning to lines (\ref{opone})-(\ref{opthr}), recall
that the orbits of the $\overline{5}_{M}$ and $\overline{5}_{H}$ contain one
of the two following possibilities:%
\begin{align}
\text{Option 1}  &  \text{: }Orb(\overline{5}_{M})\ni t_{3}+t_{5}\text{,
}Orb(10_{M})\ni t_{2}\\
\text{Option 2}  &  \text{: }Orb(\overline{5}_{M})\ni t_{2}+t_{5}\text{,
}Orb(10_{M})\ni t_{3}\\
\text{Option 3}  &  \text{: }Orb(\overline{5}_{M})\ni t_{2}+t_{3}\text{,
}Orb(10_{M})\ni t_{5}\text{.}%
\end{align}

Since the PQ charges for the $\overline{5}_{M}$ and $10_{M}$ are respectively
$+2$ and $+1$, it follows that the condition on $a_{5}$ in the three cases is:%
\begin{align}
\text{Option 1}  &  \text{: }1+a_{5}=2\\
\text{Option 2}  &  \text{: }1+a_{5}=2\\
\text{Option 3}  &  \text{: }a_{5}=1\text{.}%
\end{align}
Thus, in all three cases, we find the same requirement that $a_{5}=1$.
\ Hence, the form of $t_{PQ}^{\ast}$ is uniquely fixed to be of the form:%
\begin{equation}
t_{PQ}^{\ast}=t_{1}^{\ast}+t_{2}^{\ast}+t_{3}^{\ast}-4t_{4}^{\ast}+t_{5}%
^{\ast}\text{.}%
\end{equation}

By inspection of $t_{PQ}^{\ast}$, $t_{4}$ is fixed by the monodromy group, and
so the monodromy group must be a subgroup of $S_{4}$, the permutation group on
the letters $t_{1}$, $t_{2}$, $t_{3}$ and $t_{5}$.\ Since $t_{1}+t_{4}$ is in
the orbit of the monodromy group, and moreover, since there exists an element
of the monodromy group which sends $t_{1}$ to $t_{2}$ with $t_{4}$ fixed, we
also deduce that the orbits for $\overline{5}_{H}$ must include both
$t_{1}+t_{4}$ and $t_{2}+t_{4}$. Similar reasoning implies that the orbit for
$X^{\dag}$ must include both $t_{2}-t_{4}$ and $t_{1}-t_{4}$. Hence, the
orbits under the monodromy group must be enlarged so that:%
\begin{align}
Orb(10_{M})  &  =t_{1},t_{2},...\\
Orb(\overline{5}_{M})  &  =t_{2}+t_{3},...\\
Orb(5_{H})  &  =-t_{1}-t_{2},...\\
Orb(\overline{5}_{H})  &  =t_{1}+t_{4},t_{2}+t_{4}...\\
Orb(X^{\dag})  &  =t_{2}-t_{4},t_{1}-t_{4},...\\
Orb(N_{R})  &  =t_{1}-t_{3},....
\end{align}

The interaction term $\overline{5}_{H}\times\overline{5}_{M}\times10_{M}$
requires a further enlargement in the orbits.\ Indeed, fixing the weight of
$\overline{5}_{M}$ as $t_{2}+t_{3}$, we obtain the constraint:%
\begin{equation}
(t_{i}+t_{j})+(t_{2}+t_{3})+t_{k}=0\text{.}%
\end{equation}
Since $t_{4}$ is not in the orbit of $10_{M}$, the only available weights for
$10_{M}$ are then $t_{1}$ and $t_{5}$. The available orbits consistent with
this constraint are:%
\begin{align}
\text{Option 1}\text{: }  &  Orb(\overline{5}_{H})\ni t_{4}+t_{5}\text{,
}Orb(10_{M})\ni t_{1}\\
\text{Option 2}\text{: }  &  Orb(\overline{5}_{H})\ni t_{1}+t_{4}\text{,
}Orb(10_{M})\ni t_{5}\text{.}%
\end{align}
Note that in either case, the fact that $t_{4}$ is fixed under the monodromy
group then forces the existence of an element in the monodromy group which
sends $t_{1}$ to $t_{5}$. \ In particular, it follows that \textit{both}
weights must be included in a consistent orbit. \ Enlarging the orbits further
to include maps from $t_{1}$ to $t_{5}$, $t_{1}$ to $t_{2}$ and $t_{2}$ to
$t_{5}$ (consistent with the orbit of the $10_{M}$) we have:%
\begin{align}
Orb(10_{M})  &  =t_{1},t_{2},t_{5},...\\
Orb(\overline{5}_{M})  &  =t_{2}+t_{3},t_{5}+t_{i},t_{1}+t_{j}...\\
Orb(5_{H})  &  =-t_{1}-t_{2},-t_{5}-t_{k},...\\
Orb(\overline{5}_{H})  &  =t_{1}+t_{4},t_{2}+t_{4},t_{5}+t_{4},...\\
Orb(X^{\dag})  &  =t_{2}-t_{4},t_{1}-t_{4},t_{5}-t_{4},...\\
Orb(N_{R})  &  =t_{1}-t_{3},t_{5}-t_{l},t_{2}-t_{m},...,
\end{align}
for some $t_{i}$, $t_{j}$ and $t_{k}$.

By far the most significant constraint stems from the requirement that the
Majorana mass term $N_{R}N_{R}^{c}$ be present. Since $N_{R}$ and $N_{R}^{c}$
must localize on the same curve, and must also lie in the same orbit, it
follows that the orbit for $N_{R}$ must be enlarged to include the complex
conjugate weights. Thus, we deduce that the orbits are minimally:%
\begin{align}
Orb(10_{M})  &  =t_{1},t_{2},t_{5},...\\
Orb(\overline{5}_{M})  &  =t_{2}+t_{3},t_{5}+t_{i},t_{1}+t_{j}...\\
Orb(5_{H})  &  =-t_{1}-t_{2},-t_{5}-t_{k},...\\
Orb(\overline{5}_{H})  &  =t_{1}+t_{4},t_{2}+t_{4},t_{5}+t_{4},...\\
Orb(X^{\dag})  &  =t_{2}-t_{4},t_{1}-t_{4},t_{5}-t_{4},...\\
Orb(N_{R})  &  =\pm\left(  t_{1}-t_{3}\right)  ,\pm\left(  t_{5}-t_{l}\right)
,\pm\left(  t_{2}-t_{m}\right)  ,....
\end{align}
In particular, this implies that there exists an element of the monodromy
group which acts by interchanging $t_{1}-t_{3}$ with $t_{3}-t_{1}$. \ Hence,
there exists an element of the monodromy group which interchanges $t_{1}$ with
$t_{3}$. \ It follows that the orbit of $10_{M}$ includes $t_{1}$, $t_{2}$,
$t_{3}$, $t_{5}$ so that the length of the $10_{M}$ orbit is precisely four
(since $t_{4}$ is invariant under the monodromy group).\ Thus, there exist
elements of the monodromy group such that $t_{1}$ maps to any of $t_{2}$,
$t_{3}$ or $t_{5}$. Using the fact that $t_{4}$ remains fixed by the monodromy
group, we can thus further enlarge the orbits to:%
\begin{align}
Orb(10_{M})  &  =t_{1},t_{2},t_{3},t_{5},...\\
Orb(\overline{5}_{M})  &  =t_{2}+t_{3},t_{5}+t_{i},t_{1}+t_{j}...\\
Orb(5_{H})  &  =-t_{1}-t_{2},-t_{5}-t_{l},-t_{3}-t_{m}...\\
Orb(\overline{5}_{H})  &  =t_{1}+t_{4},t_{2}+t_{4},t_{5}+t_{4},t_{3}+t_{4}\\
Orb(X^{\dag})  &  =t_{2}-t_{4},t_{1}-t_{4},t_{5}-t_{4},t_{3}-t_{4}\\
Orb(N_{R})  &  =\pm\left(  t_{1}-t_{3}\right)  ,\pm\left(  t_{5}-t_{n}\right)
,\pm\left(  t_{2}-t_{p}\right)  ,....
\end{align}
In particular, it follows that the length of the $\overline{5}_{H}$ orbit, and
the $X^{\dag}$ orbit are both four.

We now deduce further properties of the monodromy group action. Since
$Orb(N_{R})$ contains both $t_{1}-t_{3}$ as well as $t_{3}-t_{1}$, there
exists a monodromy group element which permutes $t_{1}$ and $t_{3}$. Note that
this does not fix the action on the rest of the $t$'s. \ Indeed, the action of
the two cycle $(13)$ maps the weight $t_{2}+t_{3}$ of $\overline{5}_{M}$ to
$t_{2}+t_{1}$, which is in the conjugate orbit of $5_{H}$. Since the
$\overline{5}_{M}$ and $5_{H}$ must localize on distinct matter curves, this
implies that the group element which interchanges $t_{1}$ and $t_{3}$ also
acts on $t_{2}$. Now, the only other available $t_{i}$ is $t_{5}$, since
$t_{4}$ is invariant under the monodromy group. It thus follows that the
monodromy group contains the element $\sigma=(13)(25)$. Note that this also
excludes the two cycle $(25)$ since $(25)\cdot(13)(25)=(13)$. To summarize, we
therefore deduce that the monodromy contains neither $(13)$ nor $(25)$, but
does contain $(13)(25)$:%
\begin{align}
(13),(25)  &  \notin G_{mono}^{Maj}\\
(13)(25)  &  \in G_{mono}^{Maj}\text{.} \label{1325}%
\end{align}

Acting on the available weights with $\sigma=(13)(25)$ on the listed weights
allows us to further enlarge the available orbits. For example, acting on the
weight $t_{2}+t_{3}$ in the orbit of $\overline{5}_{M}$ with $\sigma$, it
follows that $Orb(\overline{5}_{M})$ also contains $t_{5}+t_{1}$. Moreover,
acting on the weight $-t_{1}-t_{2}$ in the orbit of $5_{H}$ with $\sigma$, it
follows that $Orb(5_{H})$ also contains the weight $-t_{3}-t_{5}$. \ Hence, we
can further fix the orbits under the monodromy group as:%
\begin{align}
Orb(10_{M})  &  =t_{1},t_{2},t_{3},t_{5}\label{orb10mbah}\\
Orb(\overline{5}_{M})  &  =t_{2}+t_{3},t_{5}+t_{1},...\\
Orb(5_{H})  &  =-t_{1}-t_{2},-t_{5}-t_{3},...\\
Orb(\overline{5}_{H})  &  =t_{1}+t_{4},t_{2}+t_{4},t_{5}+t_{4},t_{3}+t_{4}\\
Orb(X^{\dag})  &  =t_{2}-t_{4},t_{1}-t_{4},t_{5}-t_{4},t_{3}-t_{4}\\
Orb(N_{R})  &  =\pm\left(  t_{1}-t_{3}\right)  ,\pm\left(  t_{5}-t_{n}\right)
,\pm\left(  t_{2}-t_{p}\right)  ,....
\end{align}

Using the listed weights, we now show that the monodromy group does not
contain any three cycles. The essential point is that the existence of any
three cycle would lead to an identification of the $\overline{5}_{M}$ and
$5_{H}$ orbits. The three cycles in $S_{4}$ are:%
\begin{equation}
\text{Three Cycles in }S_{4}=\left\{
(123),(125),(132),(135),(152),(153),(235),(253)\right\}  \text{.}%
\end{equation}
In each case, there exists a weight of one orbit of either $\overline{5}_{M}$
or $5_{H}$ which gets mapped to a weight in the other conjugate orbit. As
explicit examples, we have three cycles which map elements of $\overline
{Orb(5_{H})}$ to $Orb(\overline{5}_{M})$:%
\begin{align}
\overline{Orb(5_{H})}  &  \rightarrow Orb(\overline{5}_{M})\\
(123)  &  :t_{1}+t_{2}\rightarrow t_{2}+t_{3}\\
(135)  &  :t_{5}+t_{3}\rightarrow t_{1}+t_{5}\\
(152)  &  :t_{1}+t_{2}\rightarrow t_{5}+t_{1}\\
(253)  &  :t_{1}+t_{2}\rightarrow t_{1}+t_{5}%
\end{align}
The remaining three cycles all map an element of $Orb(\overline{5}_{M})$ to
$\overline{Orb(5_{H})}$:%

\begin{align}
Orb(\overline{5}_{M})  &  \rightarrow\overline{Orb(5_{H})}\\
(125)  &  :t_{5}+t_{1}\rightarrow t_{1}+t_{2}\\
(132)  &  :t_{2}+t_{3}\rightarrow t_{1}+t_{2}\\
(153)  &  :t_{2}+t_{3}\rightarrow t_{2}+t_{1}\\
(235)  &  :t_{5}+t_{1}\rightarrow t_{2}+t_{1}\text{.}%
\end{align}
It therefore follows that the monodromy group contains no three cycles. In
particular, we therefore conclude that the monodromy group does not contain
any order three subgroups. By the Sylow theorem, it thus follows that the
order of the monodromy group must not be divisible by $3$ so that:%
\begin{equation}
3\nmid\#G_{mono}^{Maj}\text{.}%
\end{equation}
On the other hand, the available subgroups of $S_{4}$ have order $24$, $12$,
$8$, $6$, $4$, $3$, $2$ and $1$. This excludes order $24$, $12$, $6$ and $3$
monodromy groups as possibilities. Moreover, because the orbit of the $10_{M}$
is length $4$, the monodromy group has order at least $4$. We therefore deduce
that the monodromy group has order $4$ or $8$:%
\begin{equation}
\#G_{mono}^{Maj}=4\text{ or }\#G_{mono}^{Maj}=8\text{.}%
\end{equation}
To proceed further, we examine each possibility in turn.

\subsubsection{First Majorana Scenario: $\#G_{mono}^{Maj}=4$\label{FIRSTMAJ}}

First consider scenarios where the monodromy group is order four, which we
denote as $G_{4}^{Maj(1)}$:%
\begin{equation}
\#G_{mono}^{Maj}=4\text{.}%
\end{equation}
In this case, the maximal orbit length is 4. Returning to the parts of the
orbits fixed by prior considerations, recall that we have:%
\begin{align}
Orb(10_{M})  &  =t_{1},t_{2},t_{3},t_{5}\\
Orb(\overline{5}_{M})  &  =t_{2}+t_{3},t_{5}+t_{1},...\\
Orb(5_{H})  &  =-t_{1}-t_{2},-t_{5}-t_{3},...\\
Orb(\overline{5}_{H})  &  =t_{1}+t_{4},t_{2}+t_{4},t_{5}+t_{4},t_{3}+t_{4}\\
Orb(X^{\dag})  &  =t_{2}-t_{4},t_{1}-t_{4},t_{5}-t_{4},t_{3}-t_{4}\\
Orb(N_{R})  &  =\pm\left(  t_{1}-t_{3}\right)  ,\pm\left(  t_{5}-t_{n}\right)
,\pm\left(  t_{2}-t_{p}\right)  ,....
\end{align}
Since the orbit of $N_{R}$ has length four, it follows that this orbit
truncates to $\pm\left(  t_{1}-t_{3}\right)  $ and $\pm\left(  t_{5}%
-t_{2}\right)  $ so that:%
\begin{align}
Orb(10_{M})  &  =t_{1},t_{2},t_{3},t_{5}\\
Orb(\overline{5}_{M})  &  =t_{2}+t_{3},t_{5}+t_{1},...\\
Orb(5_{H})  &  =-t_{1}-t_{2},-t_{5}-t_{3},...\\
Orb(\overline{5}_{H})  &  =t_{1}+t_{4},t_{2}+t_{4},t_{5}+t_{4},t_{3}+t_{4}\\
Orb(X^{\dag})  &  =t_{2}-t_{4},t_{1}-t_{4},t_{5}-t_{4},t_{3}-t_{4}\\
Orb(N_{R})  &  =\pm\left(  t_{1}-t_{3}\right)  ,\pm\left(  t_{5}-t_{2}\right)
\text{.}%
\end{align}

As we now explain, the monodromy group does not contain any two cycles either.
The full list of two cycles in $S_{4}$ are:%
\begin{equation}
\text{Two Cycles in }S_{4}=\left\{  (12),(13),(15),(23),(25),(35)\right\}
\text{.}%
\end{equation}
By inspection of the action of these group elements on weights in the orbit of
$N_{R}$, the following two cycles cannot be included in $G_{4}^{Maj(1)}$:%
\begin{equation}
(12),(15),(23),(35)\notin G_{4}^{Maj(1)}\text{.}%
\end{equation}
On the other hand, we have already seen that the action of $(13)$ and $(25)$
would identify the orbits for $\overline{5}_{M}$ and $5_{H}$. Thus, the
monodromy group does not contain any two cycles:%
\begin{equation}
(12),(15),(23),(35),(13),(25)\notin G_{4}^{Maj(1)}\text{.}%
\end{equation}

The absence of two cycles now allows us to completely fix the form of the
monodromy group. Indeed, since there exists an element which maps $t_{1}%
-t_{3}$ to $t_{2}-t_{5}$, the monodromy group either contains either
$(12)(35)$ or $(1235)$. Note, however, that $(1235)$ sends the $\overline
{5}_{M}$ weight $t_{2}+t_{3}$ to $t_{3}+t_{5}$. Since $t_{3}+t_{5}$ lies in
the orbit conjugate to $5_{H}$, we therefore conclude that $(12)(35)$ is an
element of the monodromy group. Since we have already argued that the
monodromy group contains $(13)(25)$, we deduce that:%
\begin{equation}
(12)(35),(13)(25)\in G_{4}^{Maj(1)}\text{.}%
\end{equation}
Taking the product of these two elements generates a third non-trivial
element:%
\begin{equation}
(13)(25)\cdot(12)(35)=(12)(35)\cdot(13)(25)=(15)(23)\text{.}%
\end{equation}
Taking all products of group elements, it is now immediate that $G_{4}%
^{Maj(1)}$ is an abelian group. Since $G_{4}^{Maj(1)}$ is an abelian order
four group with an order two element, it follows from the Chinese remainder
theorem that $G_{4}^{Maj(1)}$ is isomorphic to $%
\mathbb{Z}
_{2}\times%
\mathbb{Z}
_{2}$. Having specified explicitly the group elements of $G_{4}^{Maj(1)}$, we
can now fix all of the orbits as:%
\begin{align}
G_{4}^{Maj(1)}  &  =\left\langle (12)(35),(13)(25)\right\rangle \simeq%
\mathbb{Z}
_{2}\times%
\mathbb{Z}
_{2}\\
Orb(10_{M})  &  =t_{1},t_{2},t_{3},t_{5}\\
Orb(\overline{5}_{M})  &  =t_{2}+t_{3},t_{5}+t_{1}\\
Orb(5_{H})  &  =-t_{1}-t_{2},-t_{5}-t_{3}\\
Orb(\overline{5}_{H})  &  =t_{1}+t_{4},t_{2}+t_{4},t_{5}+t_{4},t_{3}+t_{4}\\
Orb(X^{\dag})  &  =t_{2}-t_{4},t_{1}-t_{4},t_{5}-t_{4},t_{3}-t_{4}\\
Orb(N_{R})  &  =\pm\left(  t_{1}-t_{3}\right)  ,\pm\left(  t_{5}-t_{2}\right)
\text{.}%
\end{align}
We will return to the case where the monodromy group contains more generators
in the following subsection.

Next consider the messenger sector of the theory. First consider messengers in
the $5\oplus\overline{5}$ of $SU(5)$. Fixing the weight assignment of $X$ as
$t_{4}-t_{2}$, the presence of the interaction term $XY_{5}Y_{\overline{5}%
}^{\prime}$ requires the weights for the messengers to satisfy the constraint:%
\begin{equation}
(t_{4}-t_{2})+(-t_{i}-t_{j})+(t_{k}+t_{l})=0\text{.}%
\end{equation}
Hence, the orbits of the $Y_{5}$ and $Y_{\overline{5}}^{\prime}$ respectively
include $-t_{4}-t_{i}$ and $t_{2}+t_{i}$. Note, however, that the conjugate to
the weight for $Y_{5}$ already falls in the orbit for the $5_{H}$. Since full
GUT\ multiplets do not localize on this curve, it follows that this model does
not admit messengers in the $5\oplus\overline{5}$. Next consider messenger
fields in the $10\oplus\overline{10}$. In this case, the constraint on the
weights requires:%
\begin{equation}
(t_{4}-t_{2})+t_{i}+-t_{k}=0\text{.}%
\end{equation}
Thus, the orbit for the $Y_{10}$ contains the weight $t_{2}$ and the orbit for
the $Y_{\overline{10}}^{\prime}$ contains the weight $-t_{4}$. It thus follows
that just as for the Dirac scenario, one of the messenger fields fits inside
of the same orbit as the $10_{M}$. Summarizing, the visible sector fields
consists of the following orbits:%
\begin{align}
G_{4}^{Maj(1)}  &  =\left\langle (12)(35),(13)(25)\right\rangle \simeq%
\mathbb{Z}
_{2}\times%
\mathbb{Z}
_{2}\\
Orb(10_{M},Y_{10})  &  =t_{1},t_{2},t_{3},t_{5}\\
Orb(Y_{\overline{10}}^{\prime})  &  =-t_{4}\\
Orb(\overline{5}_{M})  &  =t_{2}+t_{3},t_{5}+t_{1}\\
Orb(5_{H})  &  =-t_{1}-t_{2},-t_{5}-t_{3}\\
Orb(\overline{5}_{H})  &  =t_{1}+t_{4},t_{2}+t_{4},t_{5}+t_{4},t_{3}+t_{4}\\
Orb(X^{\dag})  &  =t_{2}-t_{4},t_{1}-t_{4},t_{5}-t_{4},t_{3}-t_{4}\\
Orb(N_{R})  &  =\pm\left(  t_{1}-t_{3}\right)  ,\pm\left(  t_{5}-t_{2}\right)
\text{.}%
\end{align}

The action of the monodromy group manifestly identifies four directions in the
Cartan. \ Hence, of the four possible $U(1)$ gauge bosons in $SU(5)_{\bot}$,
only $U(1)_{PQ}$ remains. The PQ charge assignments for the matter fields are:%
\begin{equation}%
\begin{tabular}
[c]{|c|c|c|c|c|c|c|c|}\hline
Minimal & $10_{M},Y_{10}$ & $Y_{\overline{10}}^{\prime}$ & $\overline{5}_{M}$
& $5_{H}$ & $\overline{5}_{H}$ & $X^{\dag}$ & $N_{R}$\\\hline
$U(1)_{PQ}$ & $+1$ & $+4$ & $+2$ & $-2$ & $-3$ & $+5$ & $0$\\\hline
\end{tabular}
\ \ \ .
\end{equation}

In this case, all of the orbits of the $10$ are already exhausted. Moreover,
the available orbits for additional $5$'s are quite limited:%
\begin{align}
G_{4}^{Maj(1)}  &  =\left\langle (12)(35),(13)(25)\right\rangle \simeq%
\mathbb{Z}
_{2}\times%
\mathbb{Z}
_{2}\\
&  \text{Extra Charged}\\
Orb(\overline{5}_{(1)})  &  =t_{1}+t_{3},t_{2}+t_{5}\text{.}%
\end{align}

Next consider extra GUT\ singlets of the theory. Since there is a single
available $U(1)$ in $SU(5)_{\bot}$, only one zero weight denoted as $Z_{PQ}$
can descend from the adjoint of $SU(5)_{\bot}$. The remaining weights of the
adjoint are of the form $t_{m}-t_{n}$. Under the provided monodromy group
action, these separate into the following orbits:%
\begin{align}
G_{4}^{Maj(1)}  &  =\left\langle (12)(35),(13)(25)\right\rangle \simeq%
\mathbb{Z}
_{2}\times%
\mathbb{Z}
_{2}\\
Orb(D_{(1)})  &  =\pm(t_{1}-t_{2}),\pm(t_{3}-t_{5})\\
Orb(D_{(2)})  &  =\pm(t_{1}-t_{5}),\pm(t_{2}-t_{3})
\end{align}
The corresponding PQ charge of these fields is:%
\begin{align}
&
\begin{tabular}
[c]{|c|c|}\hline
Extra Charged & $\overline{5}_{(1)}$\\\hline
$U(1)_{PQ}$ & $+2$\\\hline
\end{tabular}
\\
&
\begin{tabular}
[c]{|c|c|c|c|}\hline
Extra Neutral & $D_{(1)}$ & $D_{(2)}$ & $Z_{PQ}$\\\hline
$U(1)_{PQ}$ & $0$ & $0$ & $0$\\\hline
\end{tabular}
\ \ \ \ \ \ .
\end{align}

\subsubsection{Second and Third Majorana Scenarios: $\#G_{mono}^{Maj}=8$}

In the previous section we completely classified the available monodromy
groups when the order of the monodromy group is four. We now turn to monodromy
groups with eight elements. Even without specifying the explicit action of the
monodromy group, it is already possible to deduce that the monodromy group
must be isomorphic to the dihedral group $Dih_{4}$, namely the group generated
by rotations and reflections of the square. At the level of abstract groups,
we have:%
\begin{equation}
Dih_{4}\simeq%
\mathbb{Z}
_{2}\ltimes%
\mathbb{Z}
_{4}\text{,}%
\end{equation}
where the $%
\mathbb{Z}
_{2}$ acts by inversion on the $%
\mathbb{Z}
_{4}$ elements.

To show that the monodromy group must be isomorphic to $Dih_{4}$, recall that
$G_{mono}^{Maj}$ is a subgroup of $S_{4}$ of order $8=2^{3}$. It now follows
from the Sylow theorems that all groups of this order are isomorphic. Further
note that $S_{4}$ is the group of symmetries of the cube. Since $Dih_{4}$ is a
symmetry which acts on one of the faces of the cube, we can already deduce
that all of the order eight monodromy groups are isomorphic to $Dih_{4}$. It
is important to note that specifying the abstract group is not enough to fix
the actual orbits.

Returning to the parts of the orbit already specified before subsection
(\ref{FIRSTMAJ}), we have:%
\begin{align}
Orb(10_{M})  &  =t_{1},t_{2},t_{3},t_{5}\\
Orb(\overline{5}_{M})  &  =t_{2}+t_{3},t_{5}+t_{1},...\\
Orb(5_{H})  &  =-t_{1}-t_{2},-t_{5}-t_{3},...\\
Orb(\overline{5}_{H})  &  =t_{1}+t_{4},t_{2}+t_{4},t_{5}+t_{4},t_{3}+t_{4}\\
Orb(X^{\dag})  &  =t_{2}-t_{4},t_{1}-t_{4},t_{5}-t_{4},t_{3}-t_{4}\\
Orb(N_{R})  &  =\pm\left(  t_{1}-t_{3}\right)  ,\pm\left(  t_{5}-t_{n}\right)
,\pm\left(  t_{2}-t_{p}\right)  ,....
\end{align}
To this end, we now proceed to classify the possible options for the
corresponding orbits. The analysis we consider splits to two cases, namely the
case where the orbit of $\overline{5}_{M}$ has length two, and the case where
the orbit of the $\overline{5}_{M}$ is bigger. In the following subsections we
consider these two cases separately.

\subsubsection{Length Two $\overline{5}_{M}$ Orbit}

First suppose that the orbit for the $\overline{5}_{M}$ truncates at exactly
two weights so that:%
\begin{equation}
Orb(\overline{5}_{M})=t_{2}+t_{3},t_{5}+t_{1}.
\end{equation}
On the other hand, we have also seen that the monodromy group must contain a
four cycle. There are six candidate four cycles of $S_{4}$, given as:%
\begin{equation}
\text{Four Cycles in }S_{4}=\left\{
(1235),(1253),(1325),(1352),(1523),(1532)\right\}  \text{.}%
\end{equation}
Of these possibilities, it is enough to focus on the cases
$(1235),(1253),(1325)$ since the remaining possibilities are inverse elements.
Note, however, that neither $(1235)$ nor $(1325)$ preserves $Orb(\overline
{5}_{M})$. It therefore follows that the four cycle of $G_{mono}^{Maj}$ must
contain $(1253).$ Combined with the analysis leading to line (\ref{1325})
showing that the group element $(13)(25)$ must always be present, we conclude
that:%
\begin{equation}
(13)(25),(1253)\in G_{8}^{Maj(2)}\text{.}%
\end{equation}
In fact, these elements generate an order eight group since $(1253)\cdot
(1253)=(15)(23)\neq$ $(13)(25)$. Taking all available products between powers
of\ these two generators, the explicit elements are:%
\begin{equation}
G_{8}^{Maj(2)}=\left\{
id,(1253),(15)(23),(1352),(13)(25),(23),(12)(35),(15)\right\}  \text{.}%
\end{equation}
The corresponding orbits are then:%
\begin{align}
G_{8}^{Maj(2)}  &  \simeq\left\langle (13)(25),(1253)\right\rangle \simeq
Dih_{4}\\
Orb(10_{M})  &  =t_{1},t_{2},t_{3},t_{5}\\
Orb(\overline{5}_{M})  &  =t_{2}+t_{3},t_{5}+t_{1}\\
Orb(5_{H})  &  =-t_{1}-t_{2},-t_{5}-t_{3},-t_{1}-t_{3},-t_{5}-t_{2}\\
Orb(\overline{5}_{H})  &  =t_{1}+t_{4},t_{2}+t_{4},t_{5}+t_{4},t_{3}+t_{4}\\
Orb(X^{\dag})  &  =t_{2}-t_{4},t_{1}-t_{4},t_{5}-t_{4},t_{3}-t_{4}\\
Orb(N_{R})  &  =\pm\left(  t_{1}-t_{3}\right)  ,\pm\left(  t_{5}-t_{2}\right)
,\pm\left(  t_{5}-t_{3}\right)  ,\pm\left(  t_{1}-t_{2}\right)  \text{.}%
\end{align}

As in subsection \ref{FIRSTMAJ}, the only available messengers can fit into
vector-like pairs in the $10\oplus\overline{10}$ of $SU(5)$. This can be
traced back to the fact that the $X^{\dag}$ field always contains a
contribution from $t_{4}$. The full list of orbits is therefore given by:%
\begin{align}
G_{8}^{Maj(2)}  &  \simeq\left\langle (13)(25),(1253)\right\rangle \simeq
Dih_{4}\\
Orb(10_{M},Y_{10})  &  =t_{1},t_{2},t_{3},t_{5}\\
Orb(Y_{\overline{10}}^{\prime})  &  =-t_{4}\\
Orb(\overline{5}_{M})  &  =t_{2}+t_{3},t_{5}+t_{1}\\
Orb(5_{H})  &  =-t_{1}-t_{2},-t_{5}-t_{3},-t_{1}-t_{3},-t_{5}-t_{2}\\
Orb(\overline{5}_{H})  &  =t_{1}+t_{4},t_{2}+t_{4},t_{5}+t_{4},t_{3}+t_{4}\\
Orb(X^{\dag})  &  =t_{2}-t_{4},t_{1}-t_{4},t_{5}-t_{4},t_{3}-t_{4}\\
Orb(N_{R})  &  =\pm\left(  t_{1}-t_{3}\right)  ,\pm\left(  t_{5}-t_{2}\right)
,\pm\left(  t_{5}-t_{3}\right)  ,\pm\left(  t_{1}-t_{2}\right)  \text{.}%
\end{align}

In this case, all available charged GUT\ matter is already part of the minimal
matter sector. Next consider extra GUT\ singlets. \ There is again a single
gauge boson, and the adjoints are either given by the one remaining zero
weight $Z_{PQ}$, or by charged weights in the adjoint. \ Tracing through all
orbits, we now have:%
\begin{align}
G_{8}^{Maj(2)}  &  \simeq\left\langle (13)(25),(1253)\right\rangle \simeq
Dih_{4}\\
&  \text{Extra Singlets}\\
Orb(D_{(1)})  &  =\pm(t_{1}-t_{5}),\pm(t_{2}-t_{3})\text{.}%
\end{align}
The corresponding charges under $U(1)_{PQ}$ are then:%
\begin{equation}%
\begin{tabular}
[c]{|c|c|c|}\hline
Extra Neutral & $D_{(1)}$ & $Z_{PQ}$\\\hline
$U(1)_{PQ}$ & $0$ & $0$\\\hline
\end{tabular}
\ \ \ \ \ .
\end{equation}

\subsubsection{Length Four $\overline{5}_{M}$ Orbit}

In the previous subsection we deduced the form of the monodromy group in the
case where the length of the $\overline{5}_{M}$ orbit is two. Next suppose
that the length of the $\overline{5}_{M}$ orbit is greater than two. In fact,
the orbit for $\overline{5}_{M}$ cannot have length three because of the
orbit-stabilizer theorem. The stabilizer of a weight $w$ is given as the set
of elements of the monodromy group leaving it invariant:

Returning to the available list of orbits:%
\begin{equation}
Stab_{G}(w)=\left\{  \sigma\in G|\sigma(w)=w\right\}  \text{.}%
\end{equation}
The orbit-stabilizer theorem establishes that the product of the length of the
orbit and the order of the stabilizer are equal to the order of the group:%
\begin{equation}
\#Orb(w)\cdot\#Stab_{G}(w)=\#G\text{.}%
\end{equation}
Since $\#G$ is not divisible by $3$, we conclude that the length of the
$\overline{5}_{M}$ orbit must be precisely $4$. On the other hand, returning
to the list of orbits already specified, we have:%
\begin{align}
Orb(10_{M})  &  =t_{1},t_{2},t_{3},t_{5}\\
Orb(\overline{5}_{M})  &  =t_{2}+t_{3},t_{5}+t_{1},...\\
Orb(5_{H})  &  =-t_{1}-t_{2},-t_{5}-t_{3},...\\
Orb(\overline{5}_{H})  &  =t_{1}+t_{4},t_{2}+t_{4},t_{5}+t_{4},t_{3}+t_{4}\\
Orb(X^{\dag})  &  =t_{2}-t_{4},t_{1}-t_{4},t_{5}-t_{4},t_{3}-t_{4}\\
Orb(N_{R})  &  =\pm\left(  t_{1}-t_{3}\right)  ,\pm\left(  t_{5}-t_{n}\right)
,\pm\left(  t_{2}-t_{p}\right)  ,....
\end{align}
Prior considerations have already fixed eight out of the ten weights available
for orbits of the $5$, leaving only $t_{1}+t_{3}$ and $t_{2}+t_{5}$ as
options. We can therefore uniquely fix the orbits for the $5$'s as:%
\begin{align}
Orb(\overline{5}_{M})  &  =t_{2}+t_{3},t_{5}+t_{1},t_{1}+t_{3},t_{2}+t_{5}\\
Orb(5_{H})  &  =-t_{1}-t_{2},-t_{5}-t_{3}\\
Orb(\overline{5}_{H})  &  =t_{1}+t_{4},t_{2}+t_{4},t_{5}+t_{4},t_{3}+t_{4}.
\end{align}

Note that the orbit for the $5_{H}$ is now of length two. Just as in the
previous subsection, we now ask which four cycles of $S_{4}$ leave this orbit
fixed. Of the available candidates:%
\begin{equation}
\text{Four Cycles in }S_{4}=\left\{
(1235),(1253),(1325),(1352),(1523),(1532)\right\}  \text{,}%
\end{equation}
only $(1325)$ and $(1523)$ preserve this orbit. Combined with the analysis
leading to line (\ref{1325}) showing that the group element $(13)(25)$ must
always be present, we conclude that:%
\begin{equation}
(13)(25),(1325)\in G_{8}^{Maj(3)}\text{.}%
\end{equation}
In fact, these elements generate an order eight group since $(1325)\cdot
(1325)=(12)(35)\neq$ $(13)(25)$. Taking all available products between powers
of\ these two generators, the explicit elements are:%
\begin{equation}
G_{8}^{Maj(3)}=\left\{
id,(1325),(12)(35),(1523),(13)(25),(12),(15)(23),(35)\right\}  \text{.}%
\end{equation}
Just as in the other Majorana scenarios, the messenger fields can only
consistently embed in the $10\oplus\overline{10}$ of $SU(5)$. This can again
be traced to the larger size of orbit lengths in the Majorana scenario. It
therefore follows that in this case as well, the full list of orbits are:%
\begin{align}
G_{8}^{Maj(3)}  &  =\left\langle (13)(25),(1325)\right\rangle \simeq Dih_{4}\\
&  \text{Minimal Matter}\\
Orb(10_{M},Y_{10})  &  =t_{1},t_{2},t_{3},t_{5}\\
Orb(\overline{Y}_{\overline{10}}^{\prime})  &  =-t_{4}\\
Orb(\overline{5}_{M})  &  =t_{2}+t_{3},t_{5}+t_{1},t_{1}+t_{3},t_{2}+t_{5}\\
Orb(5_{H})  &  =-t_{1}-t_{2},-t_{5}-t_{3}\\
Orb(\overline{5}_{H})  &  =t_{1}+t_{4},t_{2}+t_{4},t_{5}+t_{4},t_{3}+t_{4}\\
Orb(X^{\dag})  &  =t_{2}-t_{4},t_{1}-t_{4},t_{5}-t_{4},t_{3}-t_{4}\\
Orb(N_{R})  &  =\pm\left(  t_{1}-t_{3}\right)  ,\pm\left(  t_{5}-t_{2}\right)
,\pm(t_{5}-t_{1}),\pm(t_{2}-t_{3})\text{.}%
\end{align}
In this case, note that there are no additional orbits for additional matter
charged under $SU(5)_{GUT}$ to localize.

Next consider extra matter corresponding to GUT\ singlets. As for all of the
Majorana scenarios, there is a single $U(1)$ invariant under the action of the
monodromy group. The dark chiral matter is given by the zero weight of the
adjoint of $SU(5)_{\bot}$ denoted by $Z_{PQ}$, and weights of the form
$t_{m}-t_{n}$. The extra matter candidates fill out the following orbit:%
\begin{align*}
G_{8}^{Maj(3)}  &  =\left\langle (13)(25),(1325)\right\rangle \simeq Dih_{4}\\
&  \text{Extra Singlet Orbits}\\
Orb(D_{(1)})  &  =\pm\left(  t_{1}-t_{2}\right)  ,\pm\left(  t_{3}%
-t_{5}\right)  \text{.}%
\end{align*}
The PQ charge assignments are then:%

\begin{equation}%
\begin{tabular}
[c]{|c|c|c|}\hline
Extra Neutral & $D_{(1)}$ & $Z_{PQ}$\\\hline
$U(1)_{PQ}$ & $0$ & $0$\\\hline
\end{tabular}
\ \ \ \ \ \ \ .
\end{equation}
\newpage
\bibliographystyle{ssg}
\bibliography{fgutsdark}

\end{document}